\documentclass[iop]{emulateapj}
\usepackage{graphicx}

\newcommand{\ie}{i.e.,\ }
\newcommand{\eg}{e.g.,\ }

\newcommand{\etal}{et~al.\ }
\newcommand{\ltsima}{$\; \buildrel < \over \sim \;$}
\newcommand{\simlt}{\lower.5ex\hbox{\ltsima}}
\newcommand{\gtsima}{$\; \buildrel > \over \sim \;$}
\newcommand{\simgt}{\lower.5ex\hbox{\gtsima}}
\newcommand{\kms}{km~s$^{-1}$}
\newcommand{\magsec}{mag arcsec$^{-2}$}

\def\rquart{$r^{1/4}$~}

\def\muv{$\mu_V$}
\def\mub{$\mu_B$}

\def\bmv{$B-V$}
\def\Lsun{L$_\sun$}
\def\Lsunv{L$_{\sun,V}$}
\def\Msun{M$_\sun$}
\def\Wp{W$^\prime$}
\def\ficl{$f_{\rm ICL}$}
\def\brackmue#1{$\langle \mu \rangle_{e,#1}$}

\begin{document}

\title{The Burrell Schmidt Deep Virgo Survey: Tidal Debris, Galaxy Halos, and Diffuse Intracluster Light in the Virgo Cluster}

\author{J. Christopher Mihos\altaffilmark{1},
        Paul Harding\altaffilmark{1},
        John J. Feldmeier\altaffilmark{2},
        Craig Rudick\altaffilmark{3},\break
        Steven Janowiecki\altaffilmark{4},
        Heather Morrison\altaffilmark{1},
        Colin Slater\altaffilmark{5} and
        Aaron Watkins\altaffilmark{1}}

\email{mihos@case.edu, paul.harding@case.edu,
jjfeldmeier@ysu.edu,
craigrudick@gmail.com, 
steven.janowiecki@uwa.edu.au, heather@vegemite.case.edu,
ctslater@uw.edu, aew54@case.edu}

\altaffiltext{1}{Department of Astronomy, Case Western Reserve University,
Cleveland, OH 44106, USA}

\altaffiltext{2}{Department of Physics and Astronomy, Youngstown State University,
Youngstown, OH 44555, USA}

\altaffiltext{3}{University of Kentucky, Lexington, KY 40506, USA}

\altaffiltext{4}{International Centre for Radio Astronomy Research (ICRAR), University of Western Australia, 35 Stirling Highway, Crawley,
WA 6009, Australia}

\altaffiltext{5}{Department of Astronomy, University of Washington, Seattle, WA 98195, USA}

\begin{abstract}

We present the results of a deep imaging survey of the Virgo cluster of
galaxies, concentrated around the cores of Virgo subclusters A and B.
The goal of this survey was to detect and study very low surface
brightness features present in Virgo, including discrete tidal features,
the faint halos of luminous galaxies, and the diffuse intracluster
light (ICL). Our observations span
roughly 16 degrees$^{2}$ in two filters,
reaching a 3$\sigma$ limiting depth of \mub=29.5 and \muv=28.5 \magsec.
At these depths, our limiting systematic uncertainties are
astrophysical: variations in faint background sources as well as
scattered light from galactic dust. We show that this dust-scattered
light is well-traced by deep far-infrared imaging, making it possible to
separate it from true diffuse light in Virgo. We use our imaging to
trace and measure the color of the diffuse tidal streams and
intracluster light in the Virgo core near M87, in fields adjacent to the
core including the M86/M84 region, and to the south of the core around
M49 and subcluster B, along with the more distant \Wp\ cloud around
NGC~4365. Overall, the bulk of the projected ICL is found in the Virgo
core and within the \Wp\ cloud; we find little evidence for an extensive
ICL component in the field around M49. The bulk of  the ICL we detect is fairly red
in color (\bmv=0.7--0.9), indicative of old, evolved stellar
populations. Based on the luminosity of the observed ICL features in the
cluster, we estimate a total Virgo ICL fraction of 7--15\%. This value
is somewhat smaller than that expected for massive, evolved clusters,
suggesting that Virgo is still in the process of growing its extended
ICL component. We also trace the shape of M87's extremely boxy outer
halo out to $\sim$ 150 kpc, and show that the current tidal stripping rate
from low luminosity galaxies is insufficient to have built M87's outer
halo over a Hubble time. We identify a number of previously unknown low
surface brightness structures around galaxies projected close to M86 and
M84. The extensive diffuse light seen in the infalling \Wp\ cloud around
NGC~4365 is likely to be subsumed in the general Virgo ICL component
once the group enters the cluster, illustrating the importance of group
infall in generating intracluster light. Finally, we also identify
another large and extremely low surface brightness ultra-diffuse galaxy,
likely in the process of being shredded by the cluster tidal field. With
the survey complete, the full imaging dataset is now available for
public release.

\end{abstract}


\section{Introduction}

The assembly of galaxy clusters is a hierarchical process, wherein
massive clusters grow over time through the accretion of smaller groups
and clusters of galaxies. As such, it is a messy process. In the low
velocity dispersion group environment, individual galaxies interact and
merge with one another, creating discrete tidal tails and plumes of
material expelled from the galaxies by the tidal forces involved. As
these groups then merge to form clusters of galaxies, more material is
stripped from the galaxies, and the previously-liberated tails and
plumes are mixed into the cluster environment. Over time, the continued
accretion of mass adds to this diffuse stellar component, building the
intracluster light (ICL) of the growing cluster (for recent reviews, see
Arnaboldi \& Gerhard 2010, Tutukov \& Federova 2011, Mihos 2015).

A large number of numerical simulations have detailed this complex
behavior of intracluster star production (\eg Willman \etal 2004,
Sommer-Larsen \etal 2005, Conroy \etal 2007, Murante \etal 2007, Rudick
\etal 2006, 2011, Puchwein \etal 2010, Martel \etal 2012, Watson \etal
2012, Contini \etal 2014, and Cooper \etal 2015). Taken together, these
simulations imply that ICL production is an inevitable consequence of
hierarchical structure formation, and has implications in many areas of
extragalactic astronomy, including the assembly of the cluster red
sequence, the evolution of large elliptical galaxies, the metallicity of
the intracluster medium, and the overall baryon fraction of galaxy
clusters.

However, while the ICL is an important stellar component to study,
historically it has proved difficult to detect, due to its low surface
brightness, which is significantly fainter than the brightness of the
night sky. Early observations (\eg de Vaucouleurs \& de Vaucouleurs
1970, Melnick \etal 1977, Thuan \& Kormendy 1977, Struble 1988, Vilchez-Gomez \etal 1994,
Bernstein \etal 1995) were able to detect the ICL, but were unable to
place strong constraints on its spatial distribution and total
luminosity. In the last decade, deeper imaging of galaxy clusters has
clearly shown the ICL in a wide variety of environments (Feldmeier \etal
2002, 2004b, Lin \& Mohr 2004, Gonzalez \etal 2005, Zibetti \etal 2005,
Krick \& Bernstein 2007, Toledo \etal 2011, Arnaboldi \etal 2012). These
observations have shown both the smoother diffuse light that is
centrally distributed, as well as the tidal tails and plumes from recent
galaxy interactions in these galaxy clusters. At higher redshifts, the
rapid $(1+z)^4$ cosmological surface brightness dimming makes it hard to
study the ICL at the faintest levels, but recent observations have begun
to probe clusters at intermediate redshifts of $z=0.5-1.0$ (Guennou
\etal 2012, Burke \etal 2012, 2015, Presotto \etal 2014, Montes \&
Trujillo 2014, Giallongo \etal 2014). Connecting ICL properties at high
redshift to those observed locally will yield important constraints on
cluster assembly and evolution over cosmic time.

Aside from the diffuse starlight itself, many individual intracluster
objects have been detected in nearby galaxy clusters, including
intracluster globular clusters (\eg Williams \etal 2007a, Lee \etal
2010, Peng \etal 2011, Durrell \etal 2014), intracluster red giant stars
(\eg Ferguson \etal 1998, Caldwell 2006, Williams \etal 2007b),
supergiant stars (Ohyama \& Hota 2013), and planetary nebulae (\eg
Arnaboldi \etal 1996, 2002, Feldmeier \etal 2004a, Gerhard \etal 2007,
Castro-Rodriguez \etal 2009, Ventimiglia \etal 2011, Longobardi \etal
2015b), intracluster novae (Neill \etal 2005) and supernovae (Gal-Yam
\etal 2003, Dilday \etal 2010, Sand \etal 2011, Barbary \etal 2012), and
intracluster H~II regions (Gerhard \etal 2002, Cortese \etal 2004, Sun \etal 2010). These
discrete tracers of the intracluster light allow for a more detailed
view of the stellar populations and formation history of the ICL,
through studies of the metallicity, age, and kinematics of the
individual sources.

Unfortunately, connecting the wide-field view of the deep ICL imaging
studies at low and high redshift to the detailed but narrower probes of
nearby clusters using individual intracluster tracers has proved
difficult. Quantitative comparisons are often plagued by systematic
uncertainties between the various observational techniques (see \eg
Williams \etal 2007b and Mihos \etal 2009 for examples), as well as the
ambiguity in creating a clean operational definition of ICL that can be
used in all observed cases and can be directly compared with theoretical
studies (Rudick \etal 2011 and Cooper \etal 2015). To bridge these gaps,
detailed observations of nearby galaxy clusters using both deep surface
photometry {\it and} imaging of discrete ICL populations are needed,
along with kinematic followup of the discrete tracers to connect the
dynamical properties of the ICL back to computational simulations of
cluster evolution.

The Virgo Cluster of galaxies is a prime target for such studies. Given
its close proximity ($\approx$ 16.5 Mpc; Mei \etal 2007), moderate
richness (Girardi \etal 1994), and well-studied structure (Binggeli
\etal 1987), Virgo is an ideal place to search for ICL, and many
detections of individual intracluster objects have been made there.
There is also a wealth of observational data on the Virgo cluster from
many surveys over a large number of wavelengths (see the review in
Ferrarese \etal 2012), including optical surveys such as the Virgo
Cluster Catalog (VCC; Binggeli \etal 1985) and the Sloan Digital Sky
Survey (SDSS; York \etal 2000), far-infrared studies from re-calibrated
{\sl IRAS \/} data (Miville-Deschenes \& Lagache 2005) and the more
recent {\sl Herschel} Virgo Cluster Survey (HeViCS; Davies \etal 2010,
2012), deep UV imaging from {\sl GALEX} (Boselli \etal 2011), radio
continuum from the VLA FIRST and NVSS surveys (Becker \etal 1995 and
Condon \etal 1998) , 21-cm HI mapping from ALFALFA (Giovanelli \etal
2005), and ROSAT coverage of the cluster in X-ray (B\"ohringer \etal
1994). Most recently, the Next Generation Virgo Cluster Survey
(Ferrarese \etal 2012) has performed a detailed optical survey of 104
deg$^{2}$ of the Virgo cluster in the $u^{*}griz$ photometric bands,
providing a rich dataset for connecting the large scale distribution of
Virgo's diffuse ICL to the cluster's galaxy and globular cluster
populations.

The overall structure of the Virgo cluster was originally mapped out in
detail by Binggeli \etal (1987, and references therein), building on
earlier work by others (\eg Shapley \& Ames 1930; de Vaucouleurs 1961).
In their study, Binggeli \etal (1987) noted that Virgo had a complex
structure, with two major subclusters: subcluster A, centered around the
giant elliptical M87, and subcluster B to the south, centered around the
giant elliptical M49. A number of smaller subclusters were also
identified, including subcluster C, surrounding the elliptical galaxies
M59 and M60, along with the M, W, and \Wp\ ``clouds'' of galaxies. This
rich and complex dynamical structure argues that Virgo continues to
assemble today, making it an ideal target for studying the on-going
generation of intracluster light across a variety of local environments.

However, one of Virgo's strongest advantages for work like this --- its
close proximity --- also makes deep surface photometry of the cluster a
challenge. In order to reach surface photometry depths of 1\% of the
dark night sky (\muv $\sim$ 27 \magsec) over angular scales of several
degrees requires a telescope that delivers an extremely wide field of
view, along with well-controlled flat-fielding and minimal scattered
light. Furthermore, the observational demands of mapping a cluster like
Virgo that spans such a large area of the sky typically requires a large
amount of observing time on a telescope dedicated to the task. As a
result, very few attempts have been made to map the low surface
brightness features across the Virgo cluster on large angular scales.
Early studies employed photographic imaging (\eg Malin 1979; Malin \&
Hadley 1999; Katsiyannis \etal 1998) and revealed clear signs of tidal
debris around many of the luminous galaxies of Virgo, but these studies
were largely qualitative, due to the difficulty in photometric
calibration of photographic plates.

To rectify this situation and provide deep wide-field mapping of the
Virgo cluster, we have conducted a survey of diffuse light in Virgo
using Case Western Reserve University's 0.6/0.9m Burrell Schmidt
telescope. The wide field of view and closed-tube design of this
telescope makes it ideal for deep surface photometry over large areas,
and we have made extensive upgrades to the telescope and detectors to
further improve its imaging capabilities. Our main goals for the survey,
which spanned the years 2004--2011, were to detect and measure the
properties of Virgo's diffuse ICL across the main subclusters A and B,
including individual tidal streams around galaxies as well as the more
diffuse cluster ICL and stellar halos surrounding its massive
ellipticals. The survey used two photometric passbands, delivering
accurate surface photometry down to a 3$\sigma$ limiting surface brightness of
\muv=28.5 \magsec\ and \bmv\ colors to \muv=27.5 \magsec. A variety of
results from our survey have been published in earlier papers, including
the original deep imaging of ICL in the Virgo core (Mihos \etal 2005), a
comparison between the diffuse ICL and intracluster planetaries (Mihos
\etal 2009), color mapping of the core ICL (Rudick \etal 2010), a study
of accretion signatures in Virgo ellipticals (Janowiecki \etal 2010), a
photometric study of the halo of M49 (Mihos \etal 2013a), and the
discovery of large and extremely low surface brightness ultra-diffuse
galaxies (UDGs) in Virgo (Mihos \etal 2015).

In this paper, we present the imaging over the full area of the Burrell
Schmidt survey in both photometric bands, and give a detailed report on
a variety of low surface brightness features not yet reported elsewhere.
In Section 2, we outline the survey's observational strategy and data
reduction techniques (with complete details described more thoroughly in
the accompanying Appendix). Section 3 describes our methodology for
discriminating between Virgo intracluster light and a significant
source of contamination in deep imaging --- diffuse structure arising
from scattering of Milky Way starlight off of foreground Galactic
dust. Section 4 describes the properties of the diffuse starlight found
in a variety of fields throughout Virgo, including areas around M87 and
M86 in subcluster A, around M49 in subcluster B, and around NGC~4365 in
the \Wp\ cloud. We also report the discovery of a new Virgo ultradiffuse
galaxy, likely in the process of being tidal shredded by Virgo's tidal
field. Finally, in Section 5, we discuss our findings in the context of
Virgo's overall dynamical evolution. Throughout this paper, we adopt a
mean distance to Virgo of 16.5~Mpc (Mei \etal 2007); at this distance
1\arcsec\ subtends 80 parsecs, while 1\arcmin\ subtends 4.8 kiloparsecs.
We also report all magnitudes in the Vega system, and give surface brightnesses
in units of \magsec.

\section{Observations and Data Reduction}

Here we describe briefly our observing and data reduction strategies;
the interested reader can find a more thorough description of these
issues in the Appendix. The imaging was taken over the course of 107
nights, during seven observing seasons from 2004--2011, using the 24/36
inch Burrell Schmidt (hereafter, the Burrell) at KPNO. The telescope's
well-baffled, closed tube design, its aggressively anti-reflection
coated filters and dewar window, and the wide field of view of its CCD
imager (originally 0.825\degr$\times$1.65\degr, now
1.65\degr$\times$1.65\degr), all make the telescope ideal for performing
deep wide-field surface photometry, and we have made a number of
modifications to the telescope/camera system over the course of the
project to take advantage of these strengths. The telescope used a SITe
ST-002A 2048$\times$4096 pixel CCD (0.825\degr$\times$1.65\degr) during
the first four observing seasons, which was later upgraded to a STA0500A
4096$\times$4096 pixel CCD from the University of Arizona's Imaging
Technology Lab (1.65\degr$\times$1.65\degr) for the last three seasons.
Both CCDs had 15$\mu$m pixels, yielding pixel scales of 1.45 arcsec\
pixel$^{-1}$.

Our imaging was taken through two filters: Washington $M$, similar to
Johnson $V$ but $\sim$ 250\AA\ bluer (which cuts out variable night sky
emission lines in the red), along with a modified (slightly bluer)
Johnson $B$. We chose the filters to balance the signal of the ICL
(likely peaking in the red due to its expected old stellar populations)
with the noise of the sky (reduced in the blue due to the absence of
variable night sky emission lines). To further reduce sky noise and
scattered light from the moon and neighboring cities, we conducted
imaging only under moonless photometric conditions where the variation
in frame-by-frame photometric zeropoints (after correction for airmass
and nightly zeropoint changes) was typically 0.01 magnitudes. During
each season, a particular portion of the Virgo Cluster was targeted, and
a series of 30--100 images of the field was taken in an individual
filter, with the images dithered by up to half a degree relative to one
another. Flat fielding was achieved by taking a series of $\sim$ 30--70
night sky images bracketing the Virgo exposures in time and hour angle,
and combining these images to produce a ``super sky flat'' for each
season. Exposure times for both the Virgo fields and the blank night
skies were 900s in $M$ and 1200s in $B$, yielding sky backgrounds of
900--1500 ADU and 700--1000 ADU in $M$ and $B$, respectively. A total of
461 Virgo images were taken in $M$ over the first five seasons, and 173
images in $B$ over the final two seasons (with a comparable number of
blank sky images for flat fielding). A log of observations is given in
Section A.2 and Table~\ref{obslist} of the Appendix.

With the data obtained, we then began the process of carefully reducing
each image, registering each image spatially, and combining the images
into a stacked mosaic image. We did this using the IRAF\footnote{IRAF is distributed by the National Optical
Astronomy Observatory, which is operated by the Association of
Universities for Research in Astronomy (AURA) under cooperative
agreement with the National Science Foundation.} computing environment
and our own software. During data reduction, we flat field each image by
its season's night sky flat; tests show that these night sky flats are
accurate to better than 0.5\%. After flat fielding, we calculate the
photometric solution for each image based on photometry of the $\approx
100-200$ bright ($g < 16$) stars from the Sloan Digital Sky Survey
(SDSS; Data Release 9; Ahn \etal 2012) typically found in each field. By
transforming the SDSS magnitudes to Johnson $B$ and $V$ using the
prescription of Lupton (described in the Appendix), our final mosaics
are tied to the standard Johnson system, and we hereafter refer to the
data as Johnson $B$ and $V$ imaging. From each reduced image, we next
subtract the extended wings of bright stars using techniques described
in Slater \etal (2009), followed by the subtraction of a sky plane, fit
after masking bright stars and galaxies in the field using IRAF's
\textsf{objmasks} task.

At this point, the individual images are all flattened and star- and
sky-subtracted. For each filter, the images are then spatially
registered (preserving the native 1.45\arcsec\ pixel$^{-1}$ image scale)
and scaled in flux to a common photometric zero point. The individual
images are then median-combined to construct the final $B$ and $V$
mosaics, with obvious artifacts (bad columns, internal reflections, etc)
masked during the process. At this point, we also correct the mosaics
for the very slight residual background (1.125 ADU and 0.125 ADU in $B$
and $V$, respectively) still remaining. We also adopt a photometric
solution for the mosaics which uses the common photometric zeropoint
used in scaling the images for the median combine and a mean color term
averaged over all seasons of the survey (see Table~\ref{photsol}).
Finally, as a last step in the data reduction, we re-run the masking
algorithm on the final mosaics, then spatially re-bin the mosaics to
increase signal to noise at the lowest surface brightness. The IRAF
\textsf{objmasks} task is used to mask high surface brightness sources such
as stars, background galaxies, and the inner regions of bright galaxies.
We then rebin the masked mosaics into $9\times9$ pixel
(13\arcsec$\times$13\arcsec) bins, calculating the median value in each
spatial bin. Bins with more than half their pixels masked are themselves
masked.

Assessing photometric depth for surface photometry is complicated, as it
depends on a variety of noise sources that operate on different spatial
scales, such as uncertainty in flat fielding and sky subtraction, the
proximity of bright stars and galactic dust (see Section 2), and the
spatial distribution of contaminants such as faint stars and background
galaxies. We assess our limiting surface brightness in a number of ways.
First, the variance in $9\times9$ pixel intensities measured in blank
regions of the mosaics is 1.2 ADU in both our $B$ and $V$ mosaics,
leading to a 3$\sigma$ (1$\sigma$) limits of \mub=27.5 (28.7) and
\muv=27.1 (28.3). However, this is a pessimistic limit, since surface
photometry is typically done over large spatial scales averaging over
many pixels. The variation in median intensity measured in 1\arcmin\
boxes clustered in 10\arcmin\ regions (which measures uncertainty due to
local background contaminants) is much smaller: 0.25 ADU in both
mosaics, giving 3$\sigma$ (1$\sigma$) limits of \mub=29.4 (30.6) and
\muv=28.6 (29.8). Finally, a measure of the variation in median
intensity measured in 10\arcmin\ regions scattered across the mosaic ---
a measure of global sky uncertainty --- is 0.2 ADU ($B$) and 0.4 ADU
($V$), similar to the local background uncertainty. While the
photometric limits for any {\it individual} object measured from these
images will depend on its size and local background, as a fiducial
number for expressing the limiting surface brightness of the survey, we
adopt characteristic 3$\sigma$ limits of $\mu_{B,lim}=29.5$ and
$\mu_{V,lim}=28.5$

\begin{figure*}[]
\centerline{\includegraphics[height=6.5truein]{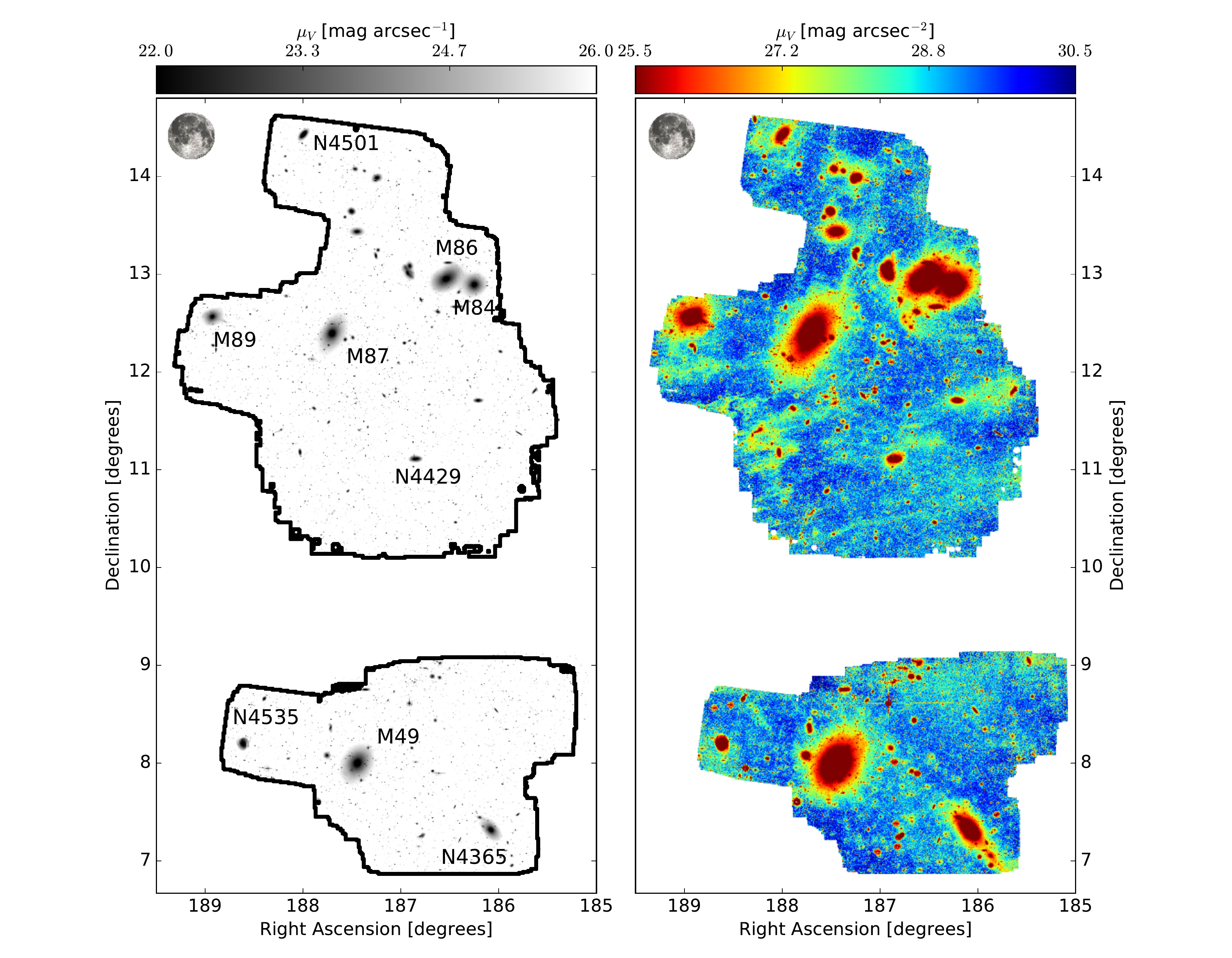}}
\caption{Final $V$ mosaic. The left panel shows a high surface
  brightness stretch of the V-band footprint of our Virgo survey, with
  several bright galaxies labeled. The right panel shows a low surface
  brightness stretch, where the extended halos of galaxies can be seen,
  along with diffuse intracluster light and galactic cirrus. North is up,
  east is to the left, and the full moon icon illustrates a half degree
  diameter scale (which projects to 145 kpc at d=16.5 Mpc).}
\label{Vmosaic}
\end{figure*}

\begin{figure*}[]
\centerline{\includegraphics[height=6.5truein]{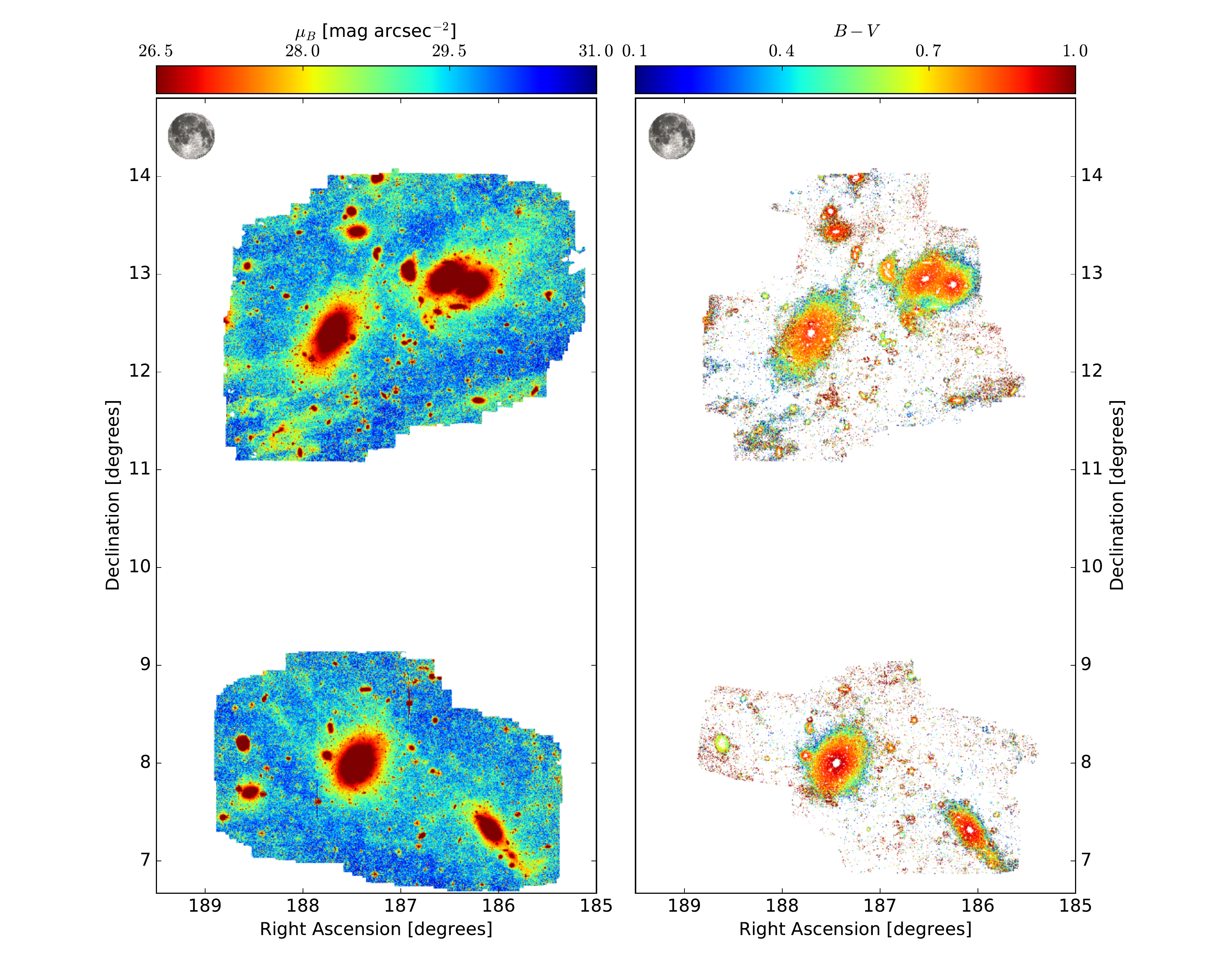}}
\caption{Left: final $B$ mosaic, stretched in intensity to show low
  surface brightness features. Right: $B-V$ colormap, showing pixel
  colors down to a surface brightness of \muv=28.25. North is up,
  east is to the left, and full moon shows a half degree diameter scale
  (144 kpc at d=16.5 Mpc).}
\label{Bcolor}
\end{figure*}

Our deep, binned Virgo mosaics are shown in Figures~\ref{Vmosaic} and
\ref{Bcolor}. In these images, the core of the cluster appears in the
upper (northern) region, with M87 near the center, M89 to the far east
(left) of the image, and the M86/M84 pair at the far west (right) of the
image. The lower (southern) region shows the bright elliptical M49 near
the center, with NGC~4365 to the southwest (lower right) and NGC~4535 to
the east (left). Because of our large scale dithering, the exposure time
varies across the mosaic; in the central regions, near M87 and M49 (the
core of the Virgo A and B subclusters respectively), the total exposure
times are typically 20 hours, but the exposure time drops towards the
edges of the mosaic. The median exposure time for the mosaic is 7.75
hours, and Figure~\ref{Vmosaic} only shows regions where we have more
than 5 images (1.25 hours) contributing to the mosaic. At this
threshold, the $V$ mosaic covers 16.7 square degrees. Our deep $B$
mosaic (Figure~\ref{Bcolor}a) surveys 15.3 square degrees with at least
5 images (1.67 hours) contributing to the mosaic; the median exposure
time is 7.7 hours, while the inner regions have typical exposure times
of 18 hours.

Figure~\ref{Bcolor}b shows the \bmv\ color map of the Virgo core where
again we restrict the map to those areas that are covered with at least
5 images in both $B$ and $V$. The total area covered in this color map
is 11.1 square degrees, and the image only shows colors in regions with
surface brightness brighter than \muv=28.5, since at lower
surface brightness the noise in the colors becomes extreme.

\begin{figure*}[]
\centerline{\includegraphics[height=6.5truein]{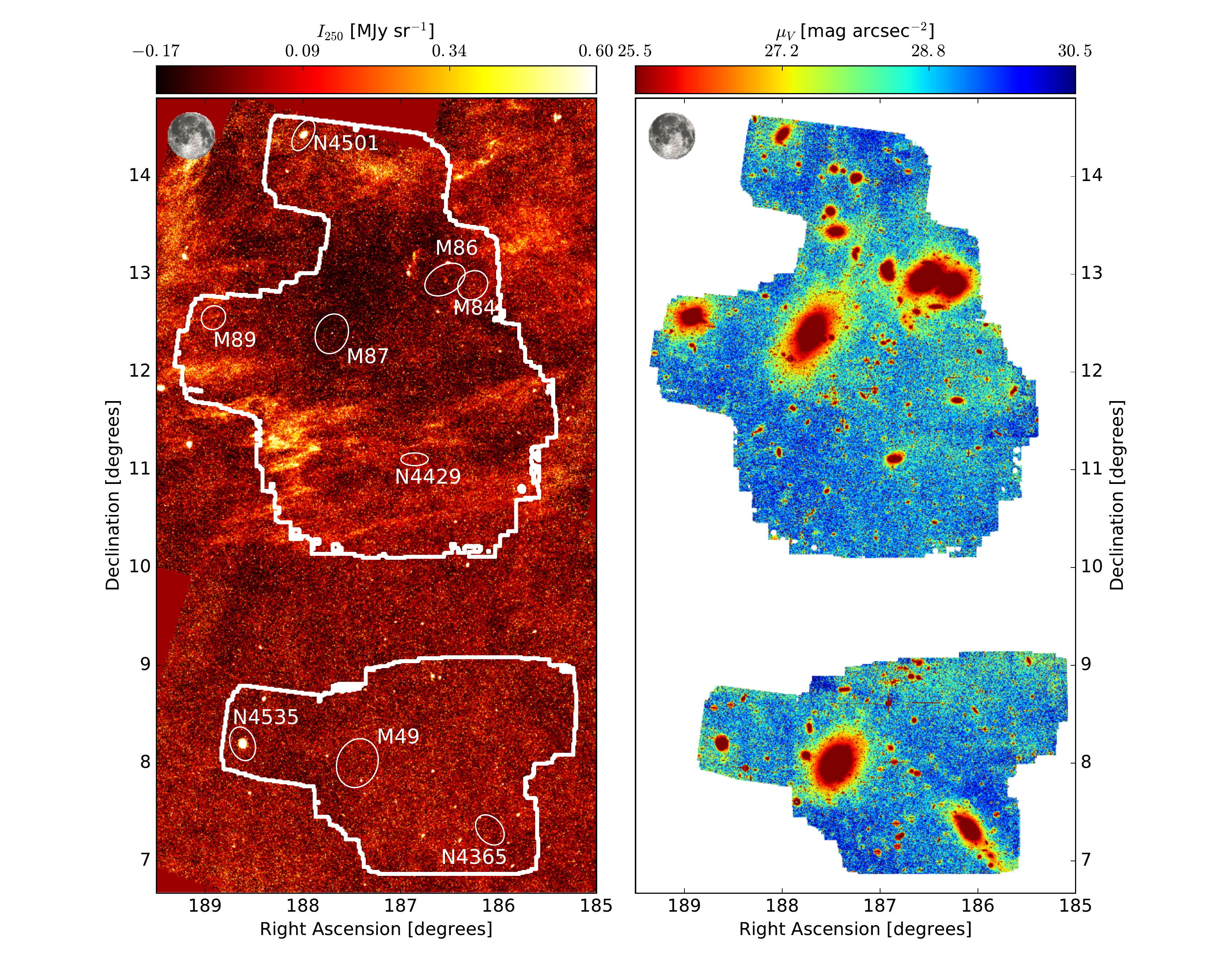}}
\caption{Left: Deep Herschel 250$\mu$ map of Virgo from the HeViCS
  project (Davies \etal 2010; 2012); the diffuse emission traces galactic 
  cirrus. The white contour shows the areal coverage of our V-band 
  imaging, and bright galaxies are marked. Right: A ``dust corrected" 
  view of Virgo, obtained by scaling and subtracting the HeViCS map 
  from our V mosaic. See text for details. North is up, east is to the left, 
  and full moon shows a half degree diameter scale
  (144 kpc at d=16.5 Mpc)..}
\label{Herschel}
\end{figure*}

\section{Detecting Contamination from Galactic Cirrus}

Figures \ref{Vmosaic} and \ref{Bcolor} show a great deal of diffuse
light spread throughout the Virgo Cluster. However, it is likely that
much of this light is not true intracluster light in the Virgo cluster,
but galactic light backscattered from dust in the disk of our own
galaxy, the so-called ``galactic cirrus'' (Low \etal 1984; see also the
discussion by Sandage 1976). Galactic cirrus apears in the images as a
diffuse, low surface brightness structure similar in many ways to the
diffuse intracluster light we are searching for, and there is
significant potential for confusion. Fortunately, in practice, two
particular properties of the cirrus may be useful in distinguishing it
from the ICL. First, with characteristic temperatures of T $= 7-20$~K
(\eg Low \etal 1984, Finkbeiner \etal 1999, Veneziani \etal 2010), the
cirrus radiates thermally in the far infrared, and can be detected in
{\sl IRAS\/} and {\sl Herschel} maps. Second, since the optical light is
scattered Milky Way starlight, it ought to be bluer on average than that
typically expected for intracluster light (Rudick \etal 2010), although
the presence or absence of extended red emission (ERE) in diffuse dust
clouds can lead to a significant scatter in optical colors (Guhathakurta
\& Tyson 1989, Witt \etal 2008, Brandt \& Draine 2012, Ieneka \etal
2013). Of course, the discriminatory power of the far infrared dust
emission is not completely unambiguous. For example, one could envision
stripping of young stars from a spiral galaxy yielding an ICL feature
with blue colors and some associated dust emission.

{\sl Herschel} imaging provides an ideal tool to probe the distribution
of galactic cirrus in the Virgo field (Bianchi \etal 2016). At 250$\mu$,
the SPIRE instrument provides a 18.1\arcsec\ beam imaged onto 6\arcsec\
pixels, comparable in spatial scale to our rebinned deep optical maps.
In Figure~\ref{Herschel}a we show the deep 250$\mu$ map of Virgo from
the {\sl Herschel} Virgo Cluster Survey (HeViCS; Davies \etal 2010;
2012), covering the same area as our deep optical images in Figures
\ref{Vmosaic} and \ref{Bcolor}. The image is dominated by a ring-like
distribution of dust in the northern half, with the core of Virgo being
viewed through a fortuitous hole in the dust. The southern half of the
image, extending down to fields surrounding M49, appears largely devoid
of dust. On smaller scales, the 250$\mu$ map shows many features which
correspond to features in our optical imaging, most noticeably in
regions south of M89, southeast of M87, southwest of M84/M86, and in the
northernmost part of the image.

Using the {\sl Herschel\/} map, we can make a simple attempt to correct
our optical image for this contamination from galactic cirrus. Studies
of high-latitude, optically thin dust clouds (Witt \etal 2008, Ienaka
\etal 2013) show a linear relationship between the optical flux and the
{\sl IRAS\/} 100$\mu$ emission. If such a model holds for the features
in our deep optical imaging, we can attempt to remove the cirrus from
our image by subtracting a suitably scaled version of the 250$\mu$ map
from our $V$-band mosaic. We do this very simply, via trial-and-error
subtraction using different scale factors for the 250$\mu$ map, until
we achieve a satisfactory subtraction. Converting our surface
brightnesses to linear flux densities (in MJy/sr) using the photometric
transformation from Johnson to AB magnitudes given in Frei \& Gunn
(1994), we find that a scale factor of $I_\nu({\rm
V})/I_\nu(250\mu) = 3.5\ (\pm 0.5)\times10^{-3}$ does a reasonable
job of removing the cirrus without significantly oversubtracting the
image. The resulting ``cirrus-subtracted'' optical map is shown in
Figure~\ref{Herschel}b.

The {\sl Herschel\/}-subtracted map shows clearly that much of the
diffuse structure we see in the raw optical maps comes from material
with associated far-infrared emission --- likely galactic cirrus.
Nonetheless, many features remain, including nearly all the diffuse
features in the core described by Mihos \etal (2005). A notable
exception is the diffuse plume north of NGC~4435/8, which has previously
been studied by Cortese \etal (2010) and shown to be associated not only
with far-infrared flux but also HI at galactic velocities --- all the
signatures expected for galactic cirrus.

As a secondary check, we also use our unsubtracted optical maps to
examine the colors of the cirrus in regions southeast of M87, where the
cirrus dominates the low surface brightness structure. We use the
differential photometry technique described in Rudick \etal (2010),
wherein we lay down polygonal apertures at various spots in this region,
and calculate the median pixel intensity values in each region after
masking out discrete sources. We do the same for surrounding background
fields to subtract off a local background level from the region
photometry. Finally, we use the background-corrected region photometry
to calculate the characteristic surface brightness and color of the
regions. We find the \bmv\ color spans values from 0.4 to 0.8, with
typical uncertainties of several tenths (see Rudick \etal 2010 for a
detailed discussion of the color uncertainties at low surface
brightness). While galactic cirrus can span a wide range of optical
colors (\eg Witt \etal 2008), the \bmv\ colors we measure are indeed
somewhat bluer than expected for the old stellar populations likely to
characterize the bulk of the ICL in Virgo (\bmv$\sim 0.7-1.0$; Rudick
\etal 2010), but by themselves do not discriminate against somewhat
younger ICL populations. However, the combination of bluer colors and
infrared emission are a likely signature of galactic cirrus.

\section{Low Surface Brightness Structures in Virgo}

In this section we identify and discuss low surface brightness features
present in the data that do {\it not} have counterparts in the {\sl
Herschel\/} 250$\mu$m map --- i.e., those that are unassociated with
galactic cirrus, and likely to be diffuse light in Virgo itself. While a
number of these features were identified in our previous imaging
releases (Mihos \etal 2005 [M+05], Janowiecki \etal 2010 [J+10], Rudick \etal 2010,
Mihos \etal 2013a), our wider imaging area presented here has revealed a
number of new features and also permits color measurements for several
of the previously identified ones. In addition, detection in both the $B$
and $V$ mosaics --- imaging taken across several observing seasons using
different instrumental setups --- gives us added confidence in
identifying structures which were marginal detections using previously
published data in a single mosaic.

\begin{figure*}[]
\centerline{\includegraphics[height=2.9truein]{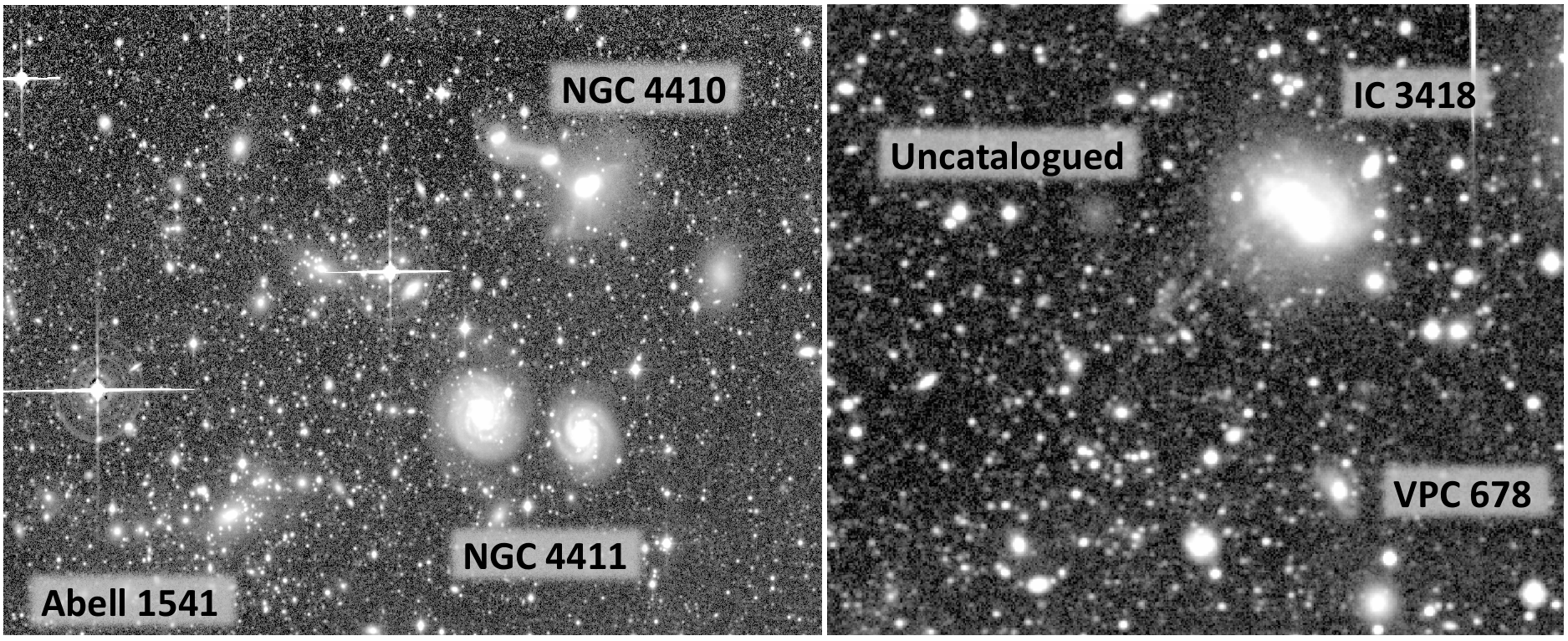}}
\caption{Two fields showing the variety of diffuse features we see in
the imaging data. Left: 35\arcmin$\times$24\arcmin\ cutout of the V mosaic
showing diffuse light in the distant cluster Abell 1541 ($z=0.09$), the
tidal plumes associated with the background galaxy group around NGC~4410
($cz=7254$ \kms), and the faint outer spiral arms of the Virgo galaxy
pair NGC~4411. Right: 12\arcmin$\times$10\arcmin\ cutout of a field near the
ram presssure stripped galaxy IC~3418 (Hester \etal 2010, Kenney \etal 
2014), showing the galaxy's diffuse outer regions and its star forming knots
extending to the southeast, an uncatalogued low surface brightness
galaxy at unknown redshift, and extensive tidal debris surrounding the
background galaxy VPC 678 ($z=0.18$). These fields represent only $\sim$
1\% of the full V mosaic; similar structures abound throughout the other
99\% of the survey area.}
\label{LSBrogues}
\end{figure*}

In examining the deep mosiacs, a wide variety of low surface brightness
features can be found on varying spatial scales, from the very extended
streams emanating to the NW of M87 (M+05, Rudick \etal 2010,
Figures~\ref{Vmosaic} and \ref{Bcolor}) to a myriad of small LSB objects
that span only a handful of pixels. We show in Figure~\ref{LSBrogues}
several examples of the variety of structures visible in our mosaics.
The left panel shows a 35\arcmin$\times$24\arcmin\ field near the top of
the southern survey area. Some of the LSB structures are associated with
known background sources, including ICL in the distant background galaxy
cluster Abell 1541 ($z=0.09$, Slinglend \etal 1998) and the tidal tails
associated with the background galaxy group around the radio galaxy
NGC~4410 ($cz=7254$ \kms, Hummel \etal 1986, Donahue \etal 2002). We
also trace the faint outer arms of the Virgo spiral NGC~4411 to a
distance of 3.6\arcmin\ (17.3 kpc). The right panel shows a
12\arcmin$\times$10\arcmin\ field near the Virgo dwarf galaxy IC~3418.
This tidally disturbed galaxy is in the process of being ram pressure
stripped, and sports a streamer of star forming knots extending to the
southeast (Hester \etal 2010, Kenney \etal 2014). To the east of IC~3418
is a small ($r=15$\arcsec), uncatalogued nucleated LSB galaxy, while to
the south, the background galaxy VPC 678 ($z=0.18$) shows extended tidal
tails. In these two cutouts, only NGC~4411 and IC~3418 are known Virgo
galaxies; the uncatalogued LSB galaxy has no redshift and may either be
one of the many LSB dwarfs in Virgo (\eg Ferrarese \etal 2016) or a
background source, while the other systems are all much more distant.
These fields represent only $\sim$ 1\% of the survey area, and similar
features are seen throughout the rest of our data. Our intent in this
paper is not to create any sort of complete surface brightness limited
catalog of structures in Virgo, but instead to identify and discuss the
properties of the large LSB structures that are most likely to trace
dynamical events in the Virgo cluster. As such, for the remainder of
this study, we focus on these larger LSB structures and leave the
smaller features for later study.
 
For ease of discussion, we break up this section to focus on specific
subregions of Virgo. In what follows, we use the {\sl Herschel} map to
discriminate between galactic cirrus and diffuse light in Virgo, but due
to the quantitative uncertainty of the cirrus subtraction using the {\sl
Herschel} data, all photometric analysis is done on the {\it
unsubtracted} optical maps alone; we simply avoid regions of
cirrus contamination in our analysis. In describing the diffuse features
we find in Virgo, we detail surface brightnesses and, where possible,
measured \bmv\ colors. As mentioned previously, detailed error modeling
at low surface brightness is complicated, and depends on the global sky
uncertainty, the total extent of the ICL feature being measured, and the
fluctuation in background sources on spatial scales comparable to that
of the feature. For purposes of the discussion here, however, we note
that typical photometric uncertainties in surface brightness are
approximately 0.1 mag at \muv = 28.5, while \bmv\ colors can be probed
to a similar 0.1 mag uncertainty down to a surface brightness of
\muv$\sim$27.5. Finally, in describing the galaxies associated with each
field, we give their observed radial velocities for context. Unless
noted otherwise, these velocities are taken from the Sloan Digital
Sky Survey DR12 (Alam \etal 2015).

\subsection{The M87 Field}

\begin{figure*}[]
\centerline{\includegraphics[height=2.35truein]{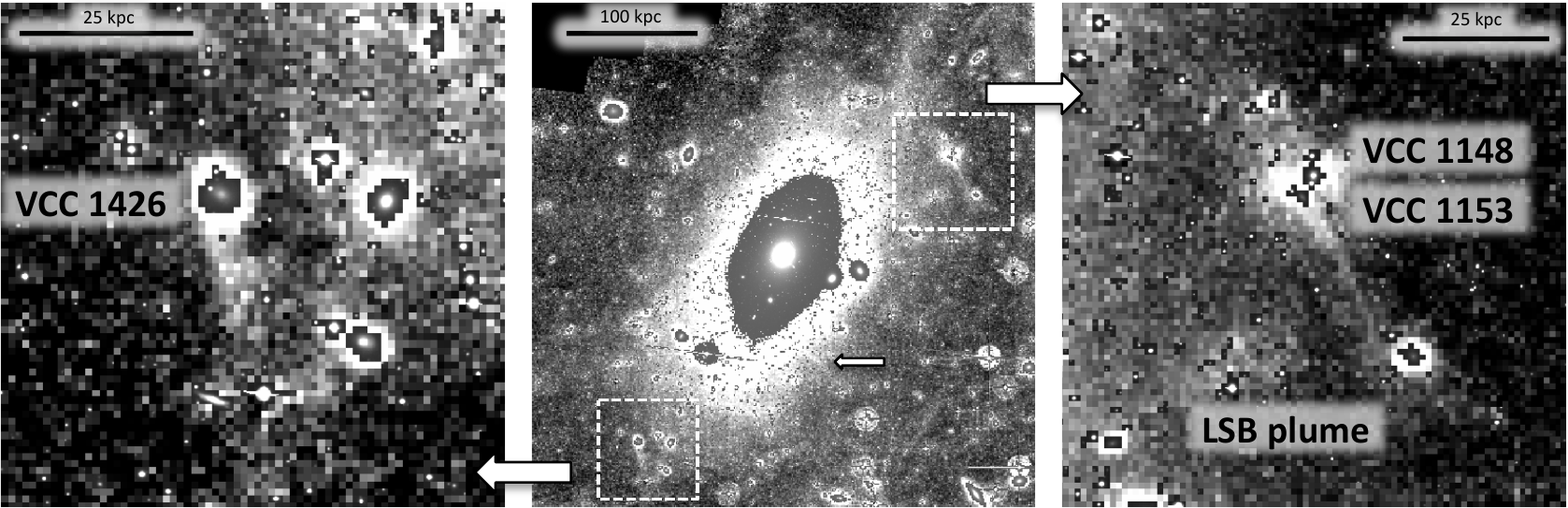}}
\caption{Streams around M87. The 1.3\degr$\times$1.3\degr\ center image
  shows M87 itself, with two subfields highlighted. The left panel shows a
  0.25\degr$\times$0.25\degr\ subfield around VCC~1426, while the right
  panel shows a 0.33\degr$\times$0.33\degr\ subfield around VCC~1153. In
  all three panels, at high surface brightnesses (\muv $<$ 26 \magsec) the
  imaging is shown at full 1.45 arcsec\ pixel{$^{-1}$}, while at lower
  surface brightnesses we show the 9$\times$9 median binned image using a
  remapped intensity scale to highlight faint structures. In the center
  figure, the arrow points to a small linear stream projecting out from
  M87's halo. See text for details.}
\label{M87streams}
\end{figure*}

In the core of the Virgo Cluster, the field around the giant elliptical
galaxy M87 (1284 \kms, Cappellari \etal 2011) hosts a rich assortment of
diffuse stellar features. Many of these were detailed in our earlier
papers, including M87's very extended and irregular stellar halo (M+05,
J+10; see also the discussion below) and the two long, linear streamers
seen projecting to the northwest from M87 (streams A and B in the
nomenclature of M+05). A closer examination of this field using our
updated mosaics shows that a number of the small galaxies in close
vicinity to the giant elliptical also sport tidal features suggestive of
ongoing tidal stripping. Figure~\ref{M87streams} shows these structures
--- all of which can be seen in both the $B$ and $V$ mosaics --- in more
detail. Lying $\sim$ 38\arcmin\ (182 kpc) southeast of M87, VCC 1426
(833 \kms) shows a low surface brightness tail (\muv=28.2, \bmv=0.6)
extending 3.9\arcmin\ (19 kpc) to the south. VCC~1426 is part of a trio
of small galaxies including IC~3461 (993 \kms) and IC~3466 (883 \kms)
which all lie within 6.5\arcmin\ of each other. The tight grouping in
velocity of these galaxies suggests that they may be a physical group,
in which case VCC~1426's tidal tail may come from slow interactions
between the group members rather than by tidal stripping from the
cluster potential or M87 itself. This stripping of material from galaxies
as they interact on group scales inside the larger cluster environment 
can enhance the production of ICL in clusters over that drawn out
by global tides and high speed encounters alone  (\eg Mihos
2004, Rudick \etal 2009).

Another set of streams can be seen 30\arcmin\ (144 kpc) northwest of
M87. One is a linear stream associated with VCC~1153 (844 \kms, Rines \&
Geller 2008), extending 5\arcmin\ SW of the galaxy. The stream has a
surface brightness of \muv=27.9 and color of \bmv=0.7. Projected very
close to this system is VCC~1148 (1417 \kms) with its own plume
(\muv=27.5, \bmv=0.8) extending 1.6\arcmin\ ENE. Unlike the situation in
the VCC 1426 group, the relatively large velocity difference between
these two galaxies makes it unlikely that these features arise from an
interaction between these galaxies; instead, it is likely due to
interactions with M87 or other galaxies in Virgo. Finally, $\sim$
7\arcmin\ SSW of the galaxy pair there is a broad, low surface
brightness plume (\muv=27.5, \bmv=0.8) without any obvious connection to
other galaxies in the field.

The streams highlighted here all come from low luminosity companion galaxies and
are themselves low luminosity streams. We also note one additional
stream (marked with a small arrow in Figure~\ref{M87streams}), due south
of M87 and extending outwards parallel to the minor axis of the galaxy. This
stream has no clear association with any of M87's existing companions
and is likely the remnant of a tidally shredded dwarf. With
characteristic surface brightnesses of \muv $\sim$ 27.5--28.0, and
covering a few square arcminutes in size, these various tidal features
contain only a few $\times 10^7$ \Lsunv\ of starlight, in contrast to the
much more extended plumes and streams identified by M+05 which have
larger luminosities of $\sim 10^8$ \Lsunv\ (J+10).

While the luminosity of the tidal streams is relatively low, in the
active dynamical environment that characterizes the core of the Virgo
cluster, their lifetimes will be short as well, such that the implied
rate of ICL production could yet be significant. Dynamical models of ICL
evolution (Rudick \etal 2009) show that coherent streams only survive
1--2 crossing times before being disrupted, which corresponds to roughly
300 Myr in the outer halo of M87 (at $r=75$ kpc). With a total stream
luminosity of $\sim 5\times10^8$ \Lsunv\ (including the features
identified in J+10), if streams are continuously created and destroyed,
this argues for an instantaneous (\ie timescales $<$ 0.5 Gyr) luminosity accretion rate
from low mass satellites onto the M87 halo of
$\sim$ 1 \Lsunv\ yr$^{-1}$. M87's halo contains appreciable luminosity
even at these radii --- integrating the measured luminosity profile
(J+10) beyond 75 kpc to the outermost measured point
(at r=175 kpc) yields a total luminosity of $2\times10^{10}$ \Lsunv. This
is significantly more than would be expected from the current accretion
rate, arguing that infall of small satellites is not the {\it dominant}
mechanism for building the outer halo of Virgo's central galaxy. This
conclusion is consistent with models for the growth of cD galaxies,
which show that central galaxies assemble late (at $z<1$) from mergers of more
massive systems (\eg Jord{\'a}n et al. 2004; De Lucia \& Blaizot 2007;
Guo \etal 2011; Tovmassian \& Andernach 2012).

\begin{figure*}[]
\centerline{
\includegraphics[height=3.4truein]{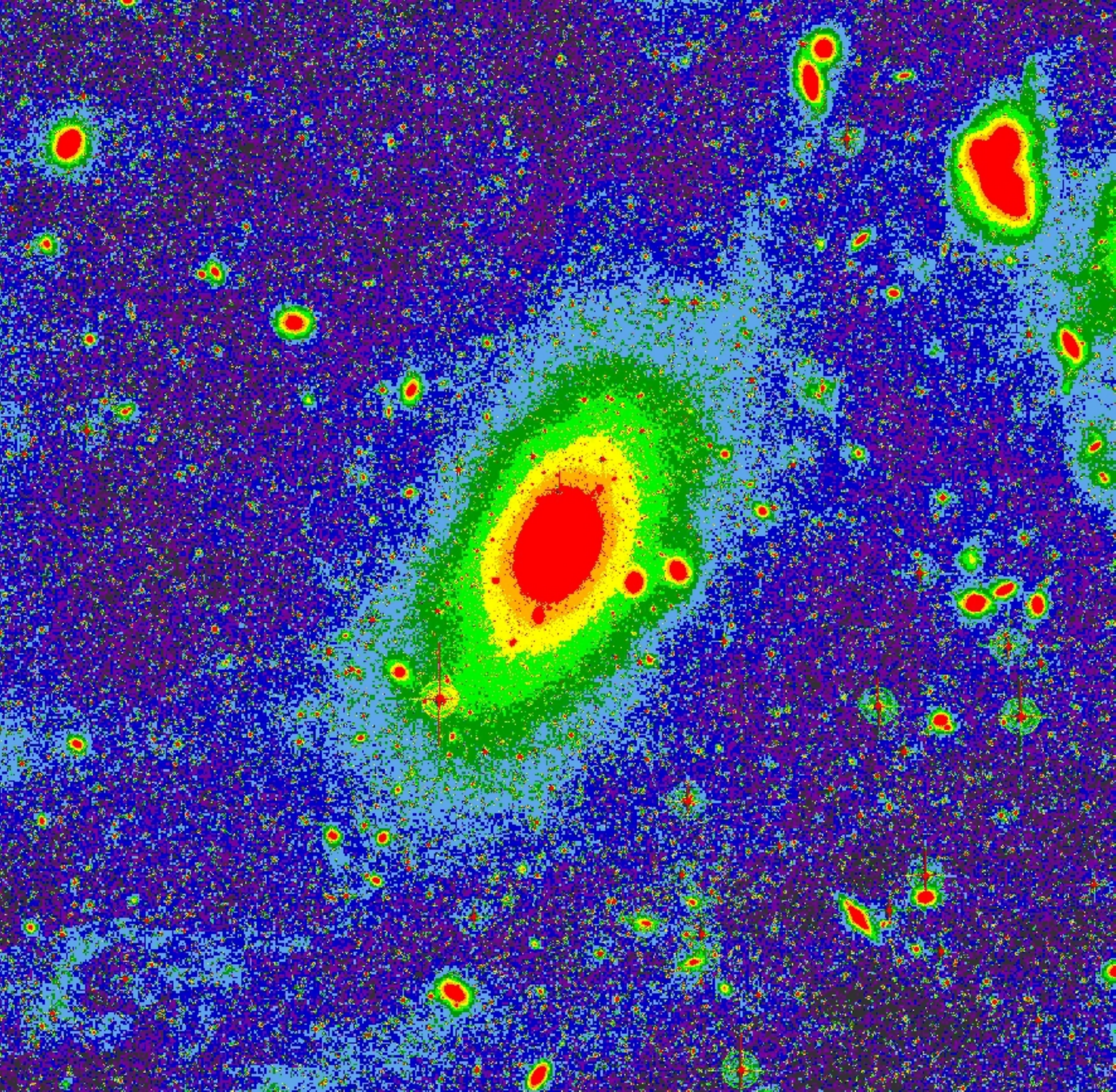}
\includegraphics[height=3.4truein]{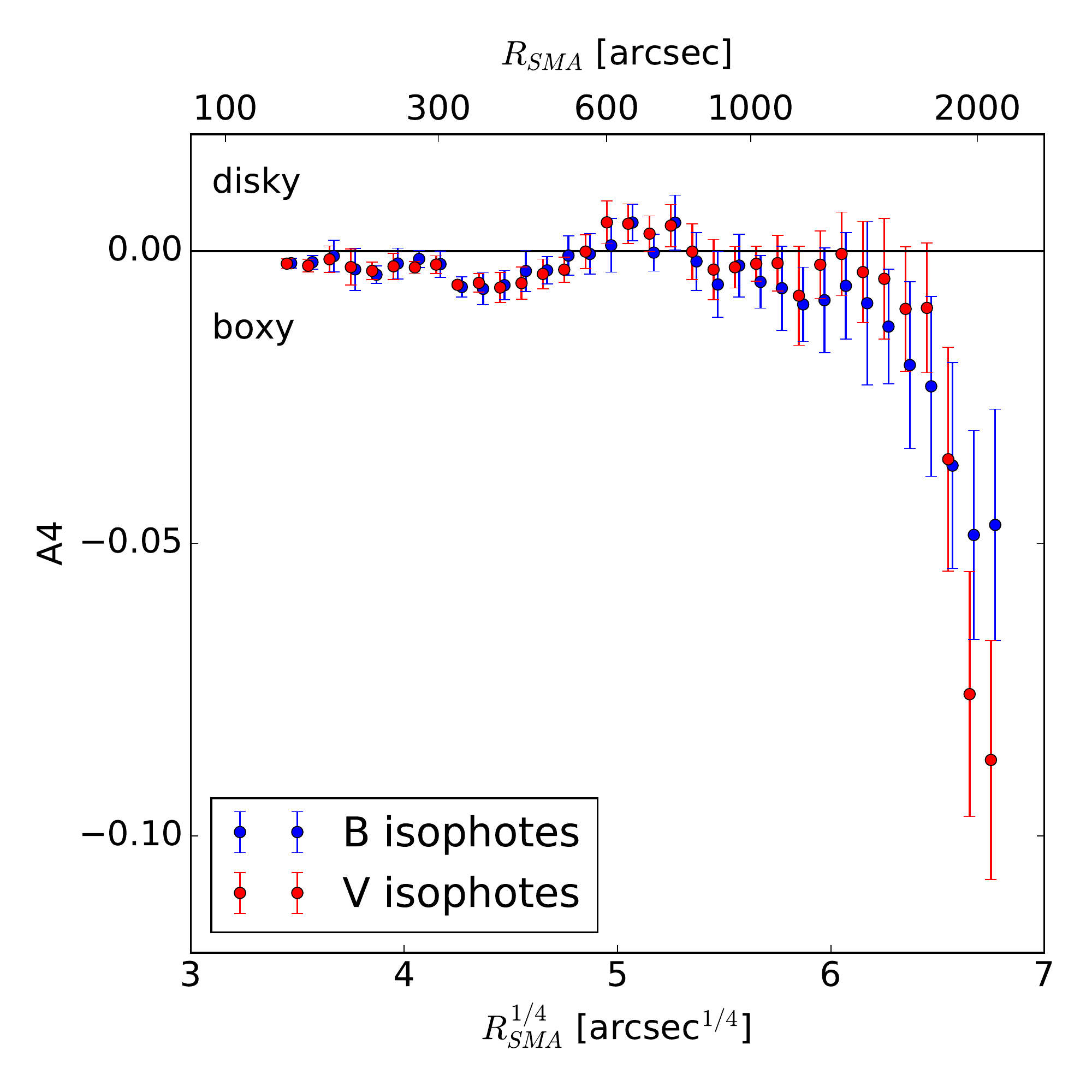}
}
\caption{Left: Deep $B$ band image of M87, 2\degr\ (550 kpc) on a side,
  color-coded by surface brightness: $\mu_B < 25$ (red), $25 < \mu_B <
  25.75$ (orange), $25.75 < \mu_B < 26.5$ (yellow), $26.5 < \mu_B < 27.25$
  (light green), $27.25 < \mu_B < 27.75$ (dark green), $27.75 < \mu_B <
  28.5$ (light blue), $\mu_B > 28.5$ (dark blue). Right: Amplitude of the
  $\cos(4\theta)$ term in the isophotal analysis, plotted as a function of
  semi-major axis distance.}
\label{M87boxy}
\end{figure*}

Indeed, signatures of more massive mergers may be lurking within
M87's halo. Kinematic studies of M87's globular clusters (Romanowsky \etal 2012) and
PNe (Longobardi \etal 2015a) suggest the recent accretion
of one or more systems with total mass $\sim$ a few $\times 10^9$ to $10^{10}$ \Msun.
Similarly, M87's isophotal structure hints at past accretion events as well.
 While the large extent of the galaxy's diffuse halo was first noted by
Arp \& Bertola (1969), Carter \& Dixon (1978) were the first to point
out the asymmetric nature of the outer isophotes. Weil, Bland-Hawthorne,
\& Malin (1997) later used deep photographic imaging to trace out a
diffuse fan of material emanating from the SE of the galaxy, as well as
a smaller ``cap'' of material to the NW, indicative of a recent
accretion event. In an earlier analysis of our Virgo imaging data, we
showed a broad but faint plume of material north of M87 (Rudick \etal
2010), while Longobardi \etal (2015a) used the updated imaging described
here to argue that the NW isophotal structure (Weil \etal's ``cap,''
referred to in Longobardi \etal 2015a as the ``crown'') at 800\arcsec --
1200\arcsec\ trace the discrete accretion event implied by the PNe
kinematics.
 
Figure~\ref{M87boxy}a shows our $B$ imaging around M87, color coded by
surface brightness, where we see large scale systematic changes with
radius in M87's isophotal structure. The mild boxiness of the inner
regions of the galaxy transitions to a more disky morphology at
10\arcmin, where the surface brightnesses has dropped to \mub $\sim$ 27
(the light green isophotes of Figure~\ref{M87boxy}a. At larger radius,
the isophotes revert back to become extremely boxy; the large NW extent
of the isophotes, traced beyond the cap/crown structure to nearly
1800\arcsec\ (140 kpc) along the major axis balances the fan of material
to the SE, and gives rise to the overall symmetric shape of the galaxy's
boxy outer isophotes at low surface brightness (\mub=27.5; the light blue isophotes
in Figure~\ref{M87boxy}a. Here in the outer halo, the boxiness of the
isophotes is echoed in the boxiness of the spatial distribution of
M87's globular cluster population as well (Durrell \etal 2014). While
we note the presence of patchy galactic cirrus in the extreme southeast
portion of the field, an examination of the Herschel 250$\mu$m map
(Figure~\ref{Herschel}) shows no evidence that this contamination
extends into the outer isophotes of M87's halo studied here.

We quantify the boxiness of the isophotes by using IRAF's
\textsf{ellipse} task (Jedrzejewski 1987, Busko 1996) to fit elliptical isophotes to
M87 and measure the harmonic deviations of the isophotes from pure
ellipses. We show the amplitude of the $\cos(4\theta)$ harmonic as a
function of semi-major axis radius. in Figure~\ref{M87boxy}b. The sign
of this A4 coefficient measures isophotal shape, with positive values
indicating diskiness and negative values indicating boxiness.
To assess the uncertainty in the A4 terms, we ran 100 iterations of the
\textsf{ellipse} algorithm, varying task parameters such as the initial
fitting radius, radial step size, and input photometric parameters;
errorbars in Figure~\ref{M87boxy}b reflect the $1\sigma$ scatter in the
extracted A4 terms. In addition, we measure the isophotal shape
independently on each of the $B$ and $V$ images, obtaining consistent
results between the two images, and note that the V-band surface
brightness, ellipticity, and position angles, while not shown, are
consistent with our earlier analysis in J+10.

This quantitative analysis confirms the visual impression of boxiness
evident in the image. Between \rquart= 4--5 (4\arcmin--10\arcmin, or
$\sim$ 20--50 kpc) the A4 terms are slightly negative, indicative of a
somewhat boxy shape, then switch sign around 50 kpc, indicating
diskiness. This behavior is also clearly seen in the M87 surface
photometry of Kormendy \etal (2009). The switch to disky isophotes can
be seen morphologically in Figure~\ref{M87boxy}a as traced by the light
green isophotes at $\mu_B \sim 27$. At even larger radius, \rquart$>6$,
or $r>100$ kpc, the A4 terms become strongly negative, reflecting the
extreme boxiness of the outermost isophotes. Such extreme boxiness is
atypical of the {\it inner} regions of bright ellipticals, which rarely
show A4 terms more negative than $-0.01$ (Bender 1989, Emsellem \etal
2011).

The connection between isophotal shape and assembly history of the
galaxy is complex. Numerical simulations of merging galaxies generally
show that boxiness is associated with major mergers (particularly dry
mergers), while unequal mass minor mergers, or mergers with significant
gaseous dissipation, tend to leave diskier remnants (\eg Khochfar \&
Burkert 2005, Naab \etal 2006). However, simple single merger models may
not be applicable to M87 --- as Virgo's central BCG it has likely
undergone a complex series of mergers with varying mass ratios, orbital
energies, and angular momenta. Simulations indicate that the growth of
cluster BCGs is marked by multiple accretions, with as much as 50\% of
the BCG mass being accreted since $z = 0.5$ in the form of intermediate
mass companions (with masses $> 10^{10}\ {\rm M}_\sun$; de Lucia \&
Blaizot 2007, Cooper \etal 2015). M87's extended halo is likely
therefore made up of an ensemble of material from past accretion events
that may not be fully relaxed, and the various structures seen at low
surface brightness (the NW streams, SE fan, N plume, and NW cap/crown)
may conspire to yield M87's varying A4 profile and extremely boxy outer
isophotes shown in Figure~\ref{M87boxy}. In this sense, M87's halo
morphology at low surface brightness is quite reminiscent of simulated
BCGs of Cooper \etal (2015), whose boxy halo morphology arises at least
in part from a superposition of variety of tidal debris structures from
many past accretion events.

\subsection{The M86/M84 Field}

\begin{figure*}[]
\centerline{\includegraphics[width=7.0truein]{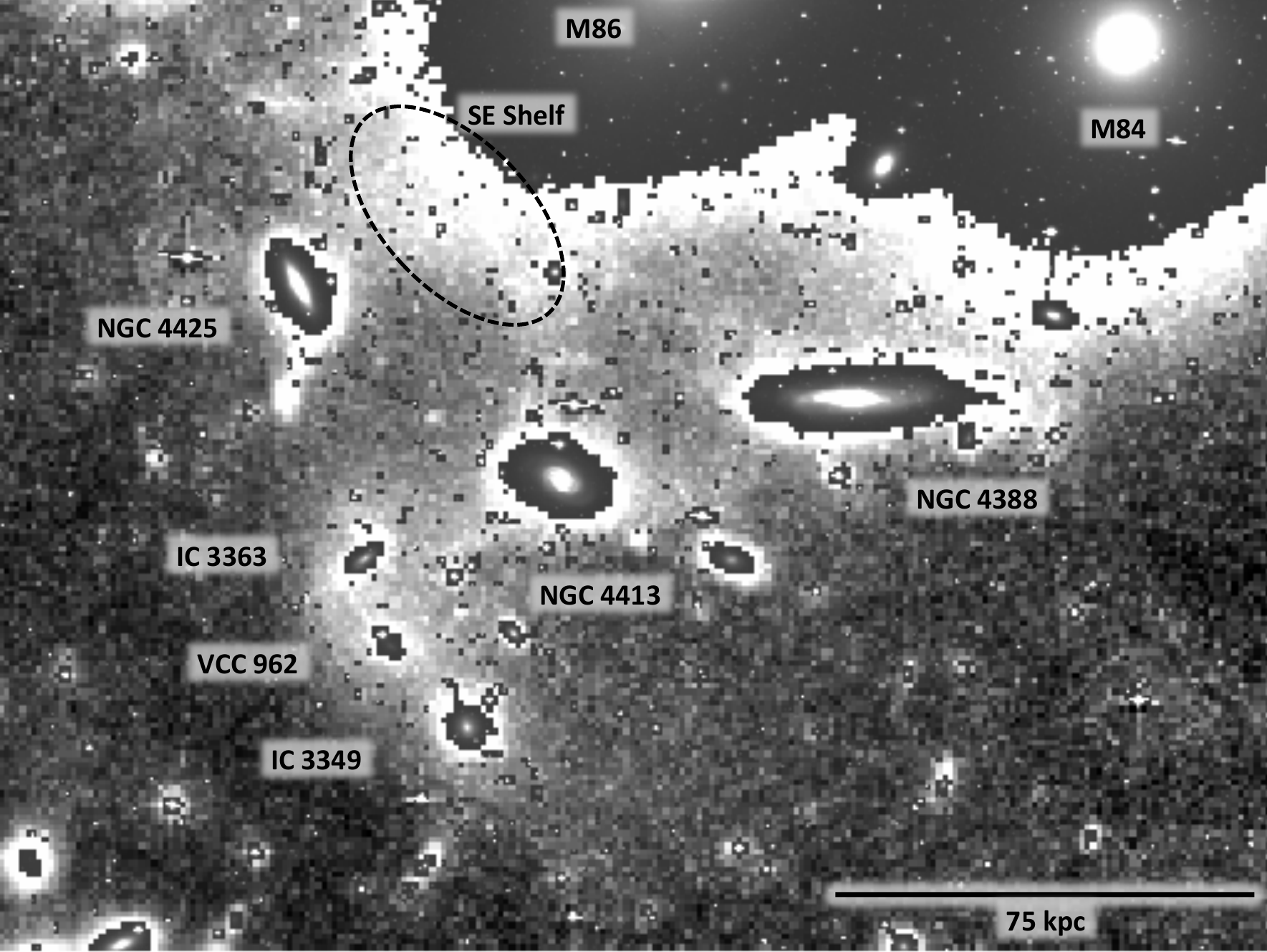}}
\caption{Deep V image of the M86/M84 field. The image covers a field of
  view 48\arcmin$\times$36\arcmin\ in size, with north up and east to the
  left. Galaxies discussed in the text are marked, as is the SE Shelf.}
\label{M86streams}
\end{figure*}

Another area of high projected galaxy density in Virgo is the region
containing the massive ellipticals M86 and M84, although the two
ellipticals themselves may be well separated along the line of sight
(\eg Mei \etal 2007). The presence of an extensive, diffuse ICL
component in this region is indicated by the large number of
intracluster PNe found in the field (Arnaboldi \etal 1996, Okamura \etal
2002), as well as by the discovery of an intracluster Type Ia supernova
(Smith 1981). As shown in J+10, the halos of both M86 and M84 contain a
number of small streams suggestive of tidal stripping of dwarf galaxies
orbiting the massive ellipticals. component in this region as well. Here
we concentrate on the features seen around other, smaller galaxies in
this field. Figure~\ref{M86streams} shows the region just south of M86
and M84, where many of the galaxies show diffuse features. While it is
tempting to interpret these features as signs of interaction with one
another, the galaxies in this field show a wide range of velocities,
from 88 \kms\ (NGC~4413; Rines \& Geller 2008) to 2513 \kms\ (NGC~4388),
and thus are not likely involved in close and slow encounters with one
another that might lead to the strongest tidal response.

The edge on disk galaxy NGC~4425 (1899 \kms) shows a bright plume
extending 3.5\arcmin\ to its south, originally identified in
photographic imaging by Malin (1994). The plume is brightest at its
southernmost tip, where it has a peak surface brightness of \muv=26.0
and color of \bmv=0.76. While tidal tails often show condensations of
stars and gas (\eg Hibbard \etal 1994, Duc \etal 2000, Weilbacher \etal 2000,
Lelli \etal 2015), in this case the tip of the plume has a fairly
constant surface brightness with no sign of any central condensation.
This morphology is somewhat suggestive of edge-brightening, where an
optically thin and curved tidal tail is observed edge on. At the point
where the tail curves back in projection along the line of sight the
observational path length through the tail is maximized and the
projected surface brightness thus increases. The position angle of the
tail is offset by 50\degr\ from the plane of the galaxy's disk, arguing
that the tail formed during an interaction that occurred out of the
galaxy's disk plane.

The other edge-on spiral galaxy in the field, NGC~4388 (2513 \kms), also
shows low surface brightness features extending from both sides of its
major axis. The system shows ionized emission extending 7\arcmin\ to the
northeast (Yoshida \etal 2002), as well as an even more extended HI tail
stretching nearly 30\arcmin\ along the same direction, likely due to ram
pressure stripping in the cluster environment (Oosterloo \& van Gorkom
2005). At the base of the HI tail, near the region of ionized emission,
we also detect diffuse light at \muv=27.5 with a broadband color
(\bmv=0.66), somewhat bluer than the tidal features we see elsewhere in
Virgo. On the other side of the disk, the western plume is even bluer,
with colors ranging from \bmv=0.5--0.6, suggestive of young stars being
stripped from the galaxy's disk, or perhaps collisionally-induced star
formation in the tidal feature. Given the intracluster HII region
projected 3.5\arcmin\ (17 kpc) to the north of the galaxy (Gerhard \etal
2002) as well, the complex structure around NGC~4388 is a good example
of how tidal stripping and ram pressure stripping can work in tandem to
deposit young stars into the intracluster environment.

The outer isophotes of the inclined spiral NGC~4413 (88 \kms, Rines \&
Geller; the galaxy is also cataloged as NGC~4407) show more complex
structure. There is a diffuse stream of starlight arcing west towards
NGC~4388, and another extending southeast towards IC~3363. Both streams
are quite red (\bmv=0.9--1.0), and are morphologically similar to the
two armed ``bridge/tail'' morphology of spiral galaxy encounters (\eg
Toomre \& Toomre 1972). While the colors of the streams are
significantly redder than the interior of NGC~4413 (\bmv=0.62), the
galaxy's disk shows a redward color gradient --- at larger radius the
colors are \bmv=0.7--0.8, and the outermost parts of many spiral disks
show even redder colors (Bakos \etal 2008, Zheng \etal 2015, Watkins
\etal 2016). Thus the colors are consistent with stripping of the disk
outskirts, although the identity of the perturbing companion remains
ambiguous. The mismatch in velocity with both NGC~4388 and IC~3363 means
that even if these galaxies are in close proximity to NGC~4413, any
interaction would be fast and less able to trigger a strong tidal
response. On the other hand, NGC~4413's velocity suggests membership in
the M86 subgroup, where an interaction with the massive elliptical might
be responsible for the observed tidal debris. However, without accurate
distances, it is impossible to determine this with any certainty.

The three small galaxies IC~3363 (851 \kms), VCC~962 (velocity unknown),
and IC~3349 (1411 \kms) are all embedded in a diffuse halo of light
which trails off in intensity to the southwest. This halo has a surface
brightness of \muv=27.5 and color \bmv=0.9, and IC~3363 itself shows a
detached, diffuse region of higher surface brightness (\muv $\sim$ 26,
\bmv=0.75) directly to its north. The large (560 \kms) velocity
difference between IC~3363 and IC~3349 makes it unlikely that this trio
of galaxies is a true physical group, and the diffuse halo surrounding
them may simply be material seen in projection.

Also extending to the SE along the major axis of M86 is a broad shelf of
light which does not trace the elliptical isophotes of the giant
elliptical. This feature, not originally identified in the analysis of
J+10, has a surface brightness of \muv=27.3 and \bmv\ color of 0.85.
Given its rather large size ($\sim 40\ {\rm arcmin}^2$) the total
luminosity in this feature is significant: $L=3.5\times10^8$ \Lsun.
Unlike the majority of the streams J+10 identified in M86's halo, which
are typically smaller and lower in luminosity than this ``SE shelf,''
this feature is more similar to the broad accretion structures seen in
M49 and M87, and may be tracing an older and more massive accretion
event in M86's past.

\subsection{The M49 Field}

\begin{figure*}[]
\centerline{\includegraphics[width=7.0truein]{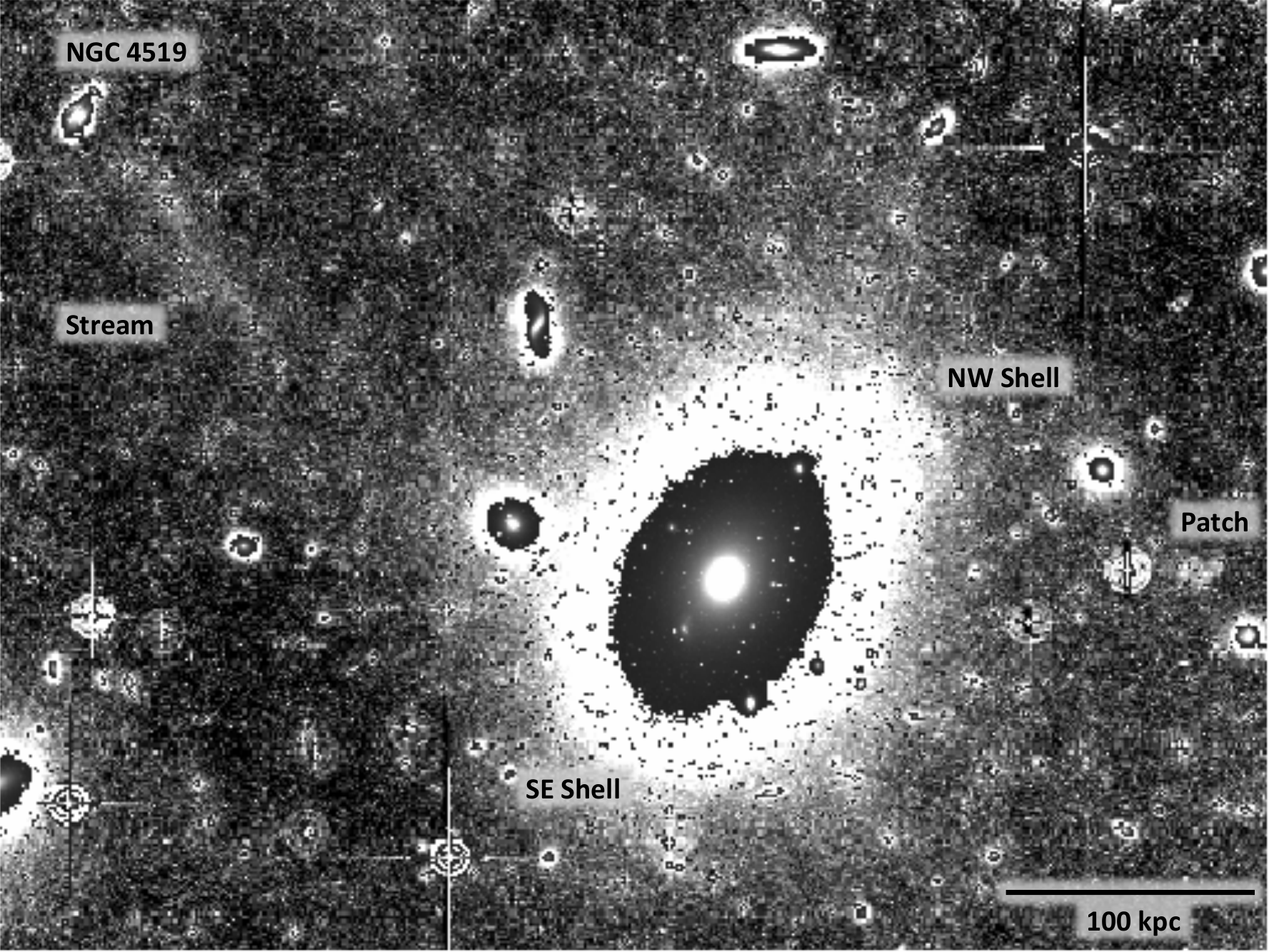}}
\caption{Deep V image of the M49 field covering a field of view 1.8\degr
  $\times$ 1.4\degr\ in size. North is up and east is to the left, and
  features discussed in the text are marked.}
\label{M49streams}
\end{figure*}

The giant elliptical M49 (981 \kms; Cappellari \etal 2011) hosts an
intricate system of debris shells, described in detail in our earlier
papers (J+10, Mihos \etal 2013a), and seen also in deep imaging by the
NGVS and VEGAS teams (Arrigoni Battaia \etal 2012 and Capaccioli \etal
2015, respectively). Figure~\ref{M49streams} shows the more extended
field around M49, where the NW and SE shells can be seen in M49's outer
halo.

At lower surface brightness we find a stream of extremely diffuse light
in the field northeast of M49, along a line roughly connecting to the
spiral galaxy NGC~4519 (1241 \kms). The stream is $\sim$ 15\arcmin\ (70
kpc) long, with a surface brightness of \muv = 28.5. There is no
corresponding feature in the HeVICS 250$\mu$ map
(Figure~\ref{Herschel}), arguing that it is not scattered light from
galactic dust. The alignment of the feature suggests instead that it may
be a stream formed from a past encounter between NGC~4519 and M49,
although the large projected separation between the galaxies (68\arcmin,
325 kpc) would place any close passage at least 1 Gyr in the past.
NGC~4519 itself is also distorted, and interacting with its close
companion NGC~4519A (1434 \kms; Binggeli \etal 1985) just to the
northwest; if this is a long lived interaction, it could have acted to
strip material from NGC~5419 that was then subsequently drawn out into
the longer diffuse stream by the tidal interaction with M49.

We also find a rather amorphous patch of diffuse light 41\arcmin\ (190
kpc) to the west of M49. It lies at a position ($\alpha, \delta$) =
(12:27:05, +8:00:38), midway between the galaxies NGC~4434 and NGC~4416
(which themselves are separated by 19\arcmin/90 kpc), but we so no
evidence of diffuse light connecting the patch with either of those
galaxies, or with M49 itself. The patch covers $\sim$ 9 square
arcminutes (200 kpc$^2$), and has a roughly uniform surface brightness
of \muv=27.5 and \bmv\ color of 0.9. With an irregular structure and no
central concentration, it is unlike known ultradiffuse galaxies in Virgo
(Mihos \etal 2015, Beasley \etal 2016) and Coma (van Dokkum \etal 2015,
Koda \etal 2015), and while it could be a localized patch of Galactic
cirrus, we see no corresponding far-infrared emission in the 250$\mu$
HeVICs map. Its origin thus remains unclear.

Aside from these two features and the shell system surrounding M49, we
find no extended diffuse light or small tidal streams in the field
brighter than \muv=29.0. Deeper inside M49's halo itself, the only
ongoing accretion event involves the stripping of the dwarf irregular
VCC 1249 (McNamara \etal 1994; Lee \etal 2000; Arrigoni Battaia \etal 2012). This is in
marked contrast to the high density fields around M87 and M86/M84 and
suggests that the current rate of ICL production around M49 is lower
than that in the cluster core, as might be expected if ICL production
traced either galaxy density or cluster potential well. Nonetheless,
while there are few ongoing interactions around M49 today, the accretion
rate onto its halo may be similar to that onto M87. Outside of the
cluster core, where it is not subject to repeated bombardment by
infalling satellites, tidal debris can survive on Gyr timescales.
If we add the observed luminosity of M49's shell system ($\sim
7.3\times10^8$ \Lsunv; J+10) and the VCC 1249 streams\footnote{If we
assume that VCC 1249 will be completely disrupted, adding in its entire
V luminosity rather than just that in its tidal streams increases our
inferred accretion rate by only $\sim$ 40\%.} (roughly another $10^8$
\Lsunv), factoring in the Gyr survival timescale implies a rough
accretion rate on the order of $\sim$ 1 \Lsunv\ yr$^{-1}$. The fact
that M49 is similar to M87 in both total luminosity and current low mass
accretion rate again argues that the halo growth of massive ellipticals
is dominated by major or past accretion history, and that present day
satellite accretions are only a minor source of additional luminosity.

Meanwhile, M49 itself is thought to be falling into the Virgo Cluster
from the south, as evidenced by the X-ray bow shock to the north of M49
(Irwin \& Sarazin 1996, Kraft \etal 2011). When the galaxy enters the
denser environment of the Virgo Cluster core, its delicate system of
shells will likely be disrupted due to encounters with other galaxies
and the cluster potential itself, adding to Virgo's growing ICL
component.

\subsection{The NGC~4365 Field}

\begin{figure}[]
\centerline{\includegraphics[height=8.5truein]{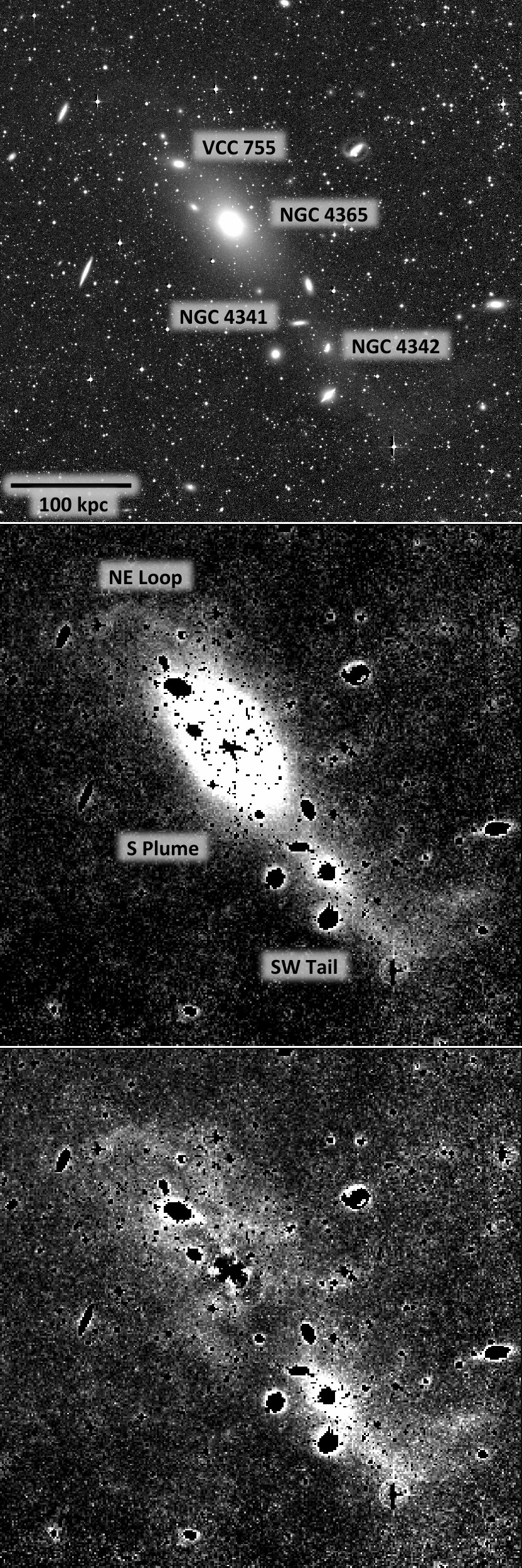}}
\caption{$B$ band imaging of NGC~4365. Each image spans 1.1\degr\ on a
  side, with north at the top and east to the left. Top: Full
  resolution image. Middle: Deep image, after masking discrete
  structures and median smoothing to 9 $\times$9 pixel (13\arcsec
  $\times$ 13\arcsec) resolution. Bottom: Masked and smoothed image after
  subtraction of an elliptical isophotal model for NGC~4365. North is up,
  east to the left. See text for details.}
\label{N4365}
\end{figure}

The giant elliptical NGC~4365 (1243 \kms; Cappellari \etal 2011) lies in
the lower right (southwest) corner of our imaging fields (Figures
\ref{Vmosaic} and \ref{Bcolor}), 5.3\degr\ SSW of M87 and 1.5\degr\ WSW
of M49. It is located $\sim$ 6 Mpc behind the Virgo Cluster (Mei \etal
2007)\footnote{For the discussion of physical sizes and luminosities for
features around NGC~4365, we adopt a distance of 23 Mpc ($m-M = 31.8$;
Mei \etal 2007) to the system, for which 1\arcsec\ = 112 pc.} and is the
dominant galaxy of the Virgo \Wp\ group (de Vaucouleurs 1961, Binggeli
\etal 1985; Mei \etal 2007), surrounded by a population of lower
luminosity galaxies (see Figure~\ref{N4365}a) within $\Delta v$ = $\pm$
350 \kms\ and a projected distance of 1\degr (400 kpc). NGC~4365 itself
has a kinematically distinct core (Bender \& Surma 1992), along with
evidence for multiple populations of globular clusters (Blom \etal
2012b). An early release of our Schmidt imaging data, shown in Bogd\'an
\etal (2012), revealed an extended tidal tail extending SW from
NGC~4365. A number of companion galaxies are projected on or near the
tail, most notably the S0 galaxy NGC~4342 (761 \kms, Cappellari \etal
2011). While Bogd\'an \etal argued the tail was unlikely to have been
stripped from NGC~4342, due to the high dark halo mass implied by the
galaxy's X-ray luminosity, subsequent observations by Blom \etal (2012a, 2014)
of globular clusters in the stream showed a convincing kinematic match
between the tidal tail and NGC~4342.

\begin{figure}[]
\centerline{\includegraphics[height=3.5truein]{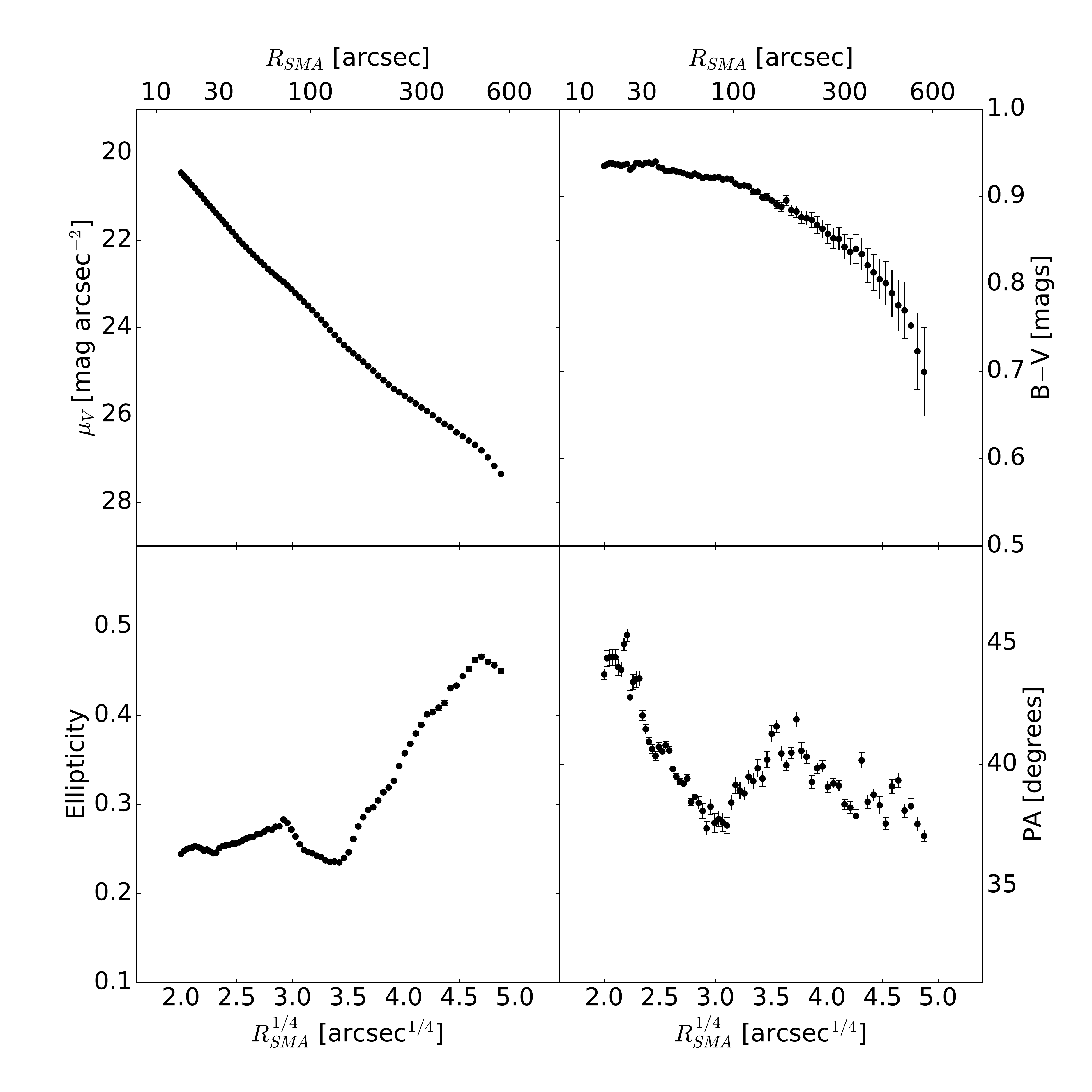}}
  \caption{Photometric profile of NGC~4365, showing the V surface brightness,
  \bmv\ color, and isophotal ellipticity and position angle. Errorbars in surface brightness
  and color are based on a residual sky uncertainty of 0.5 ADU, while errorbars on
  ellipticity and position angle are calculated as in Busko (1996). In surface brightness
  and ellipticity, the errobars are smaller than the point size.}
\label{N4365prof}
\end{figure}

Figure~\ref{N4365}b shows a deep stretch of our $B$ band image, after
masking of bright sources. The long tidal tail can easily be seen
projecting 35\arcmin\ to the southwest, with both NGC~4341 (922 \kms)
and NGC~4342 embedded within. The brightest part of the tail, southwest
of NGC~4342, reaches a peak surface brightness of \muv=27.25 before
making a 90\degr\ dogleg turn at its southwest limit, becoming more
diffuse and extending another 14\arcmin\ to the northwest. The tail has
a \bmv\ color of $\sim$ 0.75--0.85, somewhat bluer than the integrated
color of either NGC~4365 or NGC~4342 (\bmv=0.96 and 0.95, respectively,
RC3). However, both galaxies show a marked color gradient: at 40\arcsec,
NGC~4342 has a mean color of \bmv=0.8, as do NGC~4365's outer isophotes
at 500\arcsec\ (Figure~\ref{N4365prof}). While the color of the tail is
consistent with tidal stripping from the outer regions of either galaxy,
morphological arguments favor a scenario involving stripping from
NGC~4342. The outer isophotes of NGC~4342 are elongated along the
direction of the tidal tail, and the thickness of the tail (5\arcmin) is
similar to the isophotal size of NGC~4342. Also, stripping of the outer
regions of a dynamically hot, massive elliptical like NGC~4365 would
likely lead to a more diffuse plume of starlight rather than the long,
thin tail seen in Figure~\ref{N4365}. Given its size and surface
brightness, we estimate a total magnitude for the tail of $m_V\sim12.5$,
comparable to that of NGC~4342 itself (RC3). As such, the interaction
between the galaxies has led to a significant loss of stellar material
from NGC~4342, as also traced by the kinematics of the globular clusters
in the tidal tail (Blom \etal 2012a, 2014). How the galaxy could have been so
severely stripped of its stellar material, while retaining its massive
dark halo (as traced by its X-ray emission; Bogd\'an \etal 2012) remains
unclear, although as suggested by Blom \etal, resonant stripping of
NGC~4342's rotating disk may preferentially strip the galaxy's stars
while leaving its dark halo relatively intact (D'onghia \etal 2009).

Also visible in Figure~\ref{N4365} is a curved stream of diffuse
starlight to the northeast of NGC~4365, coincident with the companion
galaxy VCC~755 (1241 \kms). The stream is relatively thin (2\arcmin) and
most clearly delineated over a 12.5\arcmin\ segment where it emerges
from the northeast of VCC~755 and is then lost in the halo of a nearby
bright star. Here the stream has a surface brightness of \muv=28.0,
\bmv\ color of 0.8, and total magnitude $m_V$=15.6. To the southwest of
VCC~755 we see more diffuse light apparently connecting the galaxy to
NGC~4365, but this component is broader and higher in surface brightness
(\muv=27.0) making it unclear whether this is a continuation of the
northeast stream or an independent structure embedded in the halo of
NGC~4365. VCC~755 itself is a low surface brightness dwarf (\muv=22.8,
$m_V$=16.2, \bmv=0.85); the similarity in color between the galaxy and
the stream supports the idea that the stream originates from tidal
stripping of VCC~755, which must be significant given that the stream
contains significantly more starlight than the galaxy itself.

Finally, we identify an asymmetric plume of light in the southern
isophotes of NGC~4365's halo. This feature is broad and diffuse
(\muv=27.5), and is not obviously associated with any nearby companion
galaxy. It is similar in color to the outskirts of NGC~4365 (\bmv=0.85)
and may trace an older accretion event that has deposited starlight into
the galaxy's halo.

To better understand the connection between these various diffuse
features around NGC~4365, we also fit and subtract a smooth elliptical
isophotal model for NGC~4365, using the procedure described in J+10.
Briefly, we mask bright stars, neighboring galaxies, and faint discrete
sources in the field using IRAF's \textsf{objmask} task, then run its
\textsf{ellipse} task to fit an elliptical model to NGC~4365's light
profile. This model profile (see Figure~\ref{N4365prof}) is then
subtracted from the original image, and the residual image is then
median binned in 9$\times$9 pixel blocks to bring out the faintest
features, shown in \ref{N4365}c. All the structures discussed in this
section are visible in the residual map, demonstrating that they are
discrete structures distinct from the smooth halo of NGC~4365. However,
we see no evidence of continuity between the SW tail and either the NE
stream or S plume, arguing that these are independent structures tracing
different accretion and interaction events around NGC~4365.

The field around NGC~4365 thus illustrates an important first step in
generating diffuse intracluster light: the ``pre-processing'' of
galaxies via close encounters in the group environment. The total amount
of diffuse light and complex structure in the tidal debris is reflective
of the multiple close interactions that take place in dense group
environments (\eg Weil \& Hernquist 1996, Rudick \etal 2006, Durbala
\etal 2008). In observed compact groups, the fraction of diffuse light
is thought to be very high, 10\% or more (Da Rocha \& Mendes de Oliveira 2005, 
Da Rocha \etal 2008).
While the diffuse light fraction we observe around NGC~4365 is lower
than this ($\sim$ 3\%), it is higher than we observe in most other
massive Virgo ellipticals (typically $\sim$ 0.5\% or less; J+10). With a
single massive elliptical and a swarm of smaller companions, the
NGC~4365 group is likely more evolved than typical compact groups which
have multiple comparably massive galaxies, and much of the material
stripped early in the group's history has likely now been incorporated
into NGC~4365's luminous halo. As NGC~4365 and its companions eventually
fall into the core of Virgo, much of this diffuse material will be
stripped away from the group and dispersed throughout the cluster core
(\eg Rudick \etal 2006), further feeding Virgo's diffuse intracluster
light.

\subsection{VLSB-D: A new ultradiffuse galaxy}
    
Aside from the streams and shells of tidal debris we find in Virgo, our
deep imaging data also allows us to search for extremely low surface
brightness galaxies that may populate the cluster. Of particular
interest here is finding Virgo analogues to the population of large and
faint ``ultradiffuse galaxies'' (UDGs) that have recently been
identified in deep imaging of the Coma cluster (van Dokkum \etal 2015,
Koda \etal 2015). In our imaging, we see many examples of small LSB
objects (see, for example, Figure~\ref{LSBrogues}), but without distance
information it is hard to place them uniquely in the Virgo cluster.
Furthermore, even if they are at the Virgo distance, their small sizes
means they are not truly UDGs (which have effective radii $r_e>1$ kpc,
or 12.5\arcsec\ at Virgo), but rather part of the well-known population
of LSB dwarfs that teem inside galaxy clusters (\eg Impey \etal 1988;
Ferrarese \etal 2016). However, we also find a handful of very large
($r_e >$ 30\arcsec) and diffuse ($\langle \mu \rangle_{e,B}>$ 28)
systems in our imaging; our earlier paper (Mihos \etal 2015) discusses
three such objects found in the deep V-band mosaic. Here we add one more
UDG to that list: an object located on the B-band mosaic only, of large
angular size and extremely low surface brightness.

\begin{figure*}[]
\centerline{\includegraphics[height=2.6truein]{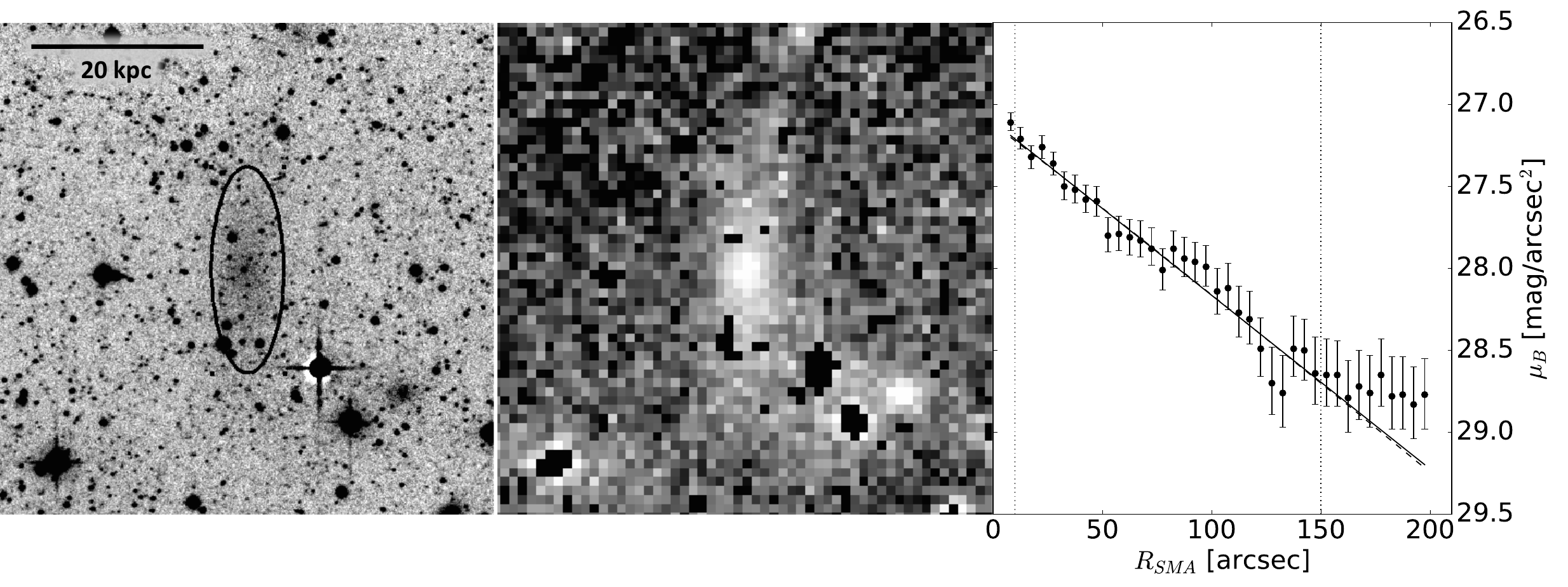}}
\caption{Left: 12\arcmin$\times$12\arcmin\ cutout of the full resolution
  $B$-band Schmidt mosaic showing a new Virgo ultradiffuse galaxy. Middle: The
  Schmidt mosaic at the same 12\arcmin$\times$12\arcmin\ scale, but
  after being masked of bright sources and median binned in 9$\times$9 pixel
  blocks to show faint structure. Right: Surface brightness profile plotted as a
  function of semi-major axis radius, with exponential (solid) and
  Ser\'sic (dashed) fits overplotted. The dotted vertical lines show
  radial range of the profile fits, and the outermost ellipse used in the
  fit is shown in the left panel.}
\label{UDG}
\end{figure*}

Figure~\ref{UDG} shows the imaging and extracted surface brightness
profile for this new object. Located at
($\alpha,\delta$)=(12:24:42.1,+13:31:02), it is projected $\sim$
38\arcmin\ (180 kpc) north of the bright elliptical M84. The system is
extremely elongated in the N--S direction, and has a compact source at
its center. While we have no color information from our data (as it
falls only in the $B$-band footprint), the compact source is detected in
SDSS with $g=20.72$ and $g-r=0.58$, a color similar to luminous Virgo
globular clusters (\eg Durrell \etal 2014) and ultra-compact dwarf
galaxies (UCDs; Liu \etal 2015). Excluding the central source, the
object has a central surface brightness of \mub=27.1 and a total
magnitude of $m_B$ = 16.3 within a 150\arcsec\ (12 kpc) extent. However,
there is additional light beyond this radius, suggestive of a tidal
stream to the north and perhaps also to the south, although the presence
of bright stars and other sources in this direction makes it difficult
to measure the profile. The surface brightness profile shows an
exponential form; a Sersic fit (shown in Figure~\ref{UDG}) is virtually
indistinguishable from an exponential fit, and yields a Sersic index
$n=0.97\pm0.14$, effective radius of $r_e = 166\pm25$\arcsec\
(13.3$\pm$2 kpc), mean surface brightness $\langle \mu \rangle_{e,B} =
28.2\pm0.2$, and total magnitude $m_{B,tot} = 15.5\pm0.4$. However,
given the object's extreme flattening and irregular, extended outer
isophotes, it is likely not a system in dynamical equilibrium, making
the interpretation of the structural properties derived from the fit
rather questionable. Nonetheless, the extremely large size and low
surface brightness of the object makes it one of the most extreme UDGs
yet discovered.

The properties of this object are similar to the Virgo UDG VLSB-A (Mihos
\etal 2015) --- a large, low surface brightness galaxy with extended
tidal debris and a compact source at its center. In both cases, we are
likely seeing the tidal destruction of a nucleated LSB galaxy, plausibly
leading to the formation of a new Virgo UCD (\eg Bekki \etal 2003;
Pfeffer \& Baumgardt 2013). The fact that two of the four very large
UDGs discovered in our deep imaging data appear in this short-lived
phase suggests that a significant fraction of cluster UDGs --- at least
in cluster cores, where our imaging is targeted --- may be actively
being stripped, and that this mechanism is a common way of producing
UCDs. However many UDGs show little sign of tidal deformation (van
Dokkum \etal 2015, Mihos \etal 2015) and at least some also appear to
possess robust globular cluster systems (Mihos \etal 2015, Beasley \etal
2016, Peng \& Lim 2016), suggesting they are rather robust against tidal
stripping. Given the loose definition of UDGs, and the wide range of
properties they display, it would not be surprising if, as a class, they
were comprised of galaxies drawn from a variety of of evolutionary
histories.

\section{Diffuse Light and the Dynamical Evolution of the Virgo Cluster}

A variety of dynamical modeling studies have shown that the build up of
intracluster light, like the growth of structure in the universe in
general, is a hierarchical process. As smaller systems interact and
merge to form larger galaxies, stars are stripped into the outskirts of
their host galaxies. When galaxies assemble into groups, this stripping
accelerates, as the loosely-bound material in the galaxies' outskirts is
shorn away during repeated encounters in the group environment.
Eventually, as the groups fall into a more massive cluster environment,
this intragroup light is easily stripped away by interactions with
massive central galaxies and the cluster potential itself (Rudick \etal
2006, Purcell \etal 2007, Contini \etal 2014), and additional stripping
is mediated both by the merging of massive galaxies in the cluster core
(Murante \etal 2007, Martel \etal 2012) and by high speed encounters
between galaxies orbiting within the cluster environment (Moore \etal
1996). In this way, both the amount and distribution of intracluster
light are intimately linked to the dynamical evolution of the cluster
(Rudick \etal 2006, 2011). Low density environments such as poor
clusters and loose groups would have less diffuse light, while dense,
highly evolved groups and clusters should possess the largest amounts of
intracluster light (\eg Rudick \etal 2006, Sommer-Larsen 2006,
Contini \etal 2014). However, at best the ICL content of a cluster
scales only weakly with cluster mass; simulated clusters show significant
scatter in their ICL fractions at fixed mass, and the driving influence
in ICL production seems to be environmental density and accretion
history rather than total mass (\eg Conroy \etal 2007, Rudick \etal
2011, Contini \etal 2014).

With this backdrop, it is useful to examine Virgo's diffuse light as a
tracer of the cluster's evolutionary state, and assess the observational
link between cluster dynamics and the generation of ICL. A cluster of
moderate virial mass ($M_{vir} \approx 1-4\times10^{14}$ \Msun; McLaughlin 1999,
Urban \etal 2011), Virgo shows signs of ongoing assembly, with ample
evidence for substructure on both large and small scales in the spatial
distribution, kinematics, and morphologies of its galaxy populations
(\eg Binggeli \etal 1985, 1987, 1993; Gavazzi \etal 1999; Mei \etal
2007, Kim \etal 2014) as well as in tracer populations such as globular
clusters (\eg Romanowsky \etal 2012, Durrell \etal 2014) and planetary
nebulae (Arnaboldi \etal 2004, Doherty \etal 2009, Longobardi \etal
2015a). As such, Virgo should be an ideal environment for probing the
connection between local density, dynamical history, and ICL populations
in galaxy clusters.

However, a possible complication in studying the ICL in Virgo is the
long line-of-sight depth of the cluster, with significant differences in
the inferred depth between the various galaxy populations in Virgo.
Early type galaxies appear to have a back-to-front depth of 2.4 Mpc (Mei
\etal 2007), while spirals and dwarfs show a much broader spread, with
depths of $\sim$ 7 Mpc (\eg Yasuda \etal 1997, Solanes \etal 2002,
Jerjen \etal 2004, although see also Cort{\'e}s \etal 2008). This argues
that some caution should be applied in comparing our projected ICL maps
to the true three dimensional structure of the Virgo Cluster. This depth
also complicates comparisons of ICL populations mapped by surface
photometry to those mapped by discrete populations (planetary nebulae,
red giant stars; see Mihos \etal 2009). Surface photometry integrates
the ICL luminosity along the full depth of the cluster, with some minor
effects due to projection (Rudick \etal 2009). In contrast, detection of
discrete tracers can suffer from incompleteness bias --- the detection
of objects will likely be incomplete before reaching the far side of
Virgo. As a result, there is a potential systematic difference between
ICL estimates using the two types of techniques, in that studies using
discrete ICL tracers may infer less intracluster light compared to those
such as ours which employ deep surface photometry. With these caveats in
mind, we now turn to a more global discussion of the intracluster light
in Virgo.

Even a casual inspection of the deep imaging in Figures~\ref{Vmosaic},
\ref{Bcolor}, and \ref{Herschel} quickly reveals that much of the
diffuse starlight in the Virgo Cluster is concentrated in the cluster
core, around M87 and in the M86/M84 region. In areas immediately north
and south of the region, there is very little diffuse light other than
what can be accounted for by scattering from Galactic dust. Around
subcluster B we also find little evidence for any extensive ICL
component, aside from the shell system inside the halo of M49 itself.
The only region in our survey area outside of Virgo's core that contains
much diffuse light is the region around NGC~4365 in the \Wp\ cloud, a
region of high {\it local} galaxy density. The fact that the Virgo ICL
is so heavily concentrated in the cluster core has also been indicated
by PNe studies of fields throughout Virgo (Castro-Rodriguez \etal 2009)
and is as expected from simulations of ICL production in massive
clusters (Murante \etal 2004, Sommer-Larsen \etal 2005, Martel \etal
2012). While this central concentration reiterates the importance of
environmental density on intracluster light production, the exact
mechanism responsible for the core ICL is not well constrained. Any of
the proposed ICL production mechanisms --- stripping by the central
galaxy, stripping by the global cluster potential, and stripping during
massive galaxy mergers --- will all be most effective near the cluster
center.

However, the concentration of diffuse light around NGC~4365 in the \Wp\
cloud does perfectly illustrate the pre-processing role played by
galaxy groups in generating the large scale diffuse intracluster light
in massive galaxy clusters. The group's high local density drives strong
interactions between galaxies, stripping stars out into the group
environment and beginning the formation of an ``intragroup halo'' of
light. NGC~4365 is the fifth most luminous galaxy in Virgo (Kormendy
\etal 2009), only 0.3 magnitudes fainter in $V$ than M87 itself; if the
\Wp\ cloud was an isolated group, it might well ultimately form a
massive field elliptical or ``fossil group'' with an extended stellar
halo (D'Onghia \etal 2005, Sommer-Larsen \etal 2006), although the
galaxy's low X-ray luminosity (Kim \etal 1992) might preclude the fossil group 
scenario. However, the distance and radial velocity of NGC~4365
and other members of the \Wp\ cloud are consistent with the group
falling into the Virgo Cluster from the cluster's far side (Mei \etal
2007); when the group eventually falls into the even denser environment
of Virgo, interactions with cluster's massive galaxies and its global
tidal field will strip this extended intragroup light and disperse it
throughout the cluster (\eg Mihos 2004, Rudick \etal 2006), further
feeding Virgo's growing population of intracluster stars.

The large amount of diffuse light around NGC~4365 ($\sim$3\% of the
total galaxy luminosity) makes the lack of ICL around M49 even more
curious. Both galaxies are the central galaxies of Virgo subgroups, but
they have very different diffuse light properties. In contrast to the
isophotal flattening and extensive tidal structure around NGC~4365 M49's
outer halo is well-behaved, showing only mild isophotal variation in
ellipticity and position angle as a function of radius (Kormendy \etal
2009, Janowiecki \etal 2010), and its faint, extended shell system
accounts for only $\sim$ 0.5\% of the total galaxy light (Janowiecki
\etal 2010). The differences in the two systems may well be tied to
their small-scale environment: NGC~4365 is attended by a close swarm of
companion galaxies. The projected radius inside which the total
companion luminosity equals half the central galaxy luminosity is 192
kpc for NGC~4365 and 300 kpc for M49 (235 kpc and 338 kpc, respectively,
if we limit to companions with velocities within 500 \kms of the
central's systemic velocity). The high density of companions around
NGC~4365 likely results in stronger interactions and more tidal
stripping from the system as compared to M49, at least at the current
epoch.

Aside from the Virgo core and the \Wp\ cloud, the bulk of the diffuse
light we find is closely coupled to the halos of Virgo's bright
ellipticals --- either the substructure embedded within the halos or the
extended, low density halos themselves. On these scales, separation of
galaxy light from intracluster light is both difficult and ill-defined.
While simulations suggest that the extended envelopes of cluster cD
galaxies are kinematically distinct from the central regions of the
galaxy (Murante \etal 2004, Dolag \etal 2010, Rudick \etal 2011) --- a
picture with growing observational support (Kelson \etal 2002, Gerhard \etal 2007,
Ventimiglia \etal 2011, Bender \etal 2015, Longobardi \etal 2015b) ---
without kinematic information, any distinction becomes much less clear,
as these components blend smoothly together photometrically. As much as
50\% of the total cD light can be contained in this outer, diffuse
envelope (Gonzalez \etal 2005) and, as illustrated by Cooper \etal
(2015), much of this light comes from accretion of galaxies
onto the central cD, a major mechanism associated with the formation of
the intracluster light itself. Thus the extended, low density halos of the
central galaxies {\it are} the accreted light that comes with the
dynamical destruction of cluster galaxies, and it is no surprise that we
see these extended halos around M87 and M49, the central galaxies of the
Virgo A and B complexes, respectively.

Information on the processes driving ICL production in Virgo is also
provided by the colors of the diffuse tidal features we see throughout
the cluster. As reported here in Section~4 and also in Rudick \etal
(2010), these streams are predominantly red, with \bmv\ colors in the
range 0.7--0.9. These colors suggest old stellar populations, with
little evidence for young stars. These colors are also are too red to
accommodate a post-starburst population --- depending on the
metallicity, a single burst population would take $\sim$ 2 Gyr or longer
to reach \bmv=0.7 (Bruzual \& Charlot 2003), longer than the survival
time of tidal tails in a cluster environment (Rudick \etal 2009).
Therefore, the reddish tails we see cannot be dominated by stars formed
during the interaction or in-situ in the tidal debris. While some bluer
streams exist --- for example, the blue tidal tails around the
interacting spiral NGC~4388 (Section~4.2) or the ram pressure stripped,
star-forming knots near IC 3418 (Figure~4; Hester \etal 2010, Kenney
\etal 2014), the predominantly red colors of the Virgo streams argues
for older, tidally stripped stellar populations as the major contributor
to the intracluster light. The observed colors do not tightly constrain
the stellar metallicities; for populations older than 4 Gyr, the color
range \bmv=0.7--0.9 covers a metallicity spread of of $-1.3 \lesssim
{\rm [Fe/H]} \lesssim 0.0$ (Bruzual \& Charlot 2003); a similar broad
spread in metallicity was inferred from studies of intracluster red
giants in Virgo (Durrell \etal 2002, Williams \etal 2007b). Given the
link between mass, metallicity, and broadband colors in galaxies, the
ICL colors and implied metallicity range argues that a variety of
progenitor systems contribute to the ICL, from lower luminosity dwarfs
to more massive systems. 

The colors of the streams are somewhat bluer than the bulk of the
stellar populations in the massive Virgo ellipticals, which typically
have \bmv\ colors $>$0.9 in their inner regions. However, negative color
gradients are common in ellipticals (as can be seen in the color maps of
Figure~\ref{Bcolor}), and the slightly bluer \bmv\ colors of the {\it outer} halos
of M87 (Rudick \etal 2009), M49 (Mihos \etal 2013a), and NGC 4365 (this
study) are similar to those of the diffuse ICL features in Virgo.
Photometric modeling indicates these color gradients are most likely
tracking metallicity gradients (Liu \etal 2005, Mihos \etal 2013a, Montes
\etal 2014), where an old metal-rich inner halo transitions to a more
metal-poor population in the galaxy outskirts and eventually into the
intracluster environment (as traced by the intracluster RGB stars;
Williams \etal 2007b). This similarity in color between the ICL streams
and the outer halos of the massive ellipticals bolsters the argument for
a common origin between these components -- mergers and tidal stripping
feed Virgo's ICL while also building up the halos of its elliptical
galaxies.

In regard to measuring the {\it total} amount of intracluster light in
our survey field, quantitative characterizations of the ICL in galaxy
clusters have proved challenging to implement. An often-quoted metric is
the ICL fraction, \ficl, defined as the fraction of the total cluster
luminosity contained in the ICL component. Unfortunately, such a number
is both hard to define and hard to measure, due largely to the fact that
there is often no clear {\bf photometric} differentiation between the
extended cluster ICL and the central galaxy's diffuse outer halo (\eg
Gonzalez \etal 2007, Rudick \etal 2011, Cooper \etal 2015). Simulations
often define the ICL as being comprised of stars within the cluster
unbound from their host galaxy (Murante \etal 2004, Dolag \etal 2010,
Rudick \etal 2011, Cui \etal 2014), but lacking kinematic information
for the diffuse light and detailed mass models for the cluster galaxies,
this definition is rather intractable to apply. Conversely,
observational definitions include measuring the total amount of light
below a given surface brightness threshold (\eg Feldmeier \etal 2004b,
Burke \etal 2015), or fitting multiple components to the observed
surface brightness profiles (Gonzalez \etal 2005, Krick \& Bernstein
2007, Seigar \etal 2007), but these definitions depend critically on the
adopted surface brightness thresholds or the functional forms for the
light profiles. Furthermore, connecting these measurements of the
intracluster {\it light} distribution back to simulations of the
stripped stellar {\it mass} in clusters requires additional knowledge of
the stellar populations involved. Much recent progress has been
made mapping out the kinematics of intracluster PNe to trace the ICL
separately from the galaxy light, although mapping this back to a {\it
stellar} ICL fraction requires knowledge of the underlying stellar
populations in the ICL (Ventimiglia \etal 2011, Longobardi \etal 2015b,
Barbosa \etal 2016). All these complications make a unique
determination of the ICL fraction extraordinarily difficult (see, for
example, the extensive discussions in Puchwein \etal 2010, Rudick \etal
2011, and Mihos 2015).

Nonetheless, we can at least attempt a rough estimate of the ICL
fraction for Virgo using our data. The simulations of Rudick \etal
(2011) show that the bulk of the cluster luminosity at surface
brightnesses fainter than $\mu_v \sim 26-27$ (for an adopted
$M/L_{*,V}=3$) is comprised of material unbound from its host galaxy.
Furthermore, of this diffuse material, roughly 5--10\% is in the form of
coherent streams (Rudick \etal 2009), largely the relatively high
surface brightness tidal features that represent material most recently
stripped from their host galaxies. Taken together, these studies suggest
that the luminosity of the observed tidal features might be used as a
rough tracer of the total amount of diffuse intracluster light in Virgo,
albeit with significant uncertainty given the caveats above. 

In Virgo, the most luminous tidal feature we see is the SW stream around
NGC~4365 in the \Wp\ cloud (Figure~\ref{N4365}), with a total luminosity
of $2.3\times10^9$ \Lsunv. In J+10, we also measured the luminosities of
most of the discrete features around the other major ellipticals in
Virgo, including M87, M86, M84, M49, and M89, which, in total, account
for another $2.3\times10^9$ \Lsunv\ of starlight. Finally, the other new
streams we catalog here have a total luminosity of another $10^9$\Lsunv,
giving a total stream luminosity in our imaging survey of $L_{\rm
stream} \sim 5.6\times10^9$ \Lsunv. If this stream luminosity represents
5--10\% of the total ICL luminosity (Rudick \etal 2009), this gives an
estimate of the ICL luminosity in our field of $L_{\rm ICL} =
0.56-1.12\times10^{11}$ \Lsunv\ in our field. To turn this into an ICL
fraction, we also need to know the total $V$ luminosity of all the
galaxies in our survey area. To calculate this, we use the Extended
Virgo Cluster Catalog (EVCC) of Kim \etal (2014), based on SDSS DR7
imaging. We transform the $ugriz$ magnitudes to Johnson $V$ 
as described in the Appendix, then sum the
$V$-band luminosities for all EVCC galaxies which fall in our total
survey area, obtaining a total galaxy luminosity of $L_{\rm gal} =
7.1\times 10^{11}$ \Lsunv. Finally, we arrive at our final estimate for
the ICL fraction of \ficl$=L_{\rm ICL} / (L_{\rm gal}+L_{\rm ICL}) = $
0.07 -- 0.15. 

Our estimate is not without significant uncertainty, of course. We
have used the simulations of Rudick \etal (2009) to estimate the rough
scaling of stream luminosity to total ICL luminosity, but those
simulations show significant variation in the stream-to-total luminosity
ratio depending on the dynamical conditions in the cluster (see their
Figure~5). If there is a relative dearth or excess of luminous ICL
streams in Virgo at present, our derived total ICL luminosity will be
mis-estimated. A second source of uncertainty is our irregular and
limited survey footprint. Virgo's virial radius covers 104 degree$^2$
(Ferrarese \etal 2012), while our survey footprint only targets
$\approx$ 16 degree$^2$ in the cores of the Virgo A and B subclusters.
How this might affect our estimate is not immediately obvious. On the
one hand, simulations suggest that the ICL in clusters is more centrally
concentrated than the galaxies (\eg Murante \etal 2004; Martel \etal
2012); if so, our survey targeting the Virgo core might be biased to
overestimate the total ICL fraction. On the other hand, there is likely
to be significant spatial variation in amount of ICL in the cluster, as
clearly demonstrated by the bright streams we observe in the \Wp\ cloud.
The fact that our survey does not cover several other substructures (for
example, cluster C around M60 to the west or the M100 group to the
north; see \eg Binggeli \etal 1987), means that we may be missing ICL
streams and thus underestimating the total Virgo ICL fraction. Deeper
imaging over a wider field of view, such as that provided by the NGVS
survey (Ferrarese \etal 2012), will be helpful in addressing this
uncertainty.

Nonetheless, even with these caveats in mind, our estimated Virgo ICL
fraction is in reasonable agreement with other estimates of the Virgo
ICL using individual luminous stars as tracers. Ferguson, Tanvir \& von
Hippel (1998) used deep {\sl HST/WFPC2} observations of intracluster red
giants to derive an ICL fraction of $\sim$ 10\% in a field 45\arcmin\
east of M87, while a similar study by Durrell \etal (2002) in a field
halfway between M87 and M86 found a somewhat higher fraction of
$15$\%$^{+7}_{-5}$. Using intracluster planetary nebulae, Feldmeier
\etal (2004a) found a global ICL fraction of $15.8\% \pm 3.0
\mbox{(statistical)} \pm 5.0 \mbox{(systematic)}$ over six fields
located in Virgo subclumps A \& B, while Castro-Rodrigu{\'e}z et
al.(2009) found a lower fraction of $\approx$ 7\% in the cluster core,
with little or no intracluster light outside of it.

Our Virgo ICL fraction is also similar to --- albeit on the low end of
--- ICL fractions derived for other galaxy clusters. In an early
compilation of estimates for nearby clusters, Ciardullo \etal (2004)
showed ICL fractions spanning the range 15--35\%, with no clear
correlation against cluster velocity dispersion or Bautz-Morgan type, a
morphological classification based on the presence of a centrally
dominant (cD) galaxy (Bautz \& Morgan 1970). However, some of the most
spectacular examples of cluster-wide, unrelaxed intracluster light are
found in clusters experiencing massive accretion events, such as Abell
1914 (Feldmeier 2004b), Hydra I (Arnaboldi \etal 2012), and Coma
(Thuan \& Kormendy 1977, Gregg \& West 1998, Gerhard \etal 2007). Recent
observations have pushed out to clusters at redshifts of $z\sim0.3-0.5$
and find somewhat smaller ICL fractions, 5--20\% (Presotto \etal 2014,
Montes \& Trujillo 2014, Giallongo \etal 2014), somewhat suggestive of
clusters building up their ICL components over time (Burke \etal 2015,
although see also Guennou \etal 2012 for a differing view). It is encouraging
that our Virgo ICL fraction is consistent with
measurements for other clusters; however, given the wide range measured
in other clusters (5--35\%), the differing observational metrics used by
these studies, and the strong variance in \ficl\ as a function of metric
used (Puchwein \etal 2010, Rudick \etal 2011) it would frankly be
difficult {\it not} to be consistent with other measured ICL fractions.

An alternative to measuring ICL fractions in individual clusters is
to photometrically ``stack'' imaging of many clusters; doing so washes
out the cluster-to-cluster variation in streams such as those we detect
in Virgo, but has the potential to push deeper in surface brightness and
measure the even more diffuse underlying ICL structure. Zibetti \etal
(2005) stacked SDSS data for 683 bright clusters, finding an ICL
fraction of 10.9\% $\pm$ 5.0\%, with another $\approx$ 20\% of the
cluster light coming from the central BCG. Given the photometric
uncertainty in decoupling the ICL from the outer BCG light, this result
is in reasonable agreement with our estimate for Virgo. D'Souza \etal
(2014) conducted a similar stacking study for {\it isolated} galaxies;
while not directly comparable to cluster BCGs, they did find that for
the most massive early type systems, comparable in luminosity to M87,
50--70\% of the galaxy light was found in a very extended, diffuse halo.
If we interpret Virgo's ICL as the extended halo of its BCG and compare
the total inferred ICL luminosity in Virgo to the luminosity of M87
itself ($1.15\pm0.5\times10^{11}$ \Lsunv; Janowiecki \etal 2010), we
find that this halo/ICL component contributes 30--50\% of the total
BCG+halo light. The common picture in all these comparisons is that the
Virgo ICL is consistent with that measured in other clusters and around
massive galaxies, although somewhat on the low end of the observed
range.

The fact that the ICL fraction we infer for Virgo is on the low end of
the scale compared to other clusters may be a result of Virgo's dynamical
state. Simulations of galaxy clusters show that ICL fractions grow as
clusters become more dynamically evolved, and can reach values as high
as 40--50\% at late times (\eg Rudick \etal 2006, 2011, Martel \etal
2012, Contini \etal 2014). However, as mentioned previously, compared to
other massive clusters such as Coma, Virgo appears not as well-evolved,
showing significant substructure in the spatial distribution,
kinematics, and morphological types of its constituent galaxies. These
various subgroups (A, B, the \Wp\ cloud, and other smaller groups; see
Binggeli \etal 1987) are reflective of a cluster still in the process of
assembly. As the smaller groups eventually accrete into the cluster
core, ICL production will increase as galaxies interact and merge in the
dynamically active environment, growing Virgo's ICL component with time.

The moderately low ICL fraction we measure is likely also related to 
Virgo's instantaneous accretion rate --- we simply do not see a {\it
large} number of {\it luminous} ICL streams. Luminous streams form
during massive accretion events and rapidly mix away to form the more
diffuse (and harder to detect) intracluster light (Rudick \etal 2006,
2009). The diffuse substructure we do observe in the extended halos of
Virgo's massive ellipticals accounts for only $\sim$ 0.5--3\% of the
total light of the host galaxy (J+10, this study). In the case of M87,
the relatively low implied rate of accretion ($\sim$ 1 \Lsun\ yr$^{-1}$;
Section 4.1) is insufficient to build up the extended cD envelope in a
Hubble time; instead, the accretion of more massive systems is required.
Evidence for more substantial recent accretion is seen in the kinematics
of globular clusters and planetary nebulae around M87 (Romanowsky \etal
2012; Longobardi \etal 2015a), and the extreme boxiness of the galaxy's
outer isophotes (Section 4.1) also suggests a more violent past history.
Interestingly, based on the population of globular clusters and UCDs in
Virgo, Ferrarese \etal (2016) argue that in the Virgo core, as much as
40\% of the present-day galaxy light may have come from disrupted
satellites, again arguing for a higher rate of accretion and tidal
stripping in the past. Clearly, the buildup of M87's halo (and
concurrent ICL formation) is an ongoing process, but also a stochastic
one --- the lack of {\it substantive} photometric substructure at the
present day argues that the cluster is currently in a rather quiescent
phase. However, with both M84 and M86 in close proximity, and M49 and
NGC~4365 both falling into the cluster, Virgo's current phase of quiet
accretion is likely to end soon.

\section{Summary}

We have conducted a deep imaging survey of diffuse light in the Virgo
Cluster, with data taken using CWRU's Burrell Schmidt telescope over the
course of seven spring observing seasons from 2004--2011. Our final
survey footprint consists of 16.7 square degrees in Washington $M$ and
15.3 square degrees in a modified $B$ filter; transformed to Johnson $B$
and $V$, our 3$\sigma$ limiting surface brightnesses are $\mu_{B,lim} =
29.5$ \magsec\ and $\mu_{V,lim} = 28.5$ \magsec. Having data taken in
two photometric bands across multiple seasons provides not only \bmv\
colors across the field, but also gives confirmation of diffuse features
detected at the faintest levels. At the depths probed by our survey we
see a wide variety of diffuse structures in Virgo, including long tidal
tails associated with individual cluster galaxies, the outer stellar
halos of massive ellipticals, a number of large ultradiffuse galaxies,
extensive intragroup light in the infalling \Wp\ cloud, and several
other diffuse structures having no clear association with individual
galaxies.

Our survey maps also clearly reveal diffuse light arising from the
scattering of Milky Way starlight by foreground galactic dust clouds, a
significant source of contamination in deep imaging studies such as
ours. We show that this diffuse scattered starlight is well traced by
deep far-infrared 250$\mu$ {\sl Herschel} imaging, and that scaling and
subtracting the 250$\mu$ map from the optical map (using a scale factor
of $I_\nu({\rm V})/I_\nu(250\mu) = 3.5\times10^{-3}$) yields a
reasonable ``dust-free'' view of diffuse starlight in the Virgo Cluster.

Using our deep imaging, we identify a number of faint, extended tidal
streams and plumes emanating from small galaxies in the Virgo core. Most
of the streams we identify are red in color (\bmv $\sim$ 0.7--0.9),
suggesting that the bulk of the starlight being added to Virgo's ICL
today is comprised of old stellar populations. These colors are similar
to those at large radius (beyond 50 kpc) in M87, M49, and NGC 4365,
suggesting a connection between the formation of the ICL and the growth
of the outer halos in massive Virgo ellipticals. These observed streams
are also low in luminosity (L $\sim 10^9$ \Lsunv), implying an
instanteous stripping rate around M87 of $\sim$ 1 \Lsunv\ yr$^{-1}$, too
low to have built the galaxy's extended outer halo via accretion of
small satellites over a Hubble time. Instead, a higher past accretion
rate, or the accretion of more massive companions (or, likely, both) is
needed, and we do find more subtle signatures of such massive accretions
in the outer halos of both M87 and M86. We also show that M87's outer
halo becomes extremely boxy beyond a radius of 100 kpc.

To the south of the Virgo core, in subcluster B surrounding M49, we find fewer
examples of active tidal stripping, aside from the complex of accretion
shells in the halo of M49 itself. We do, however, find two extended
diffuse features having no clear association with nearby galaxies. One,
a long (70 kpc) diffuse stream, may trace a past encounter between M49
and NGC~4519, while the other is an amorphous patch of diffuse light
$\approx$ 3.5\arcmin\ in size, but of unknown origin.  Aside from these
features, we see no evidence for any extensive network of diffuse ICL
in subcluster B. The lack of an extended ICL component in this region
likely reflects its lower environmental density compared to the Virgo
core around M87, as tidal stripping mechanisms operate most
efficiently in regions of high galaxy density.

In contrast, in the infalling \Wp\ cloud of galaxies, we find an
extensive system of diffuse light around the dominant elliptical NGC
4365 and its population of satellites. The most luminous feature is the
170kpc-long SW Tail, which our photometry shows to have a \bmv\ color of
0.75--0.85, similar to the outer colors of both NGC~4365 and NGC~4342,
the companion galaxy projected inside the tail. We also find a thin loop
of starlight to the NE of NGC~4365, along with a diffuse plume in the
southeast portion of galaxy's outer halo. The large extent and total
luminosity ($\sim 2.9\times10^9$ \Lsunv) of the tidal debris around NGC
4365 is suggestive of strong, slow interactions between NGC~4365 and its
satellites; when the \Wp\ cloud is accreted into the Virgo cluster,
these diffuse structures will likely be stripped from NGC~4365 and
incorporated in Virgo's extended ICL component. The \Wp\ cloud may thus
be giving us a direct view of the pre-processing stage that galaxy
groups mediate in the formation of ICL in massive clusters.

In addition to the variety of diffuse tidal structures we see within
Virgo, we also find examples of large ($r_e>3$ kpc), low surface
brightness (\brackmue{B} $> 28$) ultradiffuse galaxies within the
cluster. Three of these objects were reported in an earlier paper
(VLSB-A, -B, and -C, Mihos \etal 2015), and to that list we add here a
fourth: VLSB-D, a nucleated UDG located $\sim$ 180 kpc north of the
bright elliptical M84. The galaxy is well-fit by a Sers\'ic profile with
index $n$=0.97, $r_e$=166\arcsec\ (13.3 kpc), \brackmue{B} = 28.2, and
total magnitude $m_B$=15.5, making it one of the most extreme UDGs yet
detected in any galaxy cluster. The galaxy is quite elongated and shows
evidence of extended tidal debris, and may be in the process of being
disrupted by the cluster environment. Like its cousin VLSB-A (Mihos
\etal 2015), in VLSB-D we may be witnessing undergoing a tranformation
from a nucleated LSB galaxy to a ultracompact dwarf, once the galaxy has
been completely disrupted and only its compact nucleus remains intact.

Taken as a whole, our observations show a cluster still in the process
of growing its intracluster light component. Using the tidal streamers
as a tracer of the even more diffuse ICL below our detection limit, we
estimate an intracluster light fraction in the range 7--15\%, similar to
other estimates of the Virgo ICL fraction (Ferguson \etal 1998, Durrell
\etal 2002, Feldmeier \etal 2004a, Castro-Rodrigu\'ez \etal (2009).
However, this value falls in the lower end of measured ICL fractions in
other massive clusters, and is also smaller than expected for fully
evolved clusters, which simulations suggest should possess ICL fractions
as high as 40--50\% (\eg Rudick \etal 2009, 2011, Martel \etal 2012,
Contini \etal 2014). Virgo's relatively low ICL fraction is likely a
function of its dynamical state --- the cluster shows significant
substructure and is still in the process of assembling. As Virgo's
various subgroups are accreted into the main cluster core, the amount of
intracluster light will continue to grow, as the diffuse shells and
streams around M49 and NGC~4365 are incorporated into M87's outer halo
and Virgo's extended ICL component, while additional starlight will be
stripped from galaxies during the strong interactions that that
accompany subcluster mergers.

With the survey complete, we are making the imaging data publicly
available. The full resolution $B$ and $M$ mosaics, along with the
masked/rebinned low surface brightness maps, are available as FITS files
for download at \textsf{http://astroweb.cwru.edu/VirgoSurvey/}.

\acknowledgments

We thank Pat Durrell for useful discussions, along with Laura Ferrarese,
Pat C\^ot\'e, Eric Peng, and other members of the {\sl Next Generation
Virgo Cluster Survey} team. We also thank Charley Knox for his tireless
assistance with the mechanical, optical, and IT systems at the
observatory.  Over the course of this project, J.C.M. has been supported by the NSF
through grants AST-9876143, ASTR-0607526, AST-0707793, and AST-1108964,
as well as by the Research Corporation through a Cottrell Scholarship.
J.J.F. has been supported by the NSF through grants AST-0302030 and
AST-0807873, as well as by the Research Corporation through a Cottrell
College Science Award (7732).

This research has made use of NASA's Astrophysics Data System
Bibliographic Services. The authors further acknowledge use of the
NASA/IPAC Extragalactic Database (NED), which is operated by the Jet
Propulsion Laboratory, California Institute of Technology, under
contract with the National Aeronautics and Space Administration.

{\it Facility:} \facility{CWRU:Schmidt}

\appendix

We begin this section with a general description of the Burrell Schmidt
and the modifications made to enhance its capabilities for deep
surface photometry. We then follow with an overview of our general
observing strategy and a description of each season's observing, noting
changes to the telescope system during the course of the survey.
Finally, we detail our process of data reduction and construction of the
final photometrically-calibrated images.

\medskip

\section{A.1 The Burrell Schmidt}

The Burrell Schmidt is a Case Western Reserve University owned and
operated wide-field Schmidt telescope currently situated
at Kitt Peak National Observatory. Its 24-inch corrector and 36-inch
primary mirror deliver a f/3.5 beam resulting in a plate scale of
97.2\arcsec\ mm$^{-1}$ (Nassau 1945). Two different CCDs were in use on
the telescope over the course of the Virgo survey (see below); each
employed 15 micron pixels, yielding a pixel scale of 1.45\arcsec\
pixel$^{-1}$. Depending on the CCD in use, the field of view of the
camera was either 0.825\degr$\times$1.65\degr\ (2004--2007) or
1.65\degr$\times$1.65\degr\ (2008--2011). While the large pixel scale
undersamples the stellar point spread function of the Schmidt, for our
survey we are more interested in diffuse surface photometry rather than
point source photometry, and the f/3.5 beam works to our advantage here,
delivering a wide field of view onto a single CCD.

A second advantage for the Burrell is that it has a closed, well-baffled
telescope tube. This design minimizes scattered light from off-axis
sources, a potentially serious source of contamination in deep surface
photometry (see, \eg Feldmeier \etal 2002). We also cover the interior
of the telescope tube and the telescope baffling using black Protostar 
flocking with very low reflectivity ($<0.7$\%) to minimize off-axis light
from scattering off the interior of the telescope and reaching the CCD
(see, \eg Pompea \etal 1989).

Reflections between elements in the optical path can also be a source of
image contamination, and here again the Schmidt layout works to our
advantage. The small aperture and fast beam work to deliver a wide field
of view without the need for complicated reimaging or corrective optics,
leaving an optical path with very few elements: corrector, primary,
Newtonian flat, filter, and dewar window. In most seasons (see section
A.3), the filter and dewar window were coated with anti-reflection (AR)
coatings with reflectivities $R <$ 0.5\% (and more typically $R \sim$
0.2\%), matched to the central wavelength of the filter in use. These
reflectivities are significantly less than typical broad-band MgFl AR
coatings ($R \sim$ 1.5\%), and significantly reduce contamination in the
images. Furthermore, we can measure and remove the low level stellar
reflections that remain (see \S A.4 below). As the CCD itself is the
most reflective element in the system (other than the primary and
Newtonian mirrors), the brightest reflections involve bounces between
the CCD and the surfaces of the dewar window and filter, producing halos
of radius 5\arcmin\ and 19\arcmin, respectively. In total, these halos
typically contain $<$0.5\% of the light from the star, and we have
developed a technique for their efficient removal from the images (see
Slater \etal 2009). The only residual feature left behind is a bright
pair of rings at $r\sim$ 1\arcmin\ and 2\arcmin\ due to reflections
within the corrector itself as light first enters the telescope. More
complicated reflections involving multiple bounces between surfaces
blend into the extended profile without leaving distinct features, and
so are removed via subtraction of the PSF.

Schmidt telescopes also possess a unique reflection (the ``Schmidt
ghost'') formed by light which reflects off the CCD, goes back up the
telescope tube, hits the corrector lens, and reflects back down to the
CCD on the opposite side of the optical axis (see, for example, Yang,
Zhu, \& Song 2002). This results in complicated ghosts from bright
stars, displaced opposite the optical axis from the parent star. For
bright stars, we mask the brightest part of the ghost (an irregular
feature approximately 1.5\arcmin\ in size) in the data reduction phase.
For the more extended, low surface brightness portion of the ghost, we
take advantage of the fact that its position with respect to its star
changes as the star's position on the chip changes. Therefore, a
combination of images taken with large dithers minimizes any small
feature that it might otherwise imprint on our data.

To increase the stability of the flat fielding, over the course of the
survey we made several modifications to the support structure for the
Newtonian secondary mirror. This is important since any flexure in the
support can lead to illumination differences in the image plane,
particularly if the secondary mirror is undersized and the full beam is
not reflected over the full field of view. Prior to the start of the
survey, we re-engineered the telescope to move the Newtonian further up
the telescope tube so that the Newtonian was not overfilled by the beam,
while also moving the focal plane inwards to compensate. In 2007, we
redesigned the Newtonian support from a one-armed mount to a traditional
four-sided spider design. A larger Newtonian mirror was also installed,
supported by three Invar flexures glued to the rear of the mirror. This
mount prevents light scattering from the mirror cell onto the CCD.
Baffling is installed above the Newtonian mirror as well to prevent
incoming light from the corrector illuminating the rear of the Newtonian
mirror. All these modifications significantly improved the stability of
the support and reduced scattered light inside the tube.

As noted previously, the CCD imager went through several changes during
the lifetime of the survey. The first generation CCD, in use from
2004--2006, was a SITe ST-002A 2048$\times$4096 back-illuminated CCD
operated with a gain of 2.3 $e^-$/ADU and readnoise of 6.5 $e^-$. In
2007, we installed two of these CCDs in the image plane, but
unfortunately one failed, leaving us again with a single
2048$\times$4096 CCD. There were also grounding problems with this
system, which led to increased electronic noise --- the effective
readnoise for the CCD climbed to $\sim$ 10--20 $e^-$. This system was
replaced in 2008 with a new STA0500A 4K$\times$4K back-illuminated CCD
from the University of Arizona's Imaging Technology Lab, operated with a
gain of 2.7 $e^-$/ADU and readnoise of 4.0 $e^-$. In all seasons we use
a Leach controller (Leach \& Low 2000) and the Voodoo data acquisition
software, under which each 2K$\times$4K chip was read out through two
amplifiers, while the 4K$\times$4K chip was read out through four
amplifiers. The CCD bias frames typically show high frequency structure
(at the $\sim$ 1--2 ADU level) within 25 pixels of the CCD edges which
appear to change slightly with temperature. We mask these regions of the
chip during data process to avoid this additional noise source.

\section{A.2. Observing Strategy}

Over the course of the survey, spring observations with the Burrell
Schmidt concentrated on the Virgo Cluster. Between two and four dark
runs each season were dedicated to this project, with each run lasting
8--12 nights. At the beginning the first run of each season, the CCD
dewar was equipped with a dewar window whose AR coating was specifically
matched to wavelength of the filter used in that season's observations.
The CCD was then mounted and aligned on the telescope, and the filter
install in the filter wheel. Once this process was complete, no further
changes to the system were permitted for the rest of the observing
season. This includes filter changes --- the filter wheel position was
held fixed for the season, and all observations are made in one filter.
This limits variations in the flat fielding due to slight changes in
either the optical alignment of the CCD or the position of the filter,
and reduces dust on the optical surfaces that can adversely affect the
flat fielding.

We chose the filters used for the survey by balancing the need to
maximize signal from the intracluster stellar populations while
minimizing noise in the background sky. ICL populations are expected to
be predominantly old, such that their starlight should peak in the red.
However, the night sky in the red is both brighter and more highly
variable than it is in the blue, due to strong emission lines from
oxygen and OH molecules in the atmosphere (see, \eg Krisciunas \etal
2007, Patat 2008, Neugent \& Massey 2010). We therefore conducted survey
observations from 2004--2008 using the Washington $M$ filter, a
broadband filter similar in width to Johnson $V$, but $\sim$ 250\AA\
bluer. Washington $M$ avoids the bright and variable \ion{O}{1}
$\lambda$5577 night sky emission line present in the redder Johnson $V$
filter (see Figure~1 of Feldmeier 2002), and is thus in the ``sweet
spot'' to balance the competing demands of red starlight and a quiet
blue sky. We then switched to a modified $B$ filter for the 2009 and
2011 seasons, re-imaging the Virgo core and M49 fields to obtain \bmv\
colors for the diffuse ICL in Virgo. Because the Washington $M$ filter
is bluewards of Johnson $V$, to ensure we had a sufficient spectral
baseline to obtain meaningful colors, the $B$ filter we use is bluer
than standard Johnson $B$ by $\sim$ 200\AA. During data reduction, all
magnitudes are transformed into standard Johnson $B$ and $V$ (see the
photometric calibration discussion below), and for simplicity we refer
to the data as our $B$ and $V$ imaging with the understanding that the
filters we use are not actually standard Johnson $B$ and $V$ filters.

To achieve the $\sim$ 0.1--0.3\% accuracy we aim for in the large scale flat
fielding to allow measurements of diffuse light at low surface
brightness, dome or twilight flats are insufficient. Dome
flats illuminate the CCD with light much different in color than the
night sky, introducing color-dependant systematic uncertainties into the
flat, while twilight skies are much bluer and do not uniformly
illuminate the large field of view of the CCD due to the illumination
gradient in the twilight sky. Instead, we use night sky flats, built
from a large number of blank sky pointings taken adjacent in time and at
similar telescope position to the Virgo pointings (see, \eg Morrison
\etal 1997; Feldmeier \etal 2002; Rudick \etal 2010). A typical
observing pattern was two sky frames an hour west of Virgo, four
pointings in Virgo, and two sky pointings an hour east of Virgo which,
given our exposure times of 900s or 1200s in $M$ and $B$, respectively,
ensures that the skies are taken at the same hour angle (and
declination) as the Virgo observations. Flats built this way will thus
consist of skies taken under similar conditions to the Virgo pointing,
minimizing systematics in the flat field due to different gravity
loading of the camera/telescope. The color of the Kitt Peak night sky
(\bmv$\approx 0.9$; Neugent \& Massey 2010) is also similar to the
expected color of the ICL, further reducing systematic uncertainties in
surface photometry of the ICL.

We took data only during photometric nights with the moon below the
horizon. By observing with the moon down, we significantly reduce both
the overall sky brightness as well as the presence of sky gradients due
to scattered moonlight. The requirement for photometric conditions
guards against artificial light from the ground (most notably from
Tucson to the northeast) scattering off cirrus clouds and masquerading
as diffuse light. Furthermore, even thin cirrus will scatter starlight
and introduce extended halos around bright stars which can swamp the
diffuse ICL we are trying to measure. Therefore, our photometric
requirement was quite stringent; the scatter in frame-by-frame
photometric zeropoints in our data was typically 1\%, except for 2007
where weather conditions forced us to relax our criterion somewhat (see
Table~\ref{photsol}. We also limited all observations to be taken at
zenith distances less than 50\degr (airmasses less than 1.5), which
reduces both the absolute sky levels and any gradients in the sky even
further. We do however need sufficient counts in the blank sky images to
build accurate night sky flats. During dark time, the Kitt Peak sky
surface brightness at zenith is roughly 22.8 and 21.9 \magsec\ in $B$
and $V$ respectively (Neugent \& Massey 2010) but shows variation of
half a magnitude or more, with dependencies on time of night, telescope
pointing, and solar cycle (Neugent \& Massey 2010, see also Krisciunus
\etal 2007, Patat 2008). Our observations span the solar minimum between
cycles 24 and 25, and the observed sky brightnesses ranged from
\mub=22.4--21.8 and \muv=21.9--21.2. The exposure times for the blank
sky pointings (1200s and 900s in $B$ and $M$, identical to the Virgo
exposures) thus resulted in instrumental sky levels of 700--1000 ADU and
900--1500 ADU in $B$ and $M$ respectively, giving us a per-pixel
signal-to-noise of $\sim$ 2\%. Combining 50+ such blank sky fields
together thus delivered the $<0.3$\% accuracy we seek.

\begin{figure*}[]
\centerline{\includegraphics[height=6.5truein]{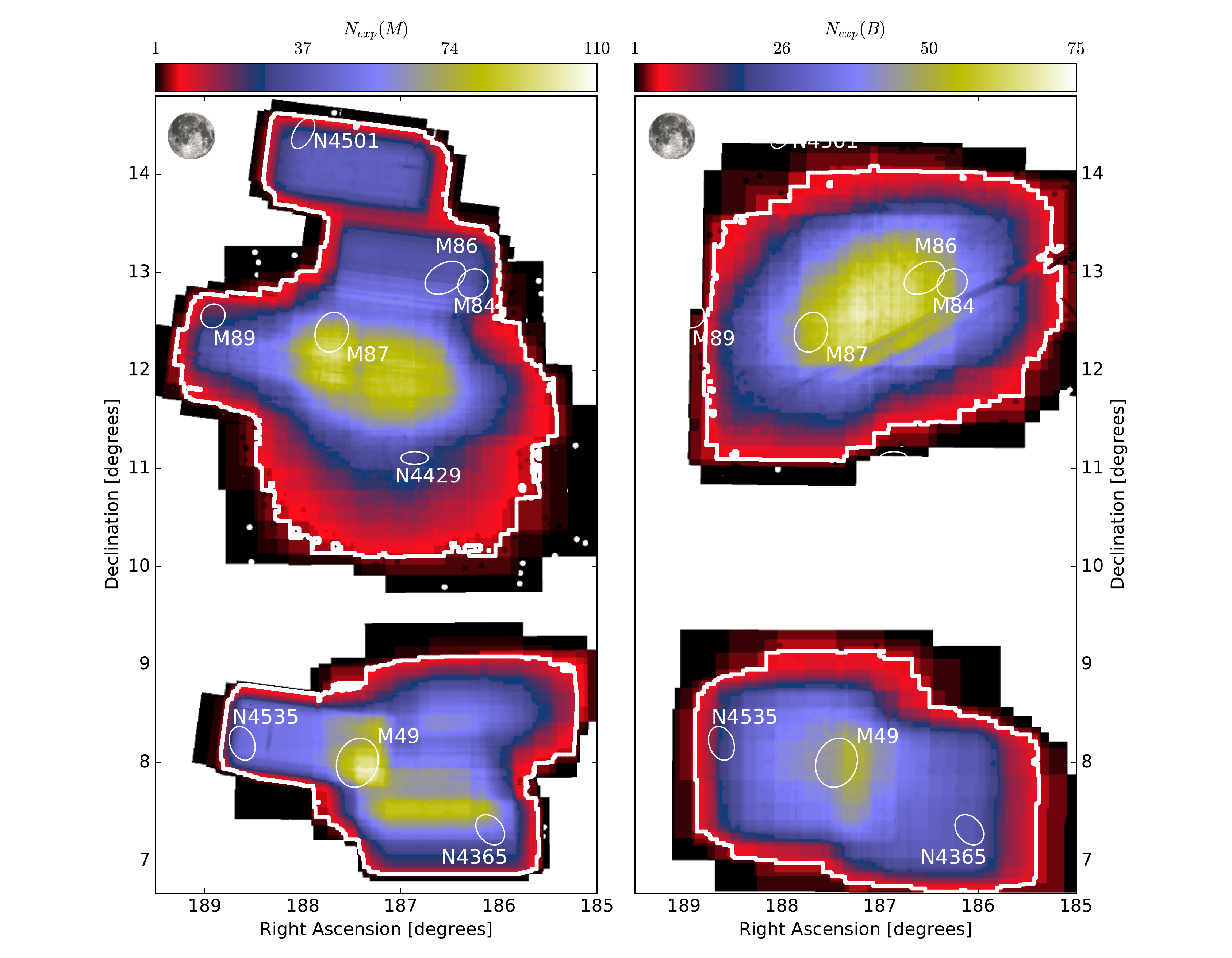}}
\caption{Exposure maps showing the number of images contributing to the
final Virgo mosaics in Washington $M$ (left) and modified Johnson $B$ (right). 
Orientation and scale are as in Figures~\ref{Vmosaic}--\ref{Herschel}. The 
white contours show the minimum exposure boundary of 5 images (100 
and 75 minutes in $B$ and $M$, respectively) that defines our official survey area.
In the $B$-band exposure map, the diagonal striping is due to the masking
of a small comet that moved through the field during the observing season.}
\label{exposuremap}
\end{figure*}

In each season we concentrated our imaging on a different subfield of
the Virgo Cluster (see Table~\ref{obslist}), chosen both for scientific
merit and to avoid, as much as possible, contamination from Galactic
cirrus. In each field, we took 30--100 images, dithering the individual
images by up to half a degree. By dithering on such large scales, we reduced
the effect of large scale flat fielding uncertainties, since in the
final mosaic any particular object will have been imaged multiple
times on a variety of positions on the CCD. The large dithering also
combats contamination from the Schmidt ghost, and widens the areal
coverage of our survey. Because of this seasonal targeting and dithering 
strategy, the full area of the survey has an effective exposure time which
varies across the field; the final exposure maps in $M$ and $B$ are shown in
Figure~\ref{exposuremap}.

\section{A.3. Season Specific Observing Notes}

\paragraph{Spring 2004}

In the first season of imaging, we imaged the core of the Virgo cluster
in the $M$ filter in two fields covering approximately 2.25 deg$^2$.
These fields included the bright ellipticals M87, M86, and M84, and a
reduction of this field was presented in Mihos \etal (2005). In 17
nights over three observing runs, we took 72 images of the Virgo core,
with 63 flanking sky fields for building the night sky flat. We also
took 70 images of the Hercules cluster for a different project, along
with 64 flanking skies. Because Hercules and Virgo are at very similar
declination, the Hercules skies are similar in pointing to the Virgo
skies, and we combined both sets of skies (127 images) to build the
final flat. In this season, both the $M$ filter and the flat dewar
window had standard MgFl antireflection coatings, giving rather strong
reflections around the stars that were later removed in the data
reduction stage.

\paragraph{Spring 2005}

In Spring 2005 we imaged three fields in the $M$ filter --- one to the
north of the Virgo core, one to the southeast of (and including) M87 and
one northeast of (and including) M49. Over two observing runs we
obtained 134 images of Virgo and 103 flanking fields in 15 nights of
observing. Prior to the start of the season, both the primary and
Newtonian mirrors were realuminized, improving throughput and reducing
scattered light.

\paragraph{Spring 2006}

In this season, we imaged a field southwest of M49 in the $M$ filter. We
had 16 nights of data in two observing runs, during which we took 93
images of Virgo and 78 flanking skies. This season also saw the
installation of a combined dewar window / field flattener to improve the
image quality across the large field of view of the CCD. Both the dewar
window and the $M$ filter were coated with low reflectivity AR coatings,
reducing stellar reflections.

\paragraph{Spring 2007}

In 2007, we upgraded the Newtonian mount and mirror and also replaced
the single 2048 $\times$ 4096 CCD with two identical 2048 $\times$ 4096
CCDs butted together. However, as noted in \S A.2, electronics problems
compromised the CCD system --- only one of the two CCDs were
operational, and it showed increased electronic noise. Nonetheless, even
though the data quality suffered, we still imaged fields southwest of
M87 and northwest of M49 in the $M$ filter. The data was taken in 16
nights spread over three observing runs, and consisted of 111 images of
Virgo and 92 flanking skies.

\paragraph{Spring 2008}

In Spring 2008, the 4096 $\times$ 4096 CCD was installed, giving us the
wider field of view. However, when cooled, this chip curved by more than
half the radius of curvature of (and in the same direction as) the
Schmidt focal plane. Better image quality was achieved by re-installing
the flat dewar window (with standard MgFl coating). The Newtonian mirror
was also realuminized at this time. This engineering work gave us less
observing time during the season, but in one science observing run using
the $M$ filter, we obtained 8 nights of data and acquired 51 images of a
field southwest of M87, along with 61 flanking skies.

\paragraph{Spring 2009}

In Spring 2009, we re-imaged the Virgo core field of Spring 2004 in the
$B$ filter. In three runs we had 24 usable nights, during which we
collected 102 Virgo images and 116 flanking skies. With the larger 4K
$\times$ 4K CCD, we were able to cover a field much larger in area than
the original Spring 2004 core field. Before this season, we also
realuminized the primary mirror. The $B$ filter was coated with a very
low reflectivity AR coating which almost completely eliminated
reflections off the filter; however we again used the flat MgFl-coated
dewar window. Finally, during this run a comet (65P/Gunn) passed through
the Virgo cluster, forcing us to aggressively mask the comet and its
tail during the data reduction phase so that it would not contaminate
our measurements of the diffuse light in Virgo.
 
\paragraph{Spring 2011}

In Spring 2011, we reimaged the region around M49 in the $B$ filter.
Over the course of three runs, we obtained 71 Virgo images and 73
flanking skies. In this season we also reinstalled a new flat dewar
window with the aggresive AR coatings matched in wavelength to our $B$
filter.

\begin{deluxetable}{llllll}
\tabletypesize{\scriptsize}
\tablewidth{0pt}
\tablecaption{Observational Datasets}
\tablehead{\colhead{Season} & \colhead{Filter} & \colhead{CCD} &
\colhead{Fields} & \colhead{N$_{\rm Virgo}$} & \colhead{N$_{\rm sky}$}}
\startdata
2004 & M & 2K $\times$ 4K & Core & 72 & 127 \\
2005 & M & 2K $\times$ 4K & North, M87SE, M49NE & 134 & 103\\
2006 & M & 2K $\times$ 4K & M49SW & 93 & 78\\
2007 & M & 2K $\times$ 4K & M87SW, M49NW & 111 & 92\\
2008 & M & 4K $\times$ 4K & M87SW & 51 & 61\\
2009 & B & 4K $\times$ 4K & Core & 102 & 116\\
2011 & B & 4K $\times$ 4K & M49 & 71 & 73
\enddata
\label{obslist}
\end{deluxetable}

\section{A.4. Data Reduction}

\subsection{A.4.1 Initial Steps} 

We reduced the data for each observing season separately, and only
combine datasets during the construction of the final mosaic. We began
data reduction for each image by subtracting overscan in both directions
along the CCD, followed by subtraction of a nightly master zero image
created from a median of 25 individual zero frames taken at the
beginning of each night. These master zeros typically showed very little
structure at the $\sim$ 1 ADU level (a level which corresponds to
surface brightnesses of \mub=30.4 and \muv=29.6 in individual images).
For observations in the 2008, 2009, and 2011 seasons, this step was
followed by a correction for crosstalk in the CCD amplifiers (at the
$\approx$ 0.04\% level), an artifact that was not present in the earlier
CCD electronics.

\subsection{A.4.2 Flat Fielding}

The next step of the process was to construct the night sky flat field
from a combination of the individual flanking sky images (see Morrison
\etal 1997, Feldmeier \etal 2002, and Rudick \etal 2010 for full details
of this process). Sky pointings are chosen to have as few bright stars
as possible; nonetheless with such a wide field of view, the presence of
some bright stars and background galaxies is inevitable. These objects,
along with large scale gradients in the night sky brightness, can
introduce noise into the flat field. To reduce the influence of these
features, we begin by masking bright or extended objects in the sky
fields. For each sky image, we use IRAF's \textsf{objmasks} task to
construct a mask of all features $3.5\sigma$ above the local sky; this
mask is then used as the input mask for a second run of
\textsf{objmasks}, which results in a better sky estimate and allows for
fainter objects to be masked. After this double run of
\textsf{objmasks}, the sky images are binned up 32$\times$32 and and
examined by eye to search for very low surface brightness extended
features such as the wings of bright stars or the presence of galactic
cirrus. These features are then masked by hand or, in cases where the
scattered light from the infrared cirrus contaminated too much of the
image, the image is removed from the process entirely.

Once masked, the sky images still contain mild illumination gradients
caused by night sky variations, which we remove in an iterative fashion.
First, the skies are median-combined to make a preliminary master flat
field. This preliminary flat is applied to each of the sky images, which
are then masked and binned 32$\times$32 and fit for any residual sky
plane. These planes are subtracted from the skies, and the
plane-subtracted skies are then median-combined to make a new flat
field, and the process of plane fitting is repeated. We repeat this
process five times, after which the residual plane fits are typically
consistent with zero.

We test the uncertainty in the seasonal flats by comparing flats
constructed of independent subsets of sky images. The largest noticeable
difference came from comparing flats constructed from data taken during
different observing runs. A division of flats from different runs often
showed an edge-to-edge gradient of \ltsima 0.5\%, which, if left
uncorrected, could imprint residual sky variation at the \muv $\sim$
27.6, \mub $\sim$ 28.0 level. However, these gradients were typically
well modeled by a plane, and our usual strategy was to create a master
flat from all the data in a given season, and then fit and apply a
planar correction term for each observing run to correct them to the
seasonal flat. In two seasons --- 2007 and 2009 --- the run-to-run flat
comparison showed more spatial variations across the chip, an effect
likely due to known problems controlling the CCD temperature during
these seasons. For these seasons, we construct run-specific flat fields
to reduce the data.

\subsection{A.4.3 Photometric calibration}

After flat-fielding the Virgo images, we calculate the photometric
zeropoint directly for each frame using SDSS photometry for the 100--200
SDSS stars brighter than g = 16 typically found in each field. SDSS
$ugriz$ photometry is transformed to Johnson $B$ and $V$ using the
prescription of Lupton, who matched SDSS Data Release 4 photometry with
Stetson's photometric standards\footnote{The Stetson standards are
closely tied to the Landolt standard system, and available at
http://www.cadc-ccda.hia-iha.nrc-cnrc.gc.ca/en/community/STETSON/standards/}.
Lupton's transformation is given by:

$$B = g + 0.3130(g - r) + 0.2271,$$
$$V = g - 0.5784(g - r) - 0.0038$$

\noindent Lupton found that the residuals from these relationships were
small ($\sigma_B=0.011, \sigma_V=0.005$), although it should be
remembered that the SDSS DR4 footprint was restricted to high-latitude
regions, so there are very few stars with \bmv\ less than 0.4 in the
data he used. Comparisons with other \bmv\ data on the Landolt system
for populations containing bluer stars (Morrison \etal, in preparation)
show that Lupton’s transformations may produce \bmv\ values which are
systematically up to 0.05 magnitudes too red for stars bluer than
\bmv=0.4. There are also some small effects due to luminosity and
metallicity which occur for redder stars (\bmv$>$1.0), but they should
be of order 0.02 magnitudes or smaller for stars in the color range
typical of intracluster light.

\begin{figure}[]
\centerline{\includegraphics[height=4.0truein]{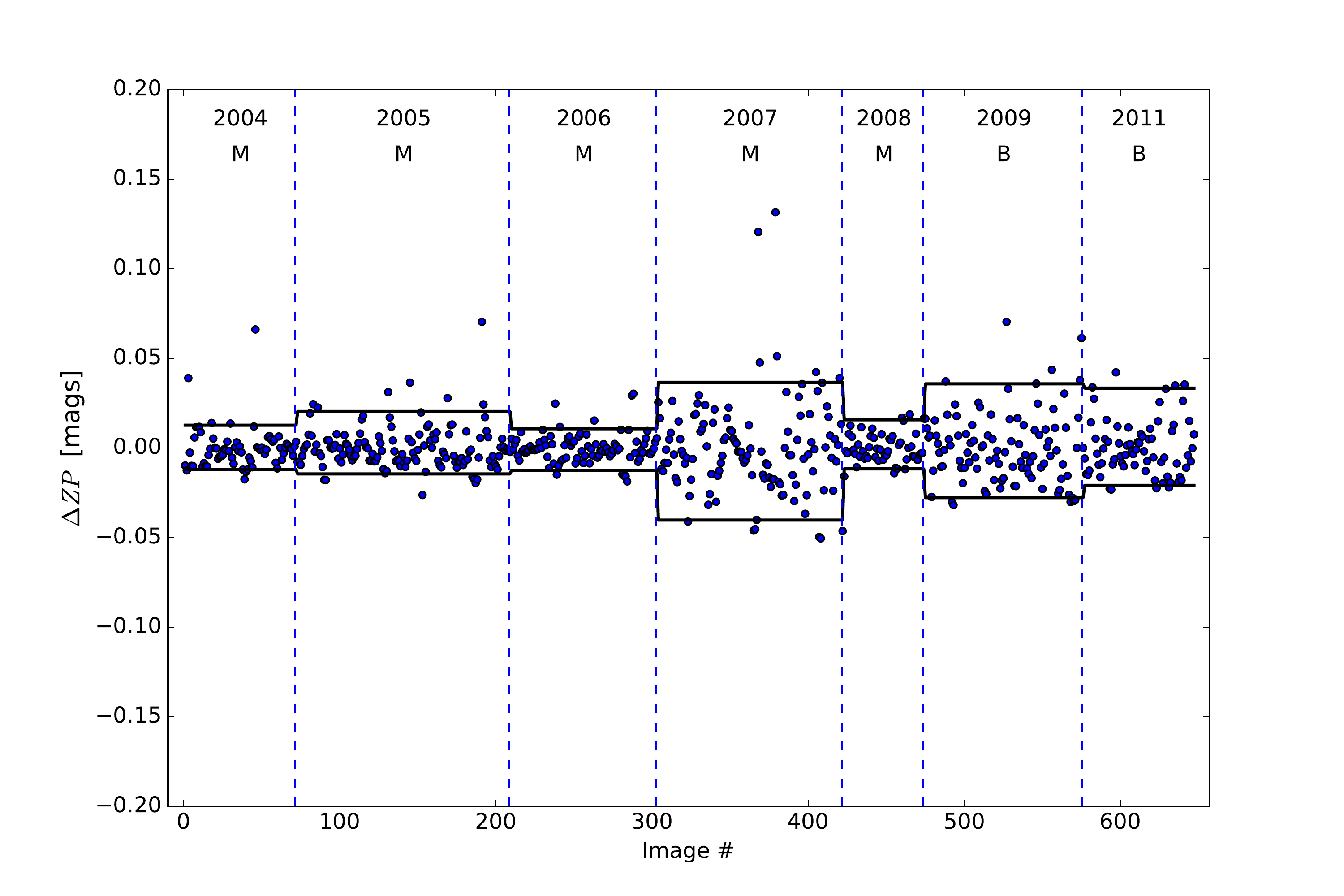}}
\caption{Frame-to-frame variation in photometric zeropoint, after
removal of airmass term and nightly zeropoint offset. Different seasons
are delineated by dashed vertical lines, and the solid horizontal lines
show the 95\% range in each season.}
\label{ZP}
\end{figure}

We then fit our instrumental aperture photometry to the SDSS-derived $B$
and $V$ magnitudes via the expressions:

$$m_i = m_{inst} + C_i(B-V) + K_iX + ZP_{frame}$$

\noindent where $i$ refers to the season-specific filter, $m_i$ and
$B-V$ are the SDSS-derived magnitude and color of the star, $m_{inst}$
is our instrumental magnitude, $C_i$ is the color term, $K_i$ is the
extinction term, $X$ is the airmass, and $ZP_{frame}$ is the frame
zeropoint. Table~\ref{photcal} gives the parameters for the photometric
solutions for each season, while Figure~\ref{ZP} shows the frame-to-frame
variation in photometric zeropoint.

\begin{deluxetable}{ccccc}
\tabletypesize{\scriptsize}
\tablewidth{0pt}
\tablecaption{Photometric Calibration\label{photcal}}
\tablehead{
\colhead{Season} & 
\colhead{Filter} & 
\colhead{$C$} &
\colhead{$K$} & 
\colhead{$\sigma(ZP)$\tablenotemark{a}}
}
\startdata
2004 & M & 0.212 & 0.131 & 0.012\\
2005 & M & 0.216 & 0.148 & 0.012\\
2006 & M & 0.219 & 0.167 & 0.008\\
2007 & M & 0.284 & 0.153 & 0.028\\
2008 & M & 0.297 & 0.174 & 0.006\\
2009 & B & 0.108 & 0.316 & 0.007\\ 
2011 & B & 0.142 & 0.274 & 0.006
\enddata
\tablenotetext{a}{$\sigma(ZP)$ is the dispersion in photometric
zeropoints for each season's images after correction for airmass 
and nightly zeropoint offsets.}
\label{photsol}
\end{deluxetable}

\subsection{A.4.4 Star Subtraction and PSF Effects}

Once the photometric solution has been established, the images are
ready for star subtraction, a process which involves removing both
the reflection halos and extended PSF wings of stars on the images.
This star subtraction must be applied on an
image-by-image basis, rather than on the final mosaic, for the
following reasons. First, the position of the reflections with respect
to their parent star changes as a function of the star's position on
the CCD, and the dithering of the pointings means that the reflections
will differ from image to image. Second, accurate sky levels must be 
modeled and removed prior to constructing the final mosaics, and
if not removed, the extended wings of stars can lead
to systematic errors in the determination of these sky levels.

Our technique for star subtraction is detailed at length in Slater
\etal (2009), and we give only an overview here. Using a series of
deep (450--1200s) images of the bright stars Regulus and Arcturus, we
make a model for the intensity and position of reflections between the
CCD, dewar window, and filter as a function of the star's position on
the CCD. After subtracting out these reflections, we characterize the
extended stellar PSF out to 1\degr\ scales using a 5th order
polynomial.  We then also take a series of short (90s, 10s, and 1s)
exposures of the Virgo fields to obtain accurate instrumental magnitudes for the
bright ($m_V=8-12$) stars in our target fields. Armed with accurate
magnitudes and a well-determined model for stellar reflections and the
PSF, we can subtract the extended halos of these stars from each of
the Virgo images as detailed in Slater \etal (2009).

One step we do not take is to attempt any deconvolution of the images
using the measured Schmidt PSF. In some telescope systems, the
instrumental PSF can scatter a significant amount of light outwards from
the inner regions of a galaxy, artificially changing the surface
brightness profile and colors of the low surface brightness outskirts
(see Sandin 2014, 2015, Duc \etal 2015, Trujillo \& Fliri 2016 for
recent discussions). A variety of arguments suggest this effect is
minimal across our survey area down to our limiting magnitudes. First,
the aggressive AR coatings used on our filters and dewar window limit
the bright halos of scattered light that plague data taken on systems
using standard MgFl coatings. Second, scattered light from the PSF would
tend to make the outer isophotes of bright ellipticals rounder at large
radius, yet we observe the opposite -- higher ellipticities at large
radius (Janowiecki \etal 2010, this paper). Third, our updated
reconstructions of the Burrell Schmidt PSF (Watkins \etal 2016) extend
to 1\degr and show that the reflections and extended wings of the PSF
are slightly brighter in $M$ than in $B$. This argues that the scattered
galaxy light would be red, in contrast to the color profiles we derive
for bright Virgo ellipticals which show modest blueward gradients
(Rudick \etal 2010, Mihos \etal 2013a, this paper). We also find no
evidence that the Burrell Schmidt produces consistently biased colors at
low surface brightness -- in other studies, we have found a variety of
color gradients in galaxy outskirts, both red and blue (Mihos \etal
2013b, Watkins \etal 2014, 2016). Fourth, detailed modeling of the effect
using the Burrell Schmidt PSF shows that color biases of 0.1 magnitudes
or more show up only below surface brightnesses of \mub $\sim$ 28.0,
where the color uncertainty is dominated by other sources of error such
as background sources and sky subtraction (Watkins \etal 2016). Finally,
since any such biases only occur in the outskirts of bright galaxies,
the bulk of the intracluster environment we survey should be
uninfluenced by scattered light from bright galaxies in the field.

\subsection{A.4.5 Sky Subtraction}

Accurate sky subtraction is critical for quantitative surface photometry
at faint surface brightnesses. The problem of sky subtraction is
particularly difficult in our imaging, for a number of reasons. First,
the angular scale of the Virgo cluster is so large that our images never
extend outside the cluster; obtaining a ``non-Virgo background'' is
impossible. Second, many of the Virgo galaxies themselves so large that
their extended low surface bright envelopes dominate the field of view
of a single image. Lastly, regions of Virgo are contaminated by galactic
cirrus, which produces an extended component of diffuse scattered
galactic light in the images. These features begin to show up at surface
brightnesses of $\mu_V \sim$ 27, corresponding to $\sim$ 8 ADU in our
imaging --- comparable to the gradients in the night sky across a single
image. With all these sources of diffuse emission in our Virgo fields,
it is often difficult to find regions of ``pure sky'' in our images with
which we can determine the absolute sky level and measure sky gradients.
As a result, sky subtraction is particularly problematic in our imaging,
and is the largest source of uncertainty in doing absolute photometry
over large angular scales in our final mosaics.

To alleviate these issues as much as possible, we first use a double
application of \textsf{objmasks} to aggressively mask sources --- Virgo
galaxies, background objects, satellite trails, and star subtraction
residuals --- in each star subtracted image. Even this masking leaves
behind the extended low surface brightness envelopes of the giant
ellipticals (most notably M87, M86, M84, and M49). We next use the
isophotal fits of Janowiecki \etal 2010 --- based on an earlier
reduction of our imaging --- to define elliptical masks around these
galaxies out to a radius where their light profiles drop below 1 ADU.
After these masks are applied, we median bin the images into 32x32 pixel
bins, and use these binned images to fit the night sky background across
the image. Because of the presence of very low surface brightness
residual structure in our images -- due to galactic cirrus, extended
galaxy halos, and possible gradients in the underlying ICL itself --
high order non-linear fits could be unduly biased by the spatial
distribution of these features, and subtracting such sky fits run the
risk of imprinting artificial structure onto our reduced images, or
possibly removing true Virgo ICL. Therefore, we take a conservative
approach and only fit planes to the sky background for removal.

Finally, we note that in Mihos \etal (2005) we used an iterative method
for fitting skies which mapped the deviations of each sky-subtracted
image from the median of the the ensemble to determine subsequent
corrections to the fitted sky plane. Subsequent testing has shown this
technique does not result in significantly improved sky fits, and under
certain conditions it can drive the plane fits away from convergence. In
the final data reduction pipeline detailed here, we take the more
conservative and simple approach of only fitting a single plane to each
image to subtract as sky.

\subsection{A.4.6 Final Mosaicing}

To build the final mosaic, we first create a series of six 6750 $\times$
6750 pixel master subimages that spatially cover the Virgo pointings and
have a pixel scale matched to that of the original images (1.45\arcsec\ 
pixel$^{-1}$). For each of these subimages, we use IRAF's
\textsf{wregister} task to register the individual images to the master
subimage, then scale them to a common photometric zeropoint. We then
median combine all the individual images together into the master
mosaics using \textsf{imcombine} with a 3$\sigma$ rejection. For this
final combine, we mask any obvious artifacts on the images, such as
glints from bright stars which fell on the edge of the CCD, and we also
conservatively mask pixels within 100 pixels of the CCD edges to avoid
the temperature-sensitive ``edge noise'' mentioned in Section~A.1.

In order to probe the faintest surface brightnesses possible, once the
final mosaics have been constructed, we spatially rebin the mosaics over
larger area, medianing the pixel values over these larger scales. Before
rebinning, however, we mask the images of discrete sources such as
stars, bright galaxies, and fainter background objects which would
otherwise bias the measurement of diffuse light on larger scales. The
masking takes place in several steps. First, we use IRAF's
\textsf{rmedian} task to run a ring median of size 10 pixels
(14.5\arcsec) across each mosaic, then subtract this ring median image
from the original to produce an image akin to an unsharp masked image
which emphasizes discrete sources on small scales. We then run the
\textsf{objmask} task on this image, detecting objects which are
5$\sigma$ above the background level. Pixels in objects detected this
way are then masked in the original mosaics. We then spatially rebin the
mosaics in 9$\times$9 pixel (13\arcsec$\times$13\arcsec) bins,
calculating the median intensity of unmasked pixels in each spatial bin
to create. During this process, spatial bins in which more than 50\% of
the pixels are masked are themselves masked. The deep masked/binned
images shown in Figure~\ref{Vmosaic},~\ref{Bcolor}, and \ref{Herschel}b
have all been processed this way.

Finally, to compute the final photometric solution for the mosaics, we
made a number of simplifying assumptions. Because the color terms for
the photometric solution varied in each season, we simply adopt an
average color term of $C_M = 0.246$ and $C_B = 0.125$. Furthermore,
because the spatial footprint of our survey is different in $B$ and $M$,
we only have color information over a subset of our survey area. In
regions where we do not having imaging in both filters, we use a \bmv\
color of 0.7 (similar to the color of the diffuse ICL; Rudick \etal
2010) in the photometric solution to calculate surface brightnesses in
$B$ and $V$. If the intrinsic colors of objects are much bluer than this
(for example, near spiral galaxies or galactic cirrus, where the colors
will be a bluer \bmv $\sim$ 0.4 or so), this will systematically
underestimate the surface brightness by about 0.1 \magsec\ at most. At
the faint surface brightnesses we are most interested in (\muv $>$ 27),
the uncertainty in sky will begin to dominate over this systematic
uncertainty in color. With these choices, our final photometric solution
is given by:

$$m_B = -2.5\log_{10}(I_B) + 0.125(B - V) + 28.973$$
$$m_V = -2.5\log_{10}(I_M) + 0.246(B - V) + 28.606$$

where $I_B$ and $I_M$ are the pixel values (in ADU) in the $B$ and $M$
mosaics, respectively.

\bigskip

\begin{thebibliography}{}


\bibitem[Ahn et al.(2012)]{2012ApJS..203...21A} Ahn, C.~P., Alexandroff, R., Allende Prieto, C., et al.\ 2012, \apjs, 203, 21 


\bibitem[Alam et al.(2015)]{2015ApJS..219...12A} Alam, S., Albareti, F.~D., Allende Prieto, C., et al.\ 2015, \apjs, 219, 12 


\bibitem[Arnaboldi et al.(1996)]{1996ApJ...472..145A} Arnaboldi, M., Freeman, K.~C., Mendez, R.~H., et al.\ 1996, \apj, 472, 145 


\bibitem[Arnaboldi et al.(2012)]{2012A&A...545A..37A} Arnaboldi, M., Ventimiglia, G., Iodice, E., Gerhard, O., \& Coccato, L.\ 2012, \aap, 545, A37 


\bibitem[Arnaboldi et al.(2002)]{2002AJ....123..760A} Arnaboldi, M., Aguerri, J.~A.~L., Napolitano, N.~R., et al.\ 2002, \aj, 123, 760 


\bibitem[Arnaboldi \& Gerhard(2010)]{2010HiA....15...97A} Arnaboldi, M., \& Gerhard, O.\ 2010, Highlights of Astronomy, 15, 97 


\bibitem[Arnaboldi et al.(2004)]{2004ApJ...614L..33A} Arnaboldi, M., Gerhard, O., Aguerri, J.~A.~L., et al.\ 2004, \apjl, 614, L33 


\bibitem[Arp \& Bertola(1969)]{1969ApL.....4...23A} Arp, H., \& Bertola, F.\ 1969, \aplett, 4, 23 


\bibitem[Arrigoni Battaia et al.(2012)]{2012A&A...543A.112A} Arrigoni Battaia, F., Gavazzi, G., Fumagalli, M., et al.\ 2012, \aap, 543, A112 


\bibitem[B{\"o}hringer et al.(1994)]{1994Natur.368..828B} B{\"o}hringer, H., Briel, U.~G., Schwarz, R.~A., et al.\ 1994, \nat, 368, 828 


\bibitem[Bakos et al.(2008)]{2008ApJ...683L.103B} Bakos, J., Trujillo, I., \& Pohlen, M.\ 2008, \apjl, 683, L103 


\bibitem[Barbary et al.(2012)]{2012ApJ...745...32B} Barbary, K., Aldering, G., Amanullah, R., et al.\ 2012, \apj, 745, 32 


\bibitem[Barbosa et al.(2016)]{2016A&A...589A.139B} Barbosa, C.~E., Arnaboldi, M., Coccato, L., et al.\ 2016, \aap, 589, A139 


\bibitem[Bautz \& Morgan(1970)]{1970ApJ...162L.149B} Bautz, L.~P., \& Morgan, W.~W.\ 1970, \apjl, 162, L149


\bibitem[Beasley et al.(2016)]{2016ApJ...819L..20B} Beasley, M.~A., Romanowsky, A.~J., Pota, V., et al.\ 2016, \apjl, 819, L20 


\bibitem[Becker et al.(1995)]{1995ApJ...450..559B} Becker, R.~H., White, R.~L., \& Helfand, D.~J.\ 1995, \apj, 450, 559 


\bibitem[Bekki et al.(2003)]{2003MNRAS.344..399B} Bekki, K., Couch, W.~J., Drinkwater, M.~J., \& Shioya, Y.\ 2003, \mnras, 344, 399 


\bibitem[Bender \& Surma(1992)]{1992A&A...258..250B} Bender, R., \& Surma, P.\ 1992, \aap, 258, 250 


\bibitem[Bender et al.(2015)]{2015ApJ...807...56B} Bender, R., Kormendy, J., Cornell, M.~E., \& Fisher, D.~B.\ 2015, \apj, 807, 56 


\bibitem[Bender et al.(1989)]{1989A&A...217...35B} Bender, R., Surma, P., Doebereiner, S., Moellenhoff, C., \& Madejsky, R.\ 1989, \aap, 217, 35 


\bibitem[Bernstein et al.(1995)]{1995AJ....110.1507B} Bernstein, G.~M., Nichol, R.~C., Tyson, J.~A., Ulmer, M.~P., \& Wittman, D.\ 1995, \aj, 110, 1507 


\bibitem[Bianchi et al.(2016)]{2016arXiv160905941B} Bianchi, S., Giovanardi, C., Smith, M.~W.~L., et al.\ 2016, arXiv:1609.05941 


\bibitem[Binggeli et al.(1993)]{1993A&AS...98..275B} Binggeli, B., Popescu, C.~C., \& Tammann, G.~A.\ 1993, \aaps, 98, 275 


\bibitem[Binggeli et al.(1985)]{1985AJ.....90.1681B} Binggeli, B., Sandage, A., \& Tammann, G.~A.\ 1985, \aj, 90, 1681 


\bibitem[Binggeli et al.(1987)]{1987AJ.....94..251B} Binggeli, B., Tammann, G.~A., \& Sandage, A.\ 1987, \aj, 94, 251 


\bibitem[Blom et al.(2012)]{2012MNRAS.426.1959B} Blom, C., Forbes, D.~A., Brodie, J.~P., et al.\ 2012, \mnras, 426, 1959 


\bibitem[Blom et al.(2014)]{2014MNRAS.439.2420B} Blom, C., Forbes, D.~A., Foster, C., Romanowsky, A.~J., \& Brodie, J.~P.\ 2014, \mnras, 439, 2420 


\bibitem[Blom et al.(2012)]{2012MNRAS.420...37B} Blom, C., Spitler, L.~R., \& Forbes, D.~A.\ 2012, \mnras, 420, 37 


\bibitem[Bogd{\'a}n et al.(2012)]{2012ApJ...753..140B} Bogd{\'a}n, {\'A}., Forman, W.~R., Zhuravleva, I., et al.\ 2012, \apj, 753, 140 


\bibitem[Boselli et al.(2011)]{2011A&A...528A.107B} Boselli, A., Boissier, S., Heinis, S., et al.\ 2011, \aap, 528, A107 


\bibitem[Brandt \& Draine(2012)]{2012ApJ...744..129B} Brandt, T.~D., \& Draine, B.~T.\ 2012, \apj, 744, 129 


\bibitem[Bruzual \& Charlot(2003)]{2003MNRAS.344.1000B} Bruzual, G., \& Charlot, S.\ 2003, \mnras, 344, 1000 


\bibitem[Burke et al.(2012)]{2012MNRAS.425.2058B} Burke, C., Collins, C.~A., Stott, J.~P., \& Hilton, M.\ 2012, \mnras, 425, 2058 


\bibitem[Burke et al.(2015)]{2015MNRAS.449.2353B} Burke, C., Hilton, M., \& Collins, C.\ 2015, \mnras, 449, 2353 


\bibitem[Busko(1996)]{1996ASPC..101..139B} Busko, I.~C.\ 1996, Astronomical Data Analysis Software and Systems V, 101, 139 


\bibitem[Caldwell(2006)]{2006ApJ...651..822C} Caldwell, N.\ 2006, \apj, 651, 822 


\bibitem[Capaccioli et al.(2015)]{2015A&A...581A..10C} Capaccioli, M., Spavone, M., Grado, A., et al.\ 2015, \aap, 581, A10 


\bibitem[Cappellari et al.(2011)]{2011MNRAS.413..813C} Cappellari, M., Emsellem, E., Krajnovi{\'c}, D., et al.\ 2011, \mnras, 413, 813 


\bibitem[Carter \& Dixon(1978)]{1978AJ.....83..574C} Carter, D., \& Dixon, K.~L.\ 1978, \aj, 83, 574 


\bibitem[Castro-Rodrigu{\'e}z et al.(2009)]{2009A&A...507..621C} Castro-Rodrigu{\'e}z, N., Arnaboldi, M., Aguerri, J.~A.~L., et al.\ 2009, \aap, 507, 621 


\bibitem[Ciardullo et al.(2004)]{2004IAUS..217...88C} Ciardullo, R., Mihos, J.~C., Feldmeier, J.~J., Durrell, P.~R., \& Sigurdsson, S.\ 2004, Recycling Intergalactic and Interstellar Matter, 217, 88 


\bibitem[Condon et al.(1998)]{1998AJ....115.1693C} Condon, J.~J., Cotton, W.~D., Greisen, E.~W., et al.\ 1998, \aj, 115, 1693 


\bibitem[Conroy et al.(2007)]{2007ApJ...668..826C} Conroy, C., Wechsler, R.~H., \& Kravtsov, A.~V.\ 2007, \apj, 668, 826 


\bibitem[Contini et al.(2014)]{2014MNRAS.437.3787C} Contini, E., De Lucia, G., Villalobos, {\'A}., \& Borgani, S.\ 2014, \mnras, 437, 3787 


\bibitem[Cooper et al.(2015)]{2015MNRAS.451.2703C} Cooper, A.~P., Gao, L., Guo, Q., et al.\ 2015, \mnras, 451, 2703 


\bibitem[Cort{\'e}s et al.(2008)]{2008ApJ...683...78C} Cort{\'e}s, J.~R., Kenney, J.~D.~P., \& Hardy, E.\ 2008, \apj, 683, 78-93 


\bibitem[Cortese et al.(2010)]{2010MNRAS.403L..26C} Cortese, L., Bendo, G.~J., Isaak, K.~G., Davies, J.~I., \& Kent, B.~R.\ 2010, \mnras, 403, L26 


\bibitem[Cortese et al.(2004)]{2004A&A...416..119C} Cortese, L., Gavazzi, G., Boselli, A., \& Iglesias-Paramo, J.\ 2004, \aap, 416, 119 


\bibitem[Cui et al.(2014)]{2014MNRAS.437..816C} Cui, W., Murante, G., Monaco, P., et al.\ 2014, \mnras, 437, 816 


\bibitem[D'Onghia et al.(2005)]{2005ApJ...630L.109D} D'Onghia, E., Sommer-Larsen, J., Romeo, A.~D., et al.\ 2005, \apjl, 630, L109 


\bibitem[D'Onghia et al.(2009)]{2009Natur.460..605D} D'Onghia, E., Besla, G., Cox, T.~J., \& Hernquist, L.\ 2009, \nat, 460, 605 


\bibitem[D'Souza et al.(2014)]{2014MNRAS.443.1433D} D'Souza, R., Kauffman, G., Wang, J., \& Vegetti, S.\ 2014, \mnras, 443, 1433


\bibitem[Da Rocha \& Mendes de Oliveira(2005)]{2005MNRAS.364.1069D} Da Rocha, C., \& Mendes de Oliveira, C.\ 2005, \mnras, 364, 1069


\bibitem[Da Rocha et al.(2008)]{2008MNRAS.388.1433D} Da Rocha, C., Ziegler, B.~L., \& Mendes de Oliveira, C.\ 2008, \mnras, 388, 1433


\bibitem[Davies et al.(2010)]{2010A&A...518L..48D} Davies, J.~I., Baes, M., Bendo, G.~J., et al.\ 2010, \aap, 518, L48 


\bibitem[Davies et al.(2012)]{2012MNRAS.419.3505D} Davies, J.~I., Bianchi, S., Cortese, L., et al.\ 2012, \mnras, 419, 3505 


\bibitem[De Lucia \& Blaizot(2007)]{2007MNRAS.375....2D} De Lucia, G., \& Blaizot, J.\ 2007, \mnras, 375, 2 


\bibitem[de Vaucouleurs \& de Vaucouleurs(1970)]{1970ApL.....5..219D} de Vaucouleurs, G., \& de Vaucouleurs, A.\ 1970, \aplett, 5, 219 


\bibitem[de Vaucouleurs(1961)]{1961ApJS....6..213D} de Vaucouleurs, G.\ 1961, \apjs, 6, 213 


\bibitem[Dilday et al.(2010)]{2010ApJ...715.1021D} Dilday, B., Bassett, B., Becker, A., et al.\ 2010, \apj, 715, 1021 


\bibitem[Doherty et al.(2009)]{2009A&A...502..771D} Doherty, M., Arnaboldi, M., Das, P., et al.\ 2009, \aap, 502, 771 


\bibitem[Dolag et al.(2010)]{2010MNRAS.405.1544D} Dolag, K., Murante, G., \& Borgani, S.\ 2010, \mnras, 405, 1544 


\bibitem[Donahue et al.(2002)]{2002AJ....123.1922D} Donahue, M., Smith, B.~J., \& Stocke, J.~T.\ 2002, \aj, 123, 1922 


\bibitem[Duc et al.(2000)]{2000AJ....120.1238D} Duc, P.-A., Brinks, E., Springel, V., et al.\ 2000, \aj, 120, 1238 


\bibitem[Duc et al.(2015)]{2015MNRAS.446..120D} Duc, P.-A., Cuillandre, J.-C., Karabal, E., et al.\ 2015, \mnras, 446, 120


\bibitem[Durbala et al.(2008)]{2008AJ....135..130D} Durbala, A., del Olmo, A., Yun, M.~S., et al.\ 2008, \aj, 135, 130 


\bibitem[Durrell et al.(2014)]{2014ApJ...794..103D} Durrell, P.~R., C{\^o}t{\'e}, P., Peng, E.~W., et al.\ 2014, \apj, 794, 103 


\bibitem[Durrell et al.(2002)]{2002ApJ...570..119D} Durrell, P.~R., Ciardullo, R., Feldmeier, J.~J., Jacoby, G.~H., \& Sigurdsson, S.\ 2002, \apj, 570, 119 


\bibitem[Emsellem et al.(2011)]{2011MNRAS.414..888E} Emsellem, E., Cappellari, M., Krajnovi{\'c}, D., et al.\ 2011, \mnras, 414, 888 


\bibitem[Feldmeier et al.(2004)]{2004ApJ...615..196F} Feldmeier, J.~J., Ciardullo, R., Jacoby, G.~H., \& Durrell, P.~R.\ 2004, \apj, 615, 196 


\bibitem[Feldmeier et al.(2004)]{2004ApJ...609..617F} Feldmeier, J.~J., Mihos, J.~C., Morrison, H.~L., et al.\ 2004, \apj, 609, 617 


\bibitem[Feldmeier et al.(2002)]{2002ApJ...575..779F} Feldmeier, J.~J., Mihos, J.~C., Morrison, H.~L., Rodney, S.~A., \& Harding, P.\ 2002, \apj, 575, 779 


\bibitem[Ferguson et al.(1998)]{1998Natur.391..461F} Ferguson, H.~C., Tanvir, N.~R., \& von Hippel, T.\ 1998, \nat, 391, 461 


\bibitem[Ferrarese et al.(2012)]{2012ApJS..200....4F} Ferrarese, L., C{\^o}t{\'e}, P., Cuillandre, J.-C., et al.\ 2012, \apjs, 200, 4 


\bibitem[Ferrarese et al.(2016)]{2016ApJ...824...10F} Ferrarese, L., C{\^o}t{\'e}, P., S{\'a}nchez-Janssen, R., et al.\ 2016, \apj, 824, 10 


\bibitem[Finkbeiner et al.(1999)]{1999ApJ...524..867F} Finkbeiner, D.~P., Davis, M., \& Schlegel, D.~J.\ 1999, \apj, 524, 867 


\bibitem[Frei \& Gunn(1994)]{1994AJ....108.1476F} Frei, Z., \& Gunn, J.~E.\ 1994, \aj, 108, 1476 


\bibitem[Gal-Yam et al.(2003)]{2003AJ....125.1087G} Gal-Yam, A., Maoz, D., Guhathakurta, P., \& Filippenko, A.~V.\ 2003, \aj, 125, 1087 


\bibitem[Gavazzi et al.(1999)]{1999MNRAS.304..595G} Gavazzi, G., Boselli, A., Scodeggio, M., Pierini, D., \& Belsole, E.\ 1999, \mnras, 304, 595 


\bibitem[Gerhard et al.(2007)]{2007A&A...468..815G} Gerhard, O., Arnaboldi, M., Freeman, K.~C., et al.\ 2007, \aap, 468, 815 


\bibitem[Gerhard et al.(2002)]{2002ApJ...580L.121G} Gerhard, O., Arnaboldi, M., Freeman, K.~C., \& Okamura, S.\ 2002, \apjl, 580, L121 


\bibitem[Giallongo et al.(2014)]{2014ApJ...781...24G} Giallongo, E., Menci, N., Grazian, A., et al.\ 2014, \apj, 781, 24 


\bibitem[Giovanelli et al.(2005)]{2005AJ....130.2598G} Giovanelli, R., Haynes, M.~P., Kent, B.~R., et al.\ 2005, \aj, 130, 2598 


\bibitem[Girardi et al.(1995)]{1995ApJ...438..527G} Girardi, M., Biviano, A., Giuricin, G., Mardirossian, F., \& Mezzetti, M.\ 1995, \apj, 438, 527 


\bibitem[Gonzalez et al.(2005)]{2005ApJ...618..195G} Gonzalez, A.~H., Zabludoff, A.~I., \& Zaritsky, D.\ 2005, \apj, 618, 195 


\bibitem[Gonzalez et al.(2007)]{2007ApJ...666..147G} Gonzalez, A.~H., Zaritsky, D., \& Zabludoff, A.~I.\ 2007, \apj, 666, 147 


\bibitem[Gregg \& West(1998)]{1998Natur.396..549G} Gregg, M.~D., \& West, M.~J.\ 1998, \nat, 396, 549 


\bibitem[Guennou et al.(2012)]{2012A&A...537A..64G} Guennou, L., Adami, C., Da Rocha, C., et al.\ 2012, \aap, 537, A64 


\bibitem[Guhathakurta \& Tyson(1989)]{1989ApJ...346..773G} Guhathakurta, P., \& Tyson, J.~A.\ 1989, \apj, 346, 773 


\bibitem[Guo et al.(2011)]{2011MNRAS.413..101G} Guo, Q., White, S., Boylan-Kolchin, M., et al.\ 2011, \mnras, 413, 101 


\bibitem[Hester et al.(2010)]{2010ApJ...716L..14H} Hester, J.~A., Seibert, M., Neill, J.~D., et al.\ 2010, \apjl, 716, L14 


\bibitem[Hibbard et al.(1994)]{1994AJ....107...67H} Hibbard, J.~E., Guhathakurta, P., van Gorkom, J.~H., \& Schweizer, F.\ 1994, \aj, 107, 67 


\bibitem[Hummel et al.(1986)]{1986A&A...155..161H} Hummel, E., Kotanyi, C.~G., \& van Gorkom, J.~H.\ 1986, \aap, 155, 161 


\bibitem[Ienaka et al.(2013)]{2013ApJ...767...80I} Ienaka, N., Kawara, K., Matsuoka, Y., et al.\ 2013, \apj, 767, 80 


\bibitem[Impey et al.(1988)]{1988ApJ...330..634I} Impey, C., Bothun, G., \& Malin, D.\ 1988, \apj, 330, 634 


\bibitem[Irwin \& Sarazin(1996)]{1996ApJ...471..683I} Irwin, J.~A., \& Sarazin, C.~L.\ 1996, \apj, 471, 683 


\bibitem[Janowiecki et al.(2010)]{2010ApJ...715..972J} Janowiecki, S., Mihos, J.~C., Harding, P., et al.\ 2010, \apj, 715, 972 


\bibitem[Jedrzejewski(1987)]{1987MNRAS.226..747J} Jedrzejewski, R.~I.\ 1987, \mnras, 226, 747 


\bibitem[Jerjen et al.(2004)]{2004AJ....127..771J} Jerjen, H., Binggeli, B., \& Barazza, F.~D.\ 2004, \aj, 127, 771 


\bibitem[Jord{\'a}n et al.(2004)]{2004AJ....127...24J} Jord{\'a}n, A., C{\^o}t{\'e}, P., West, M.~J., et al.\ 2004, \aj, 127, 24 


\bibitem[Katsiyannis et al.(1998)]{1998A&AS..132..387K} Katsiyannis, A.~C., Kemp, S.~N., Berry, D.~S., \& Meaburn, J.\ 1998, \aaps, 132, 387 


\bibitem[Kelson et al.(2002)]{2002ApJ...576..720K} Kelson, D.~D., Zabludoff, A.~I., Williams, K.~A., et al.\ 2002, \apj, 576, 720


\bibitem[Kenney et al.(2014)]{2014ApJ...780..119K} Kenney, J.~D.~P., Geha, M., J{\'a}chym, P., et al.\ 2014, \apj, 780, 119 


\bibitem[Khochfar \& Burkert(2005)]{2005MNRAS.359.1379K} Khochfar, S., \& Burkert, A.\ 2005, \mnras, 359, 1379 


\bibitem[Kim et al.(1992)]{1992ApJ...393..134K} Kim, D.-W., Fabbiano, G., \& Trinchieri, G.\ 1992, \apj, 393, 134 


\bibitem[Kim et al.(2014)]{2014ApJS..215...22K} Kim, S., Rey, S.-C., Jerjen, H., et al.\ 2014, \apjs, 215, 22 


\bibitem[Koda et al.(2015)]{2015ApJ...807L...2K} Koda, J., Yagi, M., Yamanoi, H., \& Komiyama, Y.\ 2015, \apjl, 807, L2 


\bibitem[Kormendy et al.(2009)]{2009ApJS..182..216K} Kormendy, J., Fisher, D.~B., Cornell, M.~E., \& Bender, R.\ 2009, \apjs, 182, 216 


\bibitem[Kraft et al.(2011)]{2011ApJ...727...41K} Kraft, R.~P., Forman, W.~R., Jones, C., et al.\ 2011, \apj, 727, 41 


\bibitem[Krick \& Bernstein(2007)]{2007AJ....134..466K} Krick, J.~E., \& Bernstein, R.~A.\ 2007, \aj, 134, 466 


\bibitem[Krisciunas et al.(2007)]{2007PASP..119..687K} Krisciunas, K., Semler, D.~R., Richards, J., et al.\ 2007, \pasp, 119, 687 


\bibitem[Leach \& Low(2000)]{2000SPIE.4008..337L} Leach, R.~W., \& Low, F.~J.\ 2000, \procspie, 4008, 337


\bibitem[Lee et al.(2000)]{2000ApJ...530L..17L} Lee, H., Richer, M.~G., \& McCall, M.~L.\ 2000, \apjl, 530, L17


\bibitem[Lee et al.(2010)]{2010Sci...328..334L} Lee, M.~G., Park, H.~S., \& Hwang, H.~S.\ 2010, Science, 328, 334 


\bibitem[Lelli et al.(2015)]{2015A&A...584A.113L} Lelli, F., Duc, P.-A., Brinks, E., et al.\ 2015, \aap, 584, A113 


\bibitem[Lin \& Mohr(2004)]{2004ApJ...617..879L} Lin, Y.-T., \& Mohr, J.~J.\ 2004, \apj, 617, 879 


\bibitem[Liu et al.(2015)]{2015ApJ...812...34L} Liu, C., Peng, E.~W., C{\^o}t{\'e}, P., et al.\ 2015, \apj, 812, 34 


\bibitem[Liu et al.(2005)]{2005AJ....129.2628L} Liu, Y., Zhou, X., Ma, J., et al.\ 2005, \aj, 129, 2628 


\bibitem[Longobardi et al.(2015)]{2015A&A...579L...3L} Longobardi, A., Arnaboldi, M., Gerhard, O., \& Mihos, J.~C.\ 2015, \aap, 579, L3 


\bibitem[Longobardi et al.(2015)]{2015A&A...579A.135L} Longobardi, A., Arnaboldi, M., Gerhard, O., \& Hanuschik, R.\ 2015, \aap, 579, A135 


\bibitem[Low et al.(1984)]{1984ApJ...278L..19L} Low, F.~J., Young, E., Beintema, D.~A., et al.\ 1984, \apjl, 278, L19 


\bibitem[Malin(1994)]{1994IAUS..161..567M} Malin, D.\ 1994, ``IAU Symposium 161:
  Astronomy from Wide-Field Imaging," eds. H. T. MacGillivray, E. B. Thomson, B. M. Lasker, 
  I. N. Reid,  D. F. Malin, R. M. West \& H. Lorenz.  (Dordrecht: Kluwer),  567 


\bibitem[Malin(1979)]{1979Natur.277..279M} Malin, D.~F.\ 1979, \nat, 277, 279 


\bibitem[Malin \& Hadley(1999)]{1999ASPC..182..445M} Malin, D., \& Hadley, B.\ 1999, 
  ``Galaxy Dynamics," eds. D. R. Merritt, M. Valluri, \& J. A. Sellwood (San Francisco, CA: 
  Astronomical Society of the Pacific), 445



\bibitem[Martel et al.(2012)]{2012ApJ...757...48M} Martel, H., Barai, P., \& Brito, W.\ 2012, \apj, 757, 48 


\bibitem[McLaughlin(1999)]{1999ApJ...512L...9M} McLaughlin, D.~E.\ 1999, \apjl, 512, L9 


\bibitem[McNamara et al.(1994)]{1994AJ....108..844M} McNamara, B.~R., Sancisi, R., Henning, P.~A., \& Junor, W.\ 1994, \aj, 108, 844 


\bibitem[Mei et al.(2007)]{2007ApJ...655..144M} Mei, S., Blakeslee, J.~P., C{\^o}t{\'e}, P., et al.\ 2007, \apj, 655, 144 


\bibitem[Melnick et al.(1977)]{1977MNRAS.180..207M} Melnick, J., Hoessel, J., \& White, S.~D.~M.\ 1977, \mnras, 180, 207 


\bibitem[Mihos(2015)]{2015IAUGA..2247903M} Mihos, C.\ 2015, IAU General Assembly, 22, 2247903 


\bibitem[Mihos(2004)]{2004cgpc.symp..277M} Mihos, J.~C.\ 2004, ``Clusters of Galaxies: 
  Probes of Cosmological Structure  and Galaxy Evolution,'' eds. J.S. Mulchaey, A. Dressler, \& 
  A. Oemler  (Cambridge University Press), 277



\bibitem[Mihos et al.(2015)]{2015ApJ...809L..21M} Mihos, J.~C., Durrell, P.~R., Ferrarese, L., et al.\ 2015, \apjl, 809, L21 


\bibitem[Mihos et al.(2005)]{2005ApJ...631L..41M} Mihos, J.~C., Harding, P., Feldmeier, J., \& Morrison, H.\ 2005, \apjl, 631, L41 


\bibitem[Mihos et al.(2013)]{2013ApJ...764L..20M} Mihos, J.~C., Harding, P., Rudick, C.~S., \& Feldmeier, J.~J.\ 2013, \apjl, 764, L20 


\bibitem[Mihos et al.(2013)]{2013ApJ...762...82M} Mihos, J.~C., Harding, P., Spengler, C.~E., Rudick, C.~S., \& Feldmeier, J.~J.\ 2013, \apj, 762, 82 


\bibitem[Mihos et al.(2009)]{2009ApJ...698.1879M} Mihos, J.~C., Janowiecki, S., Feldmeier, J.~J., Harding, P., \& Morrison, H.\ 2009, \apj, 698, 1879 


\bibitem[Miville-Desch{\^e}nes \& Lagache(2005)]{2005ApJS..157..302M} Miville-Desch{\^e}nes, M.-A., \& Lagache, G.\ 2005, \apjs, 157, 302 


\bibitem[Montes \& Trujillo(2014)]{2014ApJ...794..137M} Montes, M., \& Trujillo, I.\ 2014, \apj, 794, 137 


\bibitem[Montes et al.(2014)]{2014MNRAS.439..990M} Montes, M., Trujillo, I., Prieto, M.~A., \& Acosta-Pulido, J.~A.\ 2014, \mnras, 439, 990 


\bibitem[Moore et al.(1996)]{1996Natur.379..613M} Moore, B., Katz, N., Lake, G., Dressler, A., \& Oemler, A.\ 1996, \nat, 379, 613 


\bibitem[Morrison et al.(1997)]{1997AJ....113.2061M} Morrison, H.~L., Miller, E.~D., Harding, P., Stinebring, D.~R., \& Boroson, T.~A.\ 1997, \aj, 113, 2061


\bibitem[Murante et al.(2004)]{2004ApJ...607L..83M} Murante, G., Arnaboldi, M., Gerhard, O., et al.\ 2004, \apjl, 607, L83


\bibitem[Murante et al.(2007)]{2007MNRAS.377....2M} Murante, G., Giovalli, M., Gerhard, O., et al.\ 2007, \mnras, 377, 2 


\bibitem[Naab et al.(2006)]{2006MNRAS.372..839N} Naab, T., Jesseit, R., \& Burkert, A.\ 2006, \mnras, 372, 839 



\bibitem[Nassau(1945)]{1945ApJ...101..275N} Nassau, J.~J.\ 1945, \apj, 101, 275 


\bibitem[Neill et al.(2005)]{2005ApJ...618..692N} Neill, J.~D., Shara, M.~M., \& Oegerle, W.~R.\ 2005, \apj, 618, 692 


\bibitem[Neugent \& Massey(2010)]{2010PASP..122.1246N} Neugent, K.~F., \& Massey, P.\ 2010, \pasp, 122, 1246 


\bibitem[Ohyama \& Hota(2013)]{2013ApJ...767L..29O} Ohyama, Y., \& Hota, A.\ 2013, \apjl, 767, L29 


\bibitem[Okamura et al.(2002)]{2002PASJ...54..883O} Okamura, S., Yasuda, N., Arnaboldi, M., et al.\ 2002, \pasj, 54, 883 


\bibitem[Oosterloo \& van Gorkom(2005)]{2005A&A...437L..19O} Oosterloo, T., \& van Gorkom, J.\ 2005, \aap, 437, L19 



\bibitem[Patat(2008)]{2008A&A...481..575P} Patat, F.\ 2008, \aap, 481, 575 


\bibitem[Peng et al.(2011)]{2011ApJ...730...23P} Peng, E.~W., Ferguson, H.~C., Goudfrooij, P., et al.\ 2011, \apj, 730, 23 


\bibitem[Peng \& Lim(2016)]{2016ApJ...822L..31P} Peng, E.~W., \& Lim, S.\ 2016, \apjl, 822, L31 


\bibitem[Pfeffer \& Baumgardt(2013)]{2013MNRAS.433.1997P} Pfeffer, J., \& Baumgardt, H.\ 2013, \mnras, 433, 1997 


\bibitem[Pompea et al.(1989)]{1989SPIE..967..236P} Pompea, S.~M., Shepard, D.~F., \& Anderson, S.\ 1989, \procspie, 967, 236


\bibitem[Presotto et al.(2014)]{2014A&A...565A.126P} Presotto, V., Girardi, M., Nonino, M., et al.\ 2014, \aap, 565, A126 


\bibitem[Puchwein et al.(2010)]{2010MNRAS.406..936P} Puchwein, E., Springel, V., Sijacki, D., \& Dolag, K.\ 2010, \mnras, 406, 936 


\bibitem[Purcell et al.(2007)]{2007ApJ...666...20P} Purcell, C.~W., Bullock, J.~S., \& Zentner, A.~R.\ 2007, \apj, 666, 20 


\bibitem[Rines \& Geller(2008)]{2008AJ....135.1837R} Rines, K., \& Geller, M.~J.\ 2008, \aj, 135, 1837 


\bibitem[Romanowsky et al.(2012)]{2012ApJ...748...29R} Romanowsky, A.~J., Strader, J., Brodie, J.~P., et al.\ 2012, \apj, 748, 29 


\bibitem[Rudick et al.(2009)]{2009ApJ...699.1518R} Rudick, C.~S., Mihos, J.~C., Frey, L.~H., \& McBride, C.~K.\ 2009, \apj, 699, 1518 


\bibitem[Rudick et al.(2010)]{2010ApJ...720..569R} Rudick, C.~S., Mihos, J.~C., Harding, P., et al.\ 2010, \apj, 720, 569 


\bibitem[Rudick et al.(2006)]{2006ApJ...648..936R} Rudick, C.~S., Mihos, J.~C., \& McBride, C.\ 2006, \apj, 648, 936 


\bibitem[Rudick et al.(2011)]{2011ApJ...732...48R} Rudick, C.~S., Mihos, J.~C., \& McBride, C.~K.\ 2011, \apj, 732, 48 


\bibitem[Sand et al.(2011)]{2011ApJ...729..142S} Sand, D.~J., Graham, M.~L., Bildfell, C., et al.\ 2011, \apj, 729, 142 

\bibitem[Sandage(1976)]{1976AJ.....81..954S} Sandage, A.\ 1976, \aj, 81, 954 


\bibitem[Sandin(2014)]{2014A&A...567A..97S} Sandin, C.\ 2014, \aap, 567, A97 


\bibitem[Sandin(2015)]{2015A&A...577A.106S} Sandin, C.\ 2015, \aap, 577, A106 


\bibitem[Seigar et al.(2007)]{2007MNRAS.378.1575S} Seigar, M.~S., Graham, A.~W., \& Jerjen, H.\ 2007, \mnras, 378, 1575 


\bibitem[Shapley \& Ames(1930)]{1930BHarO.873....1S} Shapley, H., \& Ames, A.\ 1930, Harvard College Observatory Bulletin, 873, 1 


\bibitem[Slater et al.(2009)]{2009PASP..121.1267S} Slater, C.~T., Harding, P., \& Mihos, J.~C.\ 2009, \pasp, 121, 1267 


\bibitem[Slinglend et al.(1998)]{1998ApJS..115....1S} Slinglend, K., Batuski, D., Miller, C., et al.\ 1998, \apjs, 115, 1 


\bibitem[Smith(1981)]{1981AJ.....86..998S} Smith, H.~A.\ 1981, \aj, 86, 998 


\bibitem[Solanes et al.(2002)]{2002AJ....124.2440S} Solanes, J.~M., Sanchis, T., Salvador-Sol{\'e}, E., Giovanelli, R., \& Haynes, M.~P.\ 2002, \aj, 124, 2440 


\bibitem[Sommer-Larsen(2006)]{2006MNRAS.369..958S} Sommer-Larsen, J.\ 2006, \mnras, 369, 958 


\bibitem[Sommer-Larsen et al.(2005)]{2005MNRAS.357..478S} Sommer-Larsen, J., Romeo, A.~D., \& Portinari, L.\ 2005, \mnras, 357, 478 


\bibitem[Struble(1988)]{1988ApJ...330L..25S} Struble, M.~F.\ 1988, \apjl, 330, L25 


\bibitem[Sun et al.(2010)]{2010ApJ...708..946S} Sun, M., Donahue, M., Roediger, E., et al.\ 2010, \apj, 708, 946 


\bibitem[Thuan \& Kormendy(1977)]{1977PASP...89..466T} Thuan, T.~X., \& Kormendy, J.\ 1977, \pasp, 89, 466 


\bibitem[Toledo et al.(2011)]{2011MNRAS.414..602T} Toledo, I., Melnick, J., Selman, F., et al.\ 2011, \mnras, 414, 602 


\bibitem[Toomre \& Toomre(1972)]{1972ApJ...178..623T} Toomre, A., \& Toomre, J.\ 1972, \apj, 178, 623 


\bibitem[Tovmassian \& Andernach(2012)]{2012MNRAS.427.2047T} Tovmassian, H.~M., \& Andernach, H.\ 2012, \mnras, 427, 2047 


\bibitem[Trujillo \& Fliri(2016)]{2016ApJ...823..123T} Trujillo, I., \& Fliri, J.\ 2016, \apj, 823, 123 


\bibitem[Tutukov \& Fedorova(2011)]{2011ARep...55..383T} Tutukov, A.~V., \& Fedorova, A.~V.\ 2011, Astronomy Reports, 55, 383 


\bibitem[Urban et al.(2011)]{2011MNRAS.414.2101U} Urban, O., Werner, N., Simionescu, A., Allen, S.~W., \& B{\"o}hringer, H.\ 2011, \mnras, 414, 2101 


\bibitem[van Dokkum et al.(2015)]{2015ApJ...798L..45V} van Dokkum, P.~G., Abraham, R., Merritt, A., et al.\ 2015, \apjl, 798, L45 


\bibitem[Veneziani et al.(2010)]{2010ApJ...713..959V} Veneziani, M., Ade, P.~A.~R., Bock, J.~J., et al.\ 2010, \apj, 713, 959 


\bibitem[Ventimiglia et al.(2011)]{2011A&A...528A..24V} Ventimiglia, G., Arnaboldi, M., \& Gerhard, O.\ 2011, \aap, 528, A24 


\bibitem[Vilchez-Gomez et al.(1994)]{1994A&A...283...37V} Vilchez-Gomez, R., Pello, R., \& Sanahuja, B.\ 1994, \aap, 283, 37 


\bibitem[Watkins et al.(2014)]{2014ApJ...791...38W} Watkins, A.~E., Mihos, J.~C., Harding, P., \& Feldmeier, J.~J.\ 2014, \apj, 791, 38


\bibitem[Watkins et al.(2016)]{2016ApJ...826...59W} Watkins, A.~E., Mihos, J.~C., \& Harding, P.\ 2016, \apj, 826, 59 


\bibitem[Watson et al.(2012)]{2012ApJ...754...90W} Watson, D.~F., Berlind, A.~A., \& Zentner, A.~R.\ 2012, \apj, 754, 90 


\bibitem[Weil et al.(1997)]{1997ApJ...490..664W} Weil, M.~L., Bland-Hawthorn, J., \& Malin, D.~F.\ 1997, \apj, 490, 664 


\bibitem[Weil \& Hernquist(1996)]{1996ApJ...460..101W} Weil, M.~L., \& Hernquist, L.\ 1996, \apj, 460, 101 


\bibitem[Weilbacher et al.(2000)]{2000A&A...358..819W} Weilbacher, P.~M., Duc, P.-A., Fritze v.~Alvensleben, U., Martin, P., \& Fricke, K.~J.\ 2000, \aap, 358, 819 


\bibitem[Williams et al.(2007)]{2007ApJ...654..835W} Williams, B.~F., Ciardullo, R., Durrell, P.~R., et al.\ 2007, \apj, 654, 835 


\bibitem[Williams et al.(2007)]{2007ApJ...656..756W} Williams, B.~F., Ciardullo, R., Durrell, P.~R., et al.\ 2007, \apj, 656, 756 


\bibitem[Willman et al.(2004)]{2004MNRAS.355..159W} Willman, B., Governato, F., Wadsley, J., \& Quinn, T.\ 2004, \mnras, 355, 159 


\bibitem[Witt et al.(2008)]{2008ApJ...679..497W} Witt, A.~N., Mandel, S., Sell, P.~H., Dixon, T., \& Vijh, U.~P.\ 2008, \apj, 679, 497-511 


\bibitem[Yang et al.(2002)]{2002ChJAA...2..474Y} Yang, B., Zhu, J., \& Song, Y.-Y.\ 2002, \cjaa, 2, 474 


\bibitem[Yasuda et al.(1997)]{1997ApJS..108..417Y} Yasuda, N., Fukugita, M., \& Okamura, S.\ 1997, \apjs, 108, 417 


\bibitem[York et al.(2000)]{2000AJ....120.1579Y} York, D.~G., Adelman, J., Anderson, J.~E., Jr., et al.\ 2000, \aj, 120, 1579 


\bibitem[Yoshida et al.(2002)]{2002ApJ...567..118Y} Yoshida, M., Yagi, M., Okamura, S., et al.\ 2002, \apj, 567, 118 


\bibitem[Zheng et al.(2015)]{2015ApJ...800..120Z} Zheng, Z., Thilker, D.~A., Heckman, T.~M., et al.\ 2015, \apj, 800, 120 


\bibitem[Zibetti et al.(2005)]{2005MNRAS.358..949Z} Zibetti, S., White, S.~D.~M., Schneider, D.~P., \& Brinkmann, J.\ 2005, \mnras, 358, 949 

\end{thebibliography}
\end{document}